\documentclass[aps,nofootinbib,preprintnumbers,10pt]{revtex4}
\usepackage{eurosym}
\usepackage{CJK}
\usepackage{lipsum}
\usepackage{amsfonts}
\usepackage{amsmath}
\usepackage{amssymb}
\usepackage[english]{babel}
\usepackage{graphicx}
\usepackage{epsfig}
\usepackage{epstopdf}
\usepackage{bm}
\usepackage{float}
\usepackage{verbatim}
\usepackage[utf8]{inputenc}
\usepackage{booktabs}
\usepackage{multirow}
\usepackage{subfig}
\usepackage{slashed}
\usepackage[colorlinks=true,urlcolor=red,citecolor=red]{hyperref}
\usepackage[font=small]{caption}
\usepackage{float}
\usepackage{placeins}
\usepackage[utf8]{inputenc}
\usepackage{bbm}

\newcommand{\mathsym}[1]{{}}

\newcommand{\hs}{\hspace*{0.3cm}}

\newcommand{\be}{\begin{equation}}
\newcommand{\ee}{\end{equation}}
\newcommand{\bea}{\begin{eqnarray}}
\newcommand{\eea}{\end{eqnarray}}
\newcommand{\ben}{\begin{enumerate}}
\newcommand{\een}{\end{enumerate}}
\newcommand{\bit}{\begin{itemize}}
\newcommand{\eit}{\end{itemize}}
\newcommand{\bde}{\begin{widetext}}
\newcommand{\ede}{\end{widetext}}

\newcommand{\crn}{\nonumber \\}

\newcommand{\al}{\alpha}
\newcommand{\la}{\lambda}

\newcommand{\ga}{\gamma}

\newcommand{\fr}{\frac}

\newcommand{\bc}{\begin{center}}
	\newcommand{\ec}{\end{center}}
\newcommand{\Ga}{\Gamma}

\newcommand{\ep}{\epsilon}

\newcommand{\ka}{\kappa}

\DeclareUnicodeCharacter{2212}{-}

\input{tcilatex}
\begin{document}
	\title{Flavor phenomenology of an 
	extended 2HDM with inverse seesaw mechanism}
	\author{N. T. Duy}
	\email{ntduy@iop.vast.vn (Corresponding author)}
	\affiliation{Institute of Physics, VAST, 10 Dao Tan, Ba Dinh, Hanoi, Vietnam}
	
	\author{A. E. C\'arcamo Hern\'andez}
	\email{antonio.carcamo@usm.cl}
	\affiliation{{Universidad T\'ecnica Federico Santa Mar\'{\i}a, Casilla 110-V, Valpara%
			\'{\i}so, Chile}}
	\affiliation{{Centro Cient\'{\i}fico-Tecnol\'ogico de Valpara\'{\i}so, Casilla 110-V,
			Valpara\'{\i}so, Chile}}
	\affiliation{{Millennium Institute for Subatomic physics at high energy frontier -
			SAPHIR, Fernandez Concha 70  Santiago, Chile}}
	\author{D. T. Huong}
	\email{dthuong@iop.vast.vn}
	\affiliation{Institute of Physics, VAST, 10 Dao Tan, Ba Dinh, Hanoi, Vietnam}
	
	\date{\today }
	
	\begin{abstract}
		We perform a detailed and comprehensive study of 
		several flavor physics observables 
		in both lepton and quark sectors within the framework of an extended 2HDM theory where the inverse seesaw mechanism is implemented to generate the SM fermion mass hierarchy. In that theory, the 
		SM gauge symmetry is supplemented by the local $U(1)_X$ and discrete $Z_4\times Z_2$ groups. 
		In particular, we find that the leptonic flavor observables specifically, the branching ratios of charged lepton flavor violating decays $\mu\to e\gamma, \tau\to e(\mu)\gamma$ as well as the anomalous magnetic moments $\Delta a_{e(\mu)}$ 
	strongly depend on the couplings of the neutral CP even(odd) Higgses with exotic charged lepton $E_1$, whereas other observables involving three-body leptonic decays BR($l\to 3l'$), Mu-$\overline{\text{Mu}}$ transition and coherent conversion $\mu \to e $ in a muonic atom are predicted to be less than several orders of magnitude compared to the corresponding experimental limits. Regarding the quark flavor observables, the most stringent limits arising  from the flavor changing neutral currents (FCNC) are those involving the down type quark $d_a\to d_b$ ($a=1,2,3$) transitions and including the branching ratios of inclusive decay BR($\bar{B}\to X_s \gamma$), pure leptonic decay of $B_s$ meson BR$(B_s\to \mu^+\mu^-)$, and meson mixing $ \Delta m_{K,B_s, B_d}$.  Considering the obtained constraints from these observables, the new physics contributions to other processes such as BR$(B_s\to \tau^+\mu^-)$, BR$(B^+\to K^+\tau^+\tau^-)$, BR$(B^+\to K^+\tau^+\mu^-)$, as well as FCCC $b\to c$ transition, specifically LFUV ratios $R_{D^{(*)}}$ are shown to be remarkably small. Regarding the observables in the up type quark transitions, the FCNC top quark processes $t\to u(c)\gamma$ and $t\to u(c)h$ have branching ratios consistent with the experimental limits. Additionally, observables related to SM-like Higgs boson decays, such as LFV decays BR$(h\to \bar{l}'l)$ and modified couplings $a_{h\bar{f}f}$ are also discussed. 
	\end{abstract}
	
	\maketitle
	
	\section{\label{intro}Introduction}
	 Without a doubt,  the Standard Model (SM) is not complete, regardless of its incredible theoretical and phenomenological successes as indicated by the experimental confirmation of its predictions. However, the SM falls short in addressing 
	unsolved puzzles in both theory and experiment
	.  One such puzzle is  
	the unnatural hierarchy observed in the SM fermion masses, which is extended by five orders of magnitude from the electron mass up to the top quark mass. This mass discrepancy widens to thirteen orders of magnitude when considering the neutrino sector. Neutrino oscillation experiments have revealed that active neutrinos have very tiny masses many orders of magnitude smaller than the scale of spontaneous breaking of electroweak symmetry \cite{Super-Kamiokande:1998kpq,Super-Kamiokande:1998qwk,Super-Kamiokande:1998uiq,Super-Kamiokande:2000ywb,Super-Kamiokande:2001ljr,Super-Kamiokande:2004orf,KamLAND:2002uet,KamLAND:2004mhv,SNO:2002hgz,SNO:2002tuh,SNO:2003bmh,SNO:2005oxr}. Moreover, the SM fails to account for the substantial disparity between the patterns of quark and lepton mixings. While the mixing angles are small in the quark sector, suggesting that the quark mixing matrix is very close to the identity matrix, the lepton sector exhibits two large mixing angles are large and one small angle, which implies that the leptonic mixing matrix significantly deviates from the identity matrix \cite{Workman:2022ynf}. Additionally, the SM lacks an explanation of  
		why there are three generations of fermions and the absence of viable Dark Matter (DM) \cite{WMAP:2006bqn,Planck:2013oqw} and is unable to accommodate 
		the observed baryon asymmetry of the Universe (BAU) \cite{Planck:2018vyg,Workman:2022ynf}. Furthermore, there are some recent experimental results for low-energy processes induced by flavor-changing neutral currents (FCNCs), such as $b\to s$, $b\to c$ transitions, including angular observables, branching ratios of $b\to ll$ decays \cite{LHCb:2020gog,LHCb:2020lmf,LHCb:2021zwz,LHCb:2016ykl, LHCb:2021xxq}; as well as 
		lepton flavor universality violating (LFUV) ratios $R_{D^{(*)}}$ \cite{LHCb:2023zxo}, which exhibit 
		tensions compared with their corresponding SM values. These tensions have a confidence level less than $5\sigma$, often called flavor anomalies. \footnote{The most recent reports by LHCb for LFUV ratios in $b\to s$ transitions $R_{K^{(*)}}^{\text{LHCb}}$ show the agreement with SM predictions,  $R_{K^{(*)}}^{\text{LHCb}}\simeq R_{K^{(*)}}^{\text{SM}}\simeq 1$ \cite{LHCb:2022qnv,LHCb:2022vje}, therefore the signals of LFUV in $b\to s$ presently disappear.} Therefore, these signals could potentially indicate the emergence of new physics (NP). However, it is important to emphasize that these tensions are still controversial since they can be triggered by the lack of efficient calculation for the long distance (LD) effect in SM or issues related to the analysis background of the experiment \cite{Greljo:2022jac}. In general, we need more efforts to improve the LD calculation in SM for these processes and the analysis technique in measurements to confirm or exclude the possibility of NP in these $b\to s$ or $b\to c$ observables. In this work, we choose 
		the scenario where 
		these tensions are potentially caused by NP interactions.
	
	To address 
	the aforementioned issues, many NP models have been proposed 
	by extending the SM 
	including enlarged symmetries,  \cite{Boucenna:2016wpr},\cite{Frampton:1992wt,Foot:1992rh,Singer:1980sw, Dong:2006mg,Dong:2013wca,Dong:2005wgt,Dong:2017zxo}, expanding particle content  \cite{Dorsner:2016wpm,Bauer:2015knc,Fajfer:2015ycq,Barbieri:2015yvd,Becirevic:2016yqi,Hiller:2016kry,Crivellin:2017zlb,Alonso:2015sja}. Among these NP models, the Two-Higgs-Doublet model (2HDM) \cite{Lee:1973iz, Branco:2011iw,Crivellin:2019dun,Camara:2020efq} is one of the most motivated BSM theories. 
Recently, the authors in Ref. \cite{Hernandez:2021xet} have extended the 2HDM with the inclusion of the $U(1)_X\times Z_4\times Z_2$ symmetry and new particles including gauge singlet scalars, charged vector-like fermions, right-handed Majorana neutrinos. In that theory, the local $U(1)_X$ and the discrete $Z_4$ symmetries are spontaneously broken, whereas the $Z_2$ symmetry is preserved. Within this setup, the model can explain the hierarchy of SM fermion masses via the inverse seesaw mechanism which is implemented in both charged fermion and neutrino sectors. Specifically, 
	the second and first families of SM charged fermions obtain masses from an inverse seesaw mechanism at the tree level and one-loop level, respectively.  
	On the other hand, one of the Higgs doublets, namely $\phi_1$ generates the top quark mass, whereas the other one, i.e., $\phi_2$ contributes to the generation of the masses of the bottom quark and tau lepton. 
	The tiny masses of active neutrinos are caused by an inverse seesaw mechanism at a two-loop level, whose radiative nature is ensured by the preserved $Z_2$ symmetry. To the best of our knowledge, this model is the first one explaining the mass hierarchy of the charged fermion sector by an inverse seesaw mechanism.  Besides, the model provides stable scalar and fermionic DM candidates due to the preserved $Z_2$ symmetry. Furthermore, BAU is discussed via the leptogenesis mechanism. The flavor observables data such as meson oscillations $\Delta m_{K, B_s, B_d}$ and electron (muon) anomalous magnetic moments $\Delta a_{e},\Delta a_{\mu}$ are shown to be consistent with the model predictions. 
	
	In the model, it is important to note that both charged and neutral CP-even (odd) Higgs bosons couple with the SM and new fermions at the tree level. Furthermore, non-universal assignments between the (first and second) and third quark generations exist under the extended local $U(1)_X $symmetry. This implies  
	the existence of tree level 
	FCNC interactions between the SM up (down) type quarks, mediated by the exchange of the 
	new neutral gauge boson $Z'$. Consequently, the model provides rich sources for numerous flavor-violating processes in both lepton and quark sectors at both tree and loop levels. It is worth noting that in Ref. \cite{Hernandez:2021xet}, the authors just considered a few flavor phenomenological studies within simplified assumptions of the scalar spectrum and couplings. 
		
	In this study, we aim at investigating a variety of flavor phenomenologies related to both leptonic and quark sectors with all possible NP contributions associated with heavy neutral gauge boson $Z'$ and several Higgs bosons (CP-even, CP-odd, and charged Higgs). For the lepton flavor observables, we study the lepton flavor violating (LFV) decays of SM-like Higgs boson $h\to \bar{l}'l$ as well as the modified couplings between $h$ and SM fermions $a_{h\bar{f}f}=\fr{(a_{h\bar{f}f})_{\text{theory}}}{(a_{h\bar{f}f})_{\text{SM}}}$. We investigate the branching ratios for LFV decays of charged leptons BR$(l\to l'\ga)$,  three body leptonic decays $(l\to 3l')$, as well as the transition probability between the muonium (\text{Mu} :$\mu^+e^-$) and antimuonium $(\overline{\text{Mu}}: \mu^-e^+)$, and $\mu \to e $ conversion in a muonic atom in detail. In addition, we revise the anomalous magnetic moments for electron and muon $\Delta a_{e(\mu)}$ with all NP contributions. It should be noted that in the previous work \cite{Hernandez:2021xet}, $\Delta a_{e(\mu)}$ was only studied under a simplifying benchmark scenario, including few scalar contributions. Moving to the quark sector, we focus on the analysis of flavor observables related with FCNCs in down-type quark transitions. These contain the branching ratios of leptonic decay BR$(B_s\to \mu^+\mu^-)$, BR$(B_s\to\tau^+\mu^-)$, the inclusive decay BR$(\bar{B}\to X_s\ga)$, semileptonic decays BR$(B^+\to K^+\tau^+\tau^-)$, BR$(B^+\to K^+\tau^+\mu^-)$. The meson oscillation parameters $\Delta m_{K,B_s,B_d}$ are also reconsidered with full NP contributions, as well as 
	$\Delta a_{e(\mu)}$. Furthermore, the FCCC $b\to c$ transitions, namely the lepton flavor universality violating (LFUV) ratios $R_{D^{(*)}}$, are of great interest. We also study the up-type quark flavor observables, including FCNC top quark decays $t\to u(c)h$ and radiative decays $t \to u(c)\ga$. We would like to notice that this work is not only the update for flavor phenomenologies studies, which were not fully considered in the previous paper \cite{Hernandez:2021xet}, but also attempts to investigate in much more detail many classes of flavor observables.
	
	The paper is structured as follows: In Sec. \ref{Sec1}, we revisit the particle spectrum of the model.  Next, in Sec .\ref{Sec2}, we obtain the benchmark points that satisfy the lepton masses and mixing spectrum. Sec. \ref{Sec3} investigates the scalar potential spectrum in detail. We provide analytical expressions for lepton and quark flavor observables in the following section, and then carry out the numerical studies in Sec. \ref{Sec5}. Finally, we summarize the main results in the conclusion section.

	\section{The review of model \label{Sec1}}
	
	\label{model} In this section, we briefly review the extended 2HDM model with the universal inverse seesaw mechanism. The model is based on the 
	$SU\left( 3\right) _{C}\times SU\left( 2\right) _{L}\times U\left( 1\right)
	_{Y}\times U\left( 1\right) _{X}$ gauge symmetry supplemented by the discrete group 
	$Z_{4}\times Z_{2}$. The spontaneous breaking of the 
	$SU\left( 3\right) _{C}\times SU\left( 2\right) _{L}\times U\left( 1\right)_{Y}\times U\left( 1\right) _{X}\times Z_4$ symmetry allows the radiative inverse seesaw mechanism at two-loop level, resulting in the tiny masses of active neutrinos, while the $Z_2$ symmetry remains preserved.
	The scalar sector of the model 
consists of two Higgs doublets $\phi_1$ and $\phi_2$, with different charges under $U(1)_X$ and $Z_4$ symmetries, along with 10 gauge singlets. The 
particle spectrum of the model and their transformations under the group $SU\left( 3\right) _{C}\times SU\left(
2\right) _{L}\times U\left( 1\right) _{Y}\times U\left( 1\right) _{X}\times
Z_{2}\times Z_{4}$ are displayed in 
Tables \ref{scalars},\ref{leptons},\ref{quarks}.
	\begin{table}[tbp]
		\begin{tabular}{|c|c|c|c|c|c|c|}
			\hline
			& $SU\left( 3\right) _{C}$ & $SU\left( 2\right) _{L}$ & $U\left( 1\right)
			_{Y}$ & $U\left( 1\right) _{X}$ & $Z_{2}$ & $Z_{4}$ \\ \hline
			$\phi_{1}$ & $1$ & $2$ & $\frac{1}{2}$ & $\frac{1}{3}$ & $0$ & 
			$-1$ \\ \hline
			$\phi_{2}$ & $1$ & $2$ & $\frac{1}{2}$ & $\frac{2}{3}$ & $0$ & 
			$1$ \\ \hline
			$\sigma$ & $1$ & $1$ & $0$ & $\frac{1}{3}$ & $0$ & $-1$ \\ \hline
			$\chi$ & $1$ & $1$ & $0$ & $\frac{2}{3}$ & $0$ & $-2$ \\ \hline
			$\eta$ & $1$ & $1$ & $0$ & $-1$ & $0$ & $-1$ \\ \hline
			$\rho$ & $1$ & $1$ & $0$ & $2$ & $0$ & $0$ \\ \hline
			$\ka$ & $1$ & $1$ & $0$ & $2$ & $0$ & $-2$ \\ \hline
			$S$ & $1$ & $1$ & $0$ & $0$ & $0$ & $1$ \\ \hline
			$\zeta^+_1$ & $1$ & $1$ & $1$ & $\frac{2}{3}$ & $0$ & $1$ \\ 
			\hline
			$\zeta^+_2$ & $1$ & $1$ & $1$ & $1$ & $0$ & $0$ \\ \hline
			$\varphi _{1}$ & $1$ & $1$ & $0$ & $1$ & $1$ & $0$ \\ \hline
			$\varphi _{2}$ & $1$ & $1$ & $0$ & $0$ & $1$ & $1$ \\ \hline
		\end{tabular}
		\caption{Scalar assignments under $SU\left( 3\right) _{C}\times SU\left(
			2\right) _{L}\times U\left( 1\right) _{Y}\times U\left( 1\right) _{X}\times
			Z_{2}\times Z_{4}$.}
		\label{scalars}
	\end{table}
	\begin{table}[tbp]
		\begin{tabular}{|c|c|c|c|c|c|c|}
			\hline
			& $SU\left( 3\right) _{C}$ & $SU\left( 2\right) _{L}$ & $U\left( 1\right)
			_{Y}$ & $U\left( 1\right) _{X}$ & $Z_{2}$ & $Z_{4}$ \\ \hline
			$q_{nL}$ & $3$ & $2$ & $\frac{1}{6}$ & $0$ & $0$ & $-2$ \\ \hline
			$q_{3L}$ & $3$ & $2$ & $\frac{1}{6}$ & $\frac{1}{3}$ & $0$ & $0$ \\ \hline
			$u_{iR}$ & $3$ & $1$ & $\frac{2}{3}$ & $\frac{2}{3}$ & $0$ & $-1$ \\ \hline
			$d_{iR}$ & $3$ & $1$ & $-\frac{1}{3}$ & $-\frac{1}{3}$ & $0$ & $-1$ \\ \hline
			$U_{L}$ & $3$ & $1$ & $\frac{2}{3}$ & $\frac{1}{3}$ & $0$ & $0$ \\ \hline
			$U_{R}$ & $3$ & $1$ & $\frac{2}{3}$ & $\frac{2}{3}$ & $0$ & $-1$ \\ \hline
			$T_{L}$ & $3$ & $1$ & $\frac{2}{3}$ & $-\frac{1}{3}$ & $0$ & $2$ \\ \hline
			$T_{R}$ & $3$ & $1$ & $\frac{2}{3}$ & $-\frac{1}{3}$ & $0$ & $2$ \\ \hline
			$D_{1L}$ & $3$ & $1$ & $-\frac{1}{3}$ & $0$ & $0$ & $-2$ \\ \hline
			$D_{1R}$ & $3$ & $1$ & $-\frac{1}{3}$ & $-\frac{1}{3}$ & $0$ & $-1$ \\ \hline
			$D_{2L}$ & $3$ & $1$ & $-\frac{1}{3}$ & $0$ & $1$ & $-1$ \\ \hline
			$D_{2R}$ & $3$ & $1$ & $-\frac{1}{3}$ & $-\frac{1}{3}$ & $0$ & $0$ \\ \hline
			$B_{L}$ & $3$ & $1$ & $-\frac{1}{3}$ & $\frac{2}{3}$ & $0$ & $0$ \\ \hline
			$B_{R}$ & $3$ & $1$ & $-\frac{1}{3}$ & $\frac{2}{3}$ & $0$ & $0$ \\ \hline
		\end{tabular}%
		\caption{Quark assignments under $SU\left( 3\right) _{C}\times SU\left(
			2\right) _{L}\times U\left( 1\right) _{Y}\times U\left( 1\right) _{X}\times
			Z_{2}\times Z_{4}$. Here $i=1,2,3$ and $n=1,2$.}
		\label{quarks}
	\end{table}
	\begin{table}[tbp]
		\begin{tabular}{|c|c|c|c|c|c|c|}
			\hline
			& $SU\left( 3\right) _{C}$ & $SU\left( 2\right) _{L}$ & $U\left( 1\right)
			_{Y}$ & $U\left( 1\right) _{X}$ & $Z_{2}$ & $Z_{4}$ \\ \hline
			$l_{iL}$ & $1$ & $2$ & $-\frac{1}{2}$ & $-\frac{1}{3}$ & $0$ & $0$ \\ \hline
			$l_{nR}$ & $1$ & $1$ & $-1$ & $-1$ & $0$ & $2$ \\ \hline
			$l_{3R}$ & $1$ & $1$ & $-1$ & $-1$ & $0$ & $-1$ \\ \hline
			$E_{1L}$ & $1$ & $1$ & $-1$ & $-1$ & $0$ & $1$ \\ \hline
			$E_{1R}$ & $1$ & $1$ & $-1$ & $-1$ & $0$ & $-1$ \\ \hline
			$E_{2L}$ & $1$ & $1$ & $-1$ & $1$ & $0$ & $1$ \\ \hline
			$E_{2R}$ & $1$ & $1$ & $-1$ & $1$ & $0$ & $1$ \\ \hline
			$\nu _{iR}^{C}$ & $1$ & $1$ & $0$ & $-\frac{1}{3}$ & $0$ & $-1$ \\ \hline
			$N_{iR}$ & $1$ & $1$ & $0$ & $0$ & $0$ & $-2$ \\ \hline
			$\Psi _{nR}$ & $1$ & $1$ & $0$ & $1$ & $1$ & $2$ \\ \hline
			$\Omega _{nR}$ & $1$ & $1$ & $0$ & $-1$ & $0$ & $-1$ \\ \hline
		\end{tabular}%
		\caption{Lepton assignments under $SU\left( 3\right) _{C}\times SU\left(
			2\right) _{L}\times U\left( 1\right) _{Y}\times U\left( 1\right) _{X}\times
			Z_{2}\times Z_{4}$. Here $i=1,2,3$ and $n=1,2$.}
		\label{leptons}
	\end{table}
	With the previously specified particle content and symmetries, the following quark and leptonic Yukawa interactions arise:
	\bea 
	-\mathcal{L}_{Y}^{\left( q\right) } &=&\sum_{i=1}^{3}y_{i}^{\left( u\right) }%
		\overline{q}_{3L}\widetilde{\phi }_{1}u_{iR}+\sum_{i=1}^{3}y_{i}^{\left(
		d\right) }\overline{q}_{3L}\phi _{2}d_{iR}+\sum_{n=1}^{2}x_{n}^{\left(
		U\right) }\overline{q}_{nL}\widetilde{\phi }_{2}U_{R}+\sum_{n=1}^{2}x_{n}^{%
		\left( D\right) }\overline{q}_{nL}\phi _{1}D_{1R}  \notag \\
	&&+z_{D}\overline{D}_{2L}\sigma D_{2R}+\sum_{i=1}^{3}x_{i}^{\left( u\right) }%
	\overline{U}_{L}\sigma ^{\ast }u_{iR}+z_{T}\overline{U}_{L}\chi T_{R}+z_{U}%
	\overline{T}_{L}\eta U_{R}+m_{T}\overline{T}_{L}T_{R}  \notag \\
	&&+\sum_{i=1}^{3}x_{i}^{\left( d\right) }\overline{D}_{1L}\sigma d_{iR}+z_{B}%
	\overline{D}_{1L}\chi ^{\ast }B_{R}+z_{D}\overline{B}_{L}\eta ^{\ast
	}D_{1R}+m_{B}\overline{B}_{L}B_{R}  \notag \\
	&&+\sum_{i=1}^{3}w_{i}^{\left( u\right) }\overline{D}_{1L}\zeta
	_{1}^{-}u_{iR}+\sum_{i=1}^{3}w_{i}^{\left( d\right) }\overline{U}_{L}\zeta
	_{1}^{+}d_{iR}+H.c. ,  \label{Lyq}
	\eea 
	\bea
	-\mathcal{L}_{Y}^{\left( l\right) } &=&\sum_{i=1}^{3}y_{i}^{\left( l\right) }%
	\overline{l}_{iL}\phi _{2}l_{3R}+\sum_{i=1}^{3}y_{i}^{\left( E\right) }%
	\overline{l}_{iL}\phi _{2}E_{1R}+\sum_{n=1}^{2}x_{n}^{\left( l\right) }%
	\overline{E}_{1L}S^*l_{nR}+y_{E}\overline{E}_{1L}\rho^* E_{2R}+x_{E}\overline{E}%
		_{2L}\ka E_{1R}+m_{E}\overline{E}_{2L}E_{2R}  \notag \\
	&&+\sum_{i=1}^{3}\sum_{j=1}^{3}y_{ij}^{\left( \nu \right) }\overline{l}_{iL}%
	\widetilde{\phi }_{2}\nu _{jR}+\sum_{i=1}^{3}\sum_{n=1}^{2}z_{in}^{\left(
		l\right) }\overline{N_{iR}^{C}}\zeta
	_{2}^{+}l_{nR}+\sum_{i=1}^{3}\sum_{j=1}^{3}y_{ij}^{\left( N\right) }\nu
	_{iR}\sigma ^{\ast }\overline{N_{jR}^{C}}+\sum_{i=1}^{3}\sum_{n=1}^{2}\left(
	x_{N}\right) _{in}\overline{N}_{iR}\Psi _{nR}^{C}\varphi _{1}  \notag \\
	&&+\sum_{n=1}^{2}\sum_{m=1}^{2}\left( x_{\Psi }\right) _{nm}\overline{\Psi }%
	_{nR}\varphi _{2}\Omega _{mR}^{C}+\sum_{n=1}^{2}\sum_{m=1}^{2}\left(
	y_{\Omega }\right) _{nm}\overline{\Omega }_{nR}\Omega _{mR}^{C}\ka^{\ast}+H.c..
	\label{Lyl}
	\eea %
	We would like to note that some of Lagrangian Yukawa terms in previous work \cite{Hernandez:2021xet} are not invariant under $Z_4$ symmetry if the $Z_4$ charges of particles are assigned as shown in this reference. For {example}, the first term in Eq. (\ref{Lyq}) and the fourth, and fifth terms in Eq . (\ref{Lyl}) are not invariant concerning $Z_4$ symmetry.  Therefore, in the present work, we redefine the charges of particle content, as given in the Tables \ref{scalars},\ref{quarks} and \ref{leptons}, so that the structure of the Yukawa Lagrangian terms for quark and lepton is maintained. In particular, the $Z_4$ charge of the new singlet charged scalar $\zeta_1^{\pm}$ is assigned as $+1$, instead of $-1$ as in \cite{Hernandez:2021xet}, thus giving it the same quantum numbers as the charged singlet $\zeta_3^{\pm}$.  Consequently, one of these charged singlets can be removed; for instance, we choose to remove $\zeta_3^{\pm}$. However, the new neutral singlet $\ka$ is proposed to ensure that the fourth and fifth terms in the first line of Eq. (\ref{Lyl}) are invariant. Therefore, the total number of scalar fields remains unchanged, and the results of previous work {remain} valid.  
	
	\section{The charged lepton sector\label{Sec2}}
	In Ref. \cite{Hernandez:2021xet}, a detailed numerical analysis of the quark masses and mixing was performed, and the corresponding 
	benchmark point that successfully reproduces the quark masses and CKM  mixing parameters were 
	provided. However, the benchmark point that successfully accommodates the lepton masses and mixings was not 
	provided in \cite{Hernandez:2021xet}
	 In this section, we perform a numerical analysis of lepton masses and mixings. 
	 From the charged lepton Yukawa interactions, it is found in \cite{Hernandez:2021xet} that the mass matrix for SM charged leptons takes the form:
	\begin{eqnarray}
		\widetilde{M}_{E} &=&C_{E}+\frac{m_{E}}{X_{E}Y_{E}}F_{E}G_{E}^{T}+\Delta
		_{E}.  \label{Ml}
	\end{eqnarray}%
	where 
	\begin{eqnarray}
		M_{E} &=&\left( 
		\begin{array}{ccc}
			C_{E}+\Delta _{E} & F_{E} & 0_{3\times 1} \\ 
			G_{E}^{T} & 0 & X_{E} \\ 
			0_{1\times 3} & Y_{E} & m_{E}%
		\end{array}%
		\right) ,\hspace{0.1cm}\hspace{0.1cm}\left( F_{E}\right) _{i}=y_{i}^{\left(
			E\right) }\frac{v_{2}}{\sqrt{2}},\hspace{0.5cm}\left( G_{E}^{T}\right)
		_{n}=x_{n}^{\left( l\right) }\frac{v_{S}}{\sqrt{2}},\hspace{0.5cm}\left(
		G_{E}^{T}\right) _{3}=0,\notag \\
		C_{E} &=&\left( 
		\begin{array}{ccc}
			0 & 0 & y_{1}^{\left( l\right) } \\ 
			0 & 0 & y_{2}^{\left( l\right) } \\ 
			0 & 0 & y_{3}^{\left( l\right) }%
		\end{array}%
		\right) \frac{v_{2}}{\sqrt{2}},\hspace{0.5cm}\hspace{0.5cm}X_{E}=y_{E}\frac{%
			v_{\rho }}{\sqrt{2}},\hspace{0.5cm}\hspace{0.5cm}Y_{E}=x_{E}\frac{v_{\ka }}{%
			\sqrt{2}},\hspace{0.5cm}\hspace{0.5cm}i=1,2,3,\hspace{0.5cm}\hspace{0.5cm}%
		n=1,2,  \label{ME}
	\end{eqnarray}%
	and 
	\begin{eqnarray}
		\Delta _{E} &=&\frac{m_{\widetilde{E}}}{16\pi ^{2}}\sum_{i=1}^{3}\left( 
		\begin{array}{ccc}
			r_{1i}^{\left( E\right) }w_{1i}^{\left( E\right) } & r_{1i}^{\left( E\right)
			}w_{2i}^{\left( E\right) } & 0 \\ 
			r_{2i}^{\left( E\right) }w_{1i}^{\left( E\right) } & r_{2i}^{\left( E\right)
			}w_{2i}^{\left( E\right) } & 0\\ 
			r_{1i}^{\left( E\right) }w_{3i}^{\left(
				E\right) } & r_{2i}^{\left( E\right) }w_{3i}^{\left(
				E\right) }  & 0%
		\end{array}%
		\right) \frac{m_{H_{i}^{\pm }}^{2}}{m_{H_{i}^{\pm }}^{2}-m_{\widetilde{E}%
			}^{2}}\ln \left( \frac{m_{H_{i}^{\pm }}^{2}}{m_{\widetilde{E}}^{2}}\right). 
		\label{DeltaE}
	\end{eqnarray}%
	The first term in Eq. (\ref{Ml}) provides the dominant contribution to the SM charged lepton mass matrix and arises from the renormalizable Yukawa interaction involving the $SU\left( 2\right)_L $ scalar doublet $\phi_{2}$.  
	resulting in the tau lepton mass. 
	Conversely, the second and third terms of Eq. (\ref{Ml}) stem from the inverse seesaw mechanism at the tree and one loop levels, respectively, contributing to the masses of muon and electron. 
	By solving the eigenvalue problem of the SM charged lepton mass matrix, we successfully reproduce the experimental values for charged lepton masses, \cite{Xing:2020ijf, Workman:2022ynf} with the following benchmark point: 
		\begin{eqnarray}
		v_{S}&\simeq& 6\ \mbox{TeV},\hspace{1cm}v_{\rho }=v_{\ka }\simeq 1\ \mbox{TeV}, \hs 
		m_{E} \simeq 1.12653\ \mbox{GeV},\hspace{1cm}m_{\widetilde{E}}\simeq
		\ 4.243\ \mbox{TeV}, \\
		y_{1}^{\left( l\right) } &\simeq &0.0707,\hspace{0.8cm}y_{2}^{\left( l\right)
		}\simeq -0.0066,\hspace{0.8cm}y_{3}^{\left( l\right) }\simeq 0.0117,\hspace{%
			0.8cm}y_{1}^{\left( E\right) }\simeq 0.086,\hspace{0.8cm}y_{2}^{\left(E\right) }\simeq -0.05,\hspace{0.8cm}y_{3}^{\left( E\right) }\simeq -0.0364
		\notag \\
		x_{1}^{\left( l\right) } &\simeq &0.0645,\hspace{0.8cm}x_{2}^{\left(
			l\right) }\simeq -0.113,\hspace{0.8cm} y_E=x_E\simeq
		1.3131,\hs 
		r_{1j}^{\left( E\right) } \simeq -0.0961,\hspace{1cm}r_{2j}^{\left( E\right)
		}\simeq -0.0331,  \notag \\
		w_{1j}^{\left( E\right) } &\simeq &-0.1,\hspace{1cm}w_{2j}^{\left( E\right)
		}\simeq 0.0538,\hspace{1cm}w_{3j}^{\left( E\right) }\simeq -0.0915.\hspace{1cm}
		\label{benchmarkleptons}
	\end{eqnarray}
	\section{Scalar sector\label{Sec3}}
	The full scalar potential of the model which is invariant under the gauge and discrete symmetries can be split as follows following summation 
	\bea V=V_{\phi}+V_{\text{neutral singlets}}+V_{\text{charged singlets}}+V_{\text{odd $Z_2$ singlets}}+V_{\text{mix}}, \eea
	where $V_{\phi},V_{\text{neutral singlets}},V_{\text{charged singlets}},V_{\text{odd $Z_2$ singlets}},V_{\text{mix}}$ correspond to 
	the scalar potential of two doublets $\phi_{1,2}$, 6 neutral singlets $\sigma,\chi,\eta,\rho,\ka,S$, 2 charge singlets $\zeta_{1,2}^{\pm}$, and two odd $Z_2$ singlets $\varphi_{1,2}$ and mixing between them, respectively. These different contributions are given by: 
	\bea 
	V_{\phi}&=&\sum_{j=1}^2(|\phi_j|^2+\la_j|\phi_j|^4)+\la_{12}|\phi_1|^2|\phi_2|^2+\tilde{\la}_{12}(\phi_1^{\dagger}\phi_2)(\phi_2^{\dagger}\phi_1)+\la^{'}_{12}[\ep_{ab}\ep_{cd}(\phi_1)^a(\phi_2)^b(\phi_1^{\dagger})^c(\phi_2^{\dagger})^d+H.c.], \eea
	\bea V_{\text{neutral singlets}}&=& \sum_{j=\sigma}^{\ka}(\mu_{j}^2|s_j|^2+\la_{j}|s_j|^4)+|\sigma|^2\sum_{j=\chi}^{\ka}\la_{\sigma j}|s_j|^2+|\chi|^2\sum_{j=\eta}^\ka\la_{\chi j}|s_j|^2+|\eta|^2\sum_{j=\rho}^\ka\la_{\eta j}|s_j|^2\crn &&+|\rho|^2\sum_{j=S}^
	\ka\la_{\rho j}|s_j|^2+\la_{ S\ka}|S|^2|\ka|^2, \eea 
	\bea V_{\text{charged singlets}}&=&\sum_{j=1}^{2}[\mu_{cj}^2\zeta_i^{+}\zeta_i^{-}+\la_{cj}(\zeta_i^{+}\zeta_i^{-})^2]+\la_{\zeta_{12}}(\zeta_1^{-}\zeta_1^{+})(\zeta_2^{-}\zeta_{2}^{+}),\eea 
	\bea V_{\text{odd $Z_2$ singlets}}&=&\sum_{j=1}^2(\mu_{oj}^2|\varphi_j|^2+\la_{oj}|\varphi_j|^4)+\tilde{\la}_{o2}\varphi_2^4+2\la_{\varphi_{12}}|\varphi_1|^2|\varphi_2|^2,\eea 
	\bea V_{\text{mix}}&=&\sum_{j=1}^2\sum_{k=1}^2\al_{jk}|\phi_j|^2(\zeta_k^+\zeta_k^-)+\sum_{j=\sigma}^\ka\sum_{k=1}^2\beta_{jk}|s_j|^2(\zeta_k^+\zeta_k^-)+\sum_{j=1}^2\sum_{k=1}^2\gamma_{jk}\varphi_j^2(\zeta_k^+\zeta_k^-)\crn 
	&&+\sum_{j=1}^2\sum_{k=\sigma}^\ka\varkappa_{jk}|\phi_j|^2|s_k|^2+\sum_{j=1}^2\sum_{k=1}^2\varsigma_{jk}|\phi_j|^2||\varphi_k|^2+\sum_{j=1}^2\sum_{k=\sigma}^\ka\varrho_{jk}|\varphi_j|^2|s_k|^2
	\crn&& +B_1(\sigma\chi\eta+H.c.)+B_2(\zeta_1^+\zeta_2^-\sigma+H.c.)+B_3(\ep_{ab}\phi_1^a\phi_2^b\zeta_2^-+H.c.)+B_4(\eta^2\ka+H.c.)+B_5(\sigma^2\chi^*+H.c.)\crn
	&&+ f_1(\ep_{ab}\phi_1^a\phi_2^b\zeta_1^-\sigma^*+H.c.)+f_2(\phi_1^{\dagger}\phi_2\sigma^*S+H.c.)+f_3(\zeta_1^-\zeta_2^+\eta\chi+H.c.)+f_4(\sigma^3\eta+H.c.)\crn
	&&+f_5(\chi^2\sigma^*\eta+H.c.)+f_6(\zeta_1^-\zeta_2^+\chi^*\sigma+H.c.)+f_7(S^4+H.c.)+f_8(\sigma\chi\eta^*\ka^*+H.c.)\crn&&+f_{9}(\rho\ka^*S^2+H.c.)+f_{10}(\rho\ka^*S^{*2}+H.c.)+f_{11}(\rho^2\ka^{*2}+H.c.)+f_{12}(\chi^3\ka^*+H.c.)\crn &&+f_{13}(\varphi_1^*\varphi_2\eta\rho+H.c.),\eea 
	where $s_j$ denotes the neutral singlets $\sigma,\chi,\eta,\rho,S,\ka$, respectively. It is important to note that the couplings $B_i, (i=1,...,5)$ have mass dimension, whereas $f_{j}, (j=1,...,13)$ are dimensionless.
	Expanding the even $Z_2$ Higgs multiplets around their minimum, 
	excepting $\varphi_{1,2}$ since they are charged under $Z_2$ symmetry, we have 
	\bea &&\phi_{1,2}=\left( \begin{array}{c }
		\phi_{1,2}^{+} \\ 
		\fr{v_{1,2}+\phi^0_{1,2R}+i\phi^0_{1,2I}}{\sqrt{2}}\\ \end{array}\right), \hs  s=\fr{v_s+s_R+is_I}{\sqrt{2}}, \hs s=\sigma,..,\ka, \crn,   
	&& \varphi_{1,2}=\fr{\varphi_{9,10R}+i\varphi_{9,10I}}{\sqrt{2}}\eea 
	From the minimization conditions of the scalar potential, we obtain the following relations
	\bea
	&& \mu_{1}^2=-\fr{1}{2}\left(\fr{f_2v_2v_Sv_{\sigma}}{v_1}+2\la_1v_1^2+(\tilde{\la}_{12}+\la'_{12})v_2^2+\sum_{j=\sigma}^S\varkappa_{1j}v_j^2\right) ,\crn
	&& \mu_{2}^2=-\fr{1}{2}\left(\fr{f_2v_1v_Sv_{\sigma}}{v_2}+2\la_2v_2^2+(\tilde{\la}_{12}+\la'_{12})v_1^2+\sum_{j=\sigma}^S\varkappa_{2j}v_j^2\right),\crn 
	&& \mu_{\sigma}^2=-\fr{1}{2}\left(\fr{f_2v_2v_1v_S}{v_{\sigma}}+3f_4v_{\eta}v_{\sigma}+2\la_{\sigma}v_{\sigma}^2+\fr{v_{\eta}v_{\chi}(\sqrt{2}B_1+f_8v_{\ka}+f_5v_{\chi})}{v_{\sigma}}+2\sqrt{2}B_5v_{\chi}+\sum_{j=1}^2\varkappa_{j\sigma}v_{j}^2+\sum_{j=\chi,\eta,\rho,S,\ka}\la_{\sigma j}v_j^2\right),\crn 
	&& \mu_{\eta}^2=-\fr{1}{2}\left(2\sqrt{2}B_4v_{\ka}+\fr{f_4v_{\sigma}^3}{v_{\eta}}+2\la_{\eta}v_{\eta}^2+\la_{\sigma\eta}v_{\sigma}^2+\fr{v_{\sigma}v_{\chi}(\sqrt{2}B_1+f_8v_{\ka}+f_5v_{\chi})}{v_{\eta}}+\sum_{j=1}^2\varkappa_{j\eta}v_{j}^2+\sum_{j=\sigma,\chi,\rho,S,\ka}^S\la_{\eta j}v_j^2\right),\crn 
	&& \mu_{\chi}^2=-\fr{1}{2}\left(\fr{\sqrt{2}v_{\sigma}(B_1v_{\eta}+B_5v_{\sigma})}{v_{\chi}}+\fr{f_8v_{\sigma}v_{\eta}v_{\ka}}{v{\chi}}+2f_5v_{\eta}v_{\sigma}+3f_{12}v_{\ka}v_{\chi}+2\la_{\chi}v_{\chi}^2+\sum_{j=1}^2\varkappa_{j\chi}v_{j}^2+\sum_{j=\sigma,\eta,\rho,S,\ka}\la_{\chi j}v_j^2\right),\crn 
	&& \mu_{\rho}^2=-\fr{1}{2}\left(\fr{(f_{10}+f_9)v_S^2v_{\ka}}{v_{\rho}}+2\la_{\rho}v_{\rho}^2+2f_{11}v_{\ka}^2+\sum_{j=1}^2\varkappa_{j\rho}v_{j}^2+\sum_{j=\sigma,\chi,\eta,S,\ka}\la_{\rho j}v_j^2\right),\crn 
	&&  \mu_{S}^2=-\fr{1}{2}\left(\fr{f_2v_1v_{2}v_{\sigma}}{v_{S}}+2v_S^2(\la_S+2f_7)+2(f_9+f_{10})v_{\rho}v_{\ka}+\sum_{j=1}^2\varkappa_{jS}v_{j}^2+\sum_{j=\sigma,\chi,\eta,\rho,\ka}\la_{Sj}v_{j}^2\right),\crn 
	&&  \mu_{\ka}^2=-\fr{1}{2}\left(\fr{\sqrt{2}B_4v_{\eta}^2}{v_{\ka}}+\fr{(f_{10}+f_9)v_S^2v_{\rho}}{v_{\ka}}+\fr{f_8v_{\eta}v_{\sigma}v_{\chi}+f_{12}v_{\chi}^3}{v_{\ka}}+2\la_{\ka}v_{\ka}^2+2f_{11}v_{\rho}^2+\sum_{j=1}^2\varkappa_{j\ka}v_{j}^2+\sum_{j=\sigma,\chi,\eta,\rho,S}\la_{\ka j}v_{j}^2\right). \eea 
	\subsection{Charged sector \label{charged_scalar}}
	From the scalar potential and taking into account its minimization conditions, we find that 
	the squared mass 
	matrix for the electrically charged scalar fields in the basis $(\phi_1^+ \ \phi_2^+\ \zeta_1^+\ \zeta_2^+)^T$ is given by: 
	\bea 
	M_{\text{c}}^2=\left( 
	\begin{array}{cccc}
		-\fr{v_2(f_2v_Sv_{\sigma}+v_1v_2(\tilde{\la}_{12}-2\la'_{12}))}{2v_1}& \fr{f_2v_Sv_{\sigma}+v_1v_2(\tilde{\la}_{12}-2\la'_{12})}{2} & \fr{f_1v_2v_{\sigma}}{2} &\fr{B_3v_2}{\sqrt{2}} \\ 
		\fr{f_2v_Sv_{\sigma}+v_1v_2(\tilde{\la}_{12}-2\la'_{12})}{2} & -\fr{v_1(f_2v_Sv_{\sigma}+v_1v_2(\tilde{\la}_{12}-2\la'_{12}))}{2v_2} & -\fr{f_1v_1v_{\sigma}}{2}& -\fr{B_3v_1}{\sqrt{2}}\\ 
		\fr{f_1v_2v_{\sigma}}{2} &-\fr{f_1v_1v_{\sigma}}{2}& (M_{\text{c}}^2)_{33}& \fr{\sqrt{2}B_2v_{\sigma}+f_3v_{\eta}v_{\chi}+f_6v_{\sigma}v_{\chi}}{2}\\ 
		\fr{B_3v_2}{\sqrt{2}}  &	-\fr{B_3v_1}{\sqrt{2}} & \fr{\sqrt{2}B_2v_{\sigma}+f_3v_{\eta}v_{\chi}+f_6v_{\sigma}v_{\chi}}{2} & (M_{\text{c}}^2)_{44}
	\end{array}\right) \label{mHcharged}, 
	\eea
	where the $33$ and $44$ matrix elements are defined as 
	\bea &&  (M_{\text{c}}^2)_{33}=\mu_{\zeta_1}^2+\fr{1}{2}(\sum_{j=1}^2\al_{j1}v_j^2+\sum_{j=\sigma}^{\ka}\beta_{j1}v_j^2),\crn
	&&  (M_{\text{c}}^2)_{44}=\mu_{\zeta_2}^2+\fr{1}{2}(\sum_{j=1}^2\al_{j2}v_j^2+\sum_{j=\sigma}^{\ka}\beta_{j2}v_j^2).
	\eea
	This matrix contains one vanishing eigenvalue,  corresponding to  
	 the SM Goldstone bosons $G_{W^{\pm}}$ associated with the longitudinal components of the $W_{\mu}^{\pm}$ bosons. The remaining eigenstates 
	are massive and depend on the new physics scales $v_{\sigma,.., S}$, which generally have very complicated forms. However, we can use the perturbation method to approximately diagonalize this matrix in the limit $v_{1,2}\ll v_{\sigma,...,\ka}$ and $B_i\sim \mathcal{O}(10^{-3}-10^{-2})\  \text{TeV} \ll v=\sqrt{v_1^2+v_2^2}\sim 0.246$ TeV, $f_i \sim\mathcal{O}(10^{-3}-10^{-2})\ll 1$, {assuming} that all mass dimension(less) couplings $B_i$ and $f_i$ are the same for different indices $i$. Additionally, we assume that the couplings between the doublet scalars $\phi_{1,2}$ with charged singlets $\zeta_{1,2}^{\pm}$ satisfy 
	$\al_{11}\sim\al_{12},\al_{21}\sim\al_{22}$.  Therefore, the original matrix can be split into two parts such as $M_c^2=M_{c^0}^2+M_{c^p}^2$ where $M_{c^0}^2$ is the leading matrix obtained in the limit $B,v\to 0$, and inversely $M_{c^p}^2 $ contains $B,v$ terms.
    At the first-order in $\mathcal{O}(v^2/v_{\sigma,..,\ka}^2)$, we find that the physical masses and states of electrically charged scalars are given as follows:
	\bea 
	m^2_{G^{\pm}_W}&=& 0, \hs 
	m_{H^{\pm}}^2\simeq -\fr{fv_Sv_{\sigma}}{s_{2\al}}-\fr{v^2(\tilde{\la}_{12}-2\la'_{12})}{2},\\ 
	m_{H_{1,2}^{\pm}}^2&\simeq &\fr{1}{2}\left\{\left(\mu_{\zeta_1}^2+\mu_{\zeta_2}^2+\sum_{j=\sigma}^{\ka}\fr{(\beta_{j1}+\beta_{j2})v_j^2}{2}\right) \right .\crn  && \left  . \mp\sqrt{\left(\mu_{\zeta_1}^2-\mu_{\zeta_2}^2+\sum_{j=\sigma}^{\ka}\fr{(\beta_{j1}-\beta_{j2})v_j^2}{2}\right)^2+[fv_{\chi}(v_{\eta}+v_{\sigma})]^2}\right\}\crn && +\fr{v^2[\al_{12}+\al_{22}+(\al_{12}-\al_{22})c_{2\al}]}{4}\pm\fr{Bv_{\sigma}s_{2\theta}}{\sqrt{2}},
	\eea 
	where the coupling $f$ satisfies $f<0$ to ensure that the squared masses are positive. The physical eigenstates are related to the gauge eigenstates of charged scalars by the following relation: 
	\bea 
	(G^{\pm}_W \ H^{\pm}\ H_1^{\pm}\ H_2^{\pm})^T=U_c(\phi_1^{\pm} \ \phi_2^{\pm} \ \zeta_1^{\pm} \ \zeta_2^{\pm})^T, \label{phys_gauge_Mc}
	\eea 
	where the mixing matrix $U_c$ is given by 
	\bea 
	U_c&\simeq &\left( \begin{array}{cccc }
		c_{\al}& -s_{\al} &0&0\\ 
		s_{\al}&c_{\al}&\fr{v[fv_{\sigma}(\ep_1+\ep_2)-(\ep_1-\ep_2)(fv_{\sigma}c_{2\theta}+\sqrt{2}Bs_{\theta})]}{4\ep_1\ep_2}&\fr{v[\sqrt{2}B(\ep_1+\ep_2)+(\ep_1-\ep_2)(-fv_{\sigma}s_{2\theta}+\sqrt{2}Bc_{2\theta})]}{4\ep_1\ep_2}\\
		
		\fr{vs_{\al}(fv_{\sigma}c_{\theta}+\sqrt{2}Bs_{\theta})}{2\ep_1} & -\fr{vc_{\al}(fv_{\sigma}c_{\theta}+\sqrt{2}Bs_{\theta})}{2\ep_1}& c_{\theta}-\fr{Bv_{\sigma}c_{2\theta}s_{\theta}}{\sqrt{2}(\ep_1-\ep_2)}&s_{\theta}+\fr{Bv_{\sigma}c_{2\theta}c_{\theta}}{\sqrt{2}(\ep_1-\ep_2)}\\ 	 
\fr{vs_{\al}(-fv_{\sigma}s_{\theta}+\sqrt{2}Bc_{\theta})}{2\ep_2} & \fr{vc_{\al}(fv_{\sigma}s_{\theta}-\sqrt{2}Bc_{\theta})}{2\ep_2}& -s_{\theta}-\fr{Bv_{\sigma}c_{2\theta}c_{\theta}}{\sqrt{2}(\ep_1-\ep_2)}& c_{\theta}-\fr{Bv_{\sigma}c_{2\theta}s_{\theta}}{\sqrt{2}(\ep_1-\ep_2)}\end{array}\right), 
	\eea 
	and the mixing angles $\al,\theta$ and terms $\ep_1,\ep_2$ 
	defined as
	\bea t_{\al}&=&\fr{v_2}{v_1}, \hs t_{2\theta}=\fr{fv_{\chi}(v_{\eta}+v_{\sigma})}{\mu_{\zeta_1}^2-\mu_{\zeta_2}^2+\sum_{j=\sigma}^{\ka}(\beta_{j1}-\beta_{j2})v_j^2/2}, \\
	\ep_{1,2}&=&m_{H_{1,2}^{\pm}}^2-m_{H^{\pm}}^2=\fr{1}{2}\left\{\left(\mu_{\zeta_1}^2+\mu_{\zeta_2}^2+\sum_{j=\sigma}^{\ka}\fr{(\beta_{j1}+\beta_{j2})v_j^2}{2}\right) \right .\crn  && \left  . \mp\sqrt{\left(\mu_{\zeta_1}^2-\mu_{\zeta_2}^2+\sum_{j=\sigma}^{\ka}\fr{(\beta_{j1}-\beta_{j2})v_j^2}{2}\right)^2+[fv_{\chi}(v_{\eta}+v_{\sigma})]^2}\right\}+\fr{fv_Sv_{\sigma}}{s_{2\al}}. \label{Uc_par}
	\eea 
	\subsection{CP odd sector\label{CPodd_scalar}}
	The squared mass 
	matrix of the	CP-odd scalar fields in the basis $(\phi^0_{1I}\ \phi^0_{2I} \ \sigma_I \ \eta_I \ \chi_I \ \rho_I\ S_I\ \ka_I)^T$ can be written as: 
	\bea 
	 M_{A}^2&=&\left( \begin{array}{cc }
		M^2_{A_1} & (M^2_{A_2})^T \\ 
		M^2_{A_2}& M^2_{A_3}\\ \end{array}\right),  \eea 
	where the 
	submatrices $M^2_{A_{1,2,3}}$ are given by
	\bea 
	&& M_{A_1}^2=
	\left(\begin{array}{cc}
		-\fr{f_2v_2v_Sv_{\sigma}}{2v_1}& \fr{f_2v_Sv_{\sigma}}{2}\\ 
		\fr{f_2v_Sv_{\sigma}}{2}& 	-\fr{f_2v_1v_Sv_{\sigma}}{  2v_2}  \\ 
	\end{array}\right),\hs 
	M_{A_2}^2=\left( 
	\begin{array}{cc}
		-\fr{f_2v_2v_{S}}{2} &\fr{f_2v_1v_{S}}{2}\\ 
		0& 0 \\ 
		0& 0   \\ 
		0& 0 \\
		\fr{f_2v_2v_{\sigma}}{2}  &-\fr{f_2v_1v_{\sigma}}{2} \\ 
		0& 0   \\
	\end{array}\right), \crn 
	&& 
	M_{A_3}^2=\left(\begin{array}{cccccc}
		
		(m_{A_3}^2)_{11}& (m_{A_3}^2)_{12}&(m_{A_3}^2)_{13} &0 &\fr{f_2v_1v_2}{2}& \fr{f_8v_{\eta}v_{\chi}}{2}\\ 
		(m_{A_3}^2)_{12}  &(m_{A_3}^2)_{22}& (m_{A_3}^2)_{23}&0 &0 &(m_{A_3}^2)_{26} \\
		(m_{A_3}^2)_{13}  &(m_{A_3}^2)_{23}&(m_{A_3}^2)_{33} &0 &0 &\fr{f_8v_{\sigma}v_{\eta}+3f_{12}v_{\chi}^2}{2}\\
		0&0& 0 &(m_{A_3}^2)_{44}&(f_{10}-f_{9})v_{S}v_{\ka} &(m_{A_3}^2)_{46}\\
		\fr{f_2v_1v_2}{2}  &0 & 0 &(f_{10}-f_{9})v_{S}v_{\ka} &(M_{A_3}^2)_{55} &(-f_{10}+f_{9})v_{S}v_{\rho}\\
		\fr{f_8v_{\eta}v_{\chi}}{2}&(M_{A_3}^2)_{26}& \fr{f_8v_{\sigma}v_{\eta}+3f_{12}v_{\chi}^2}{2}&(m_{A_3}^2)_{46} &(-f_{10}+f_{9})v_{S}v_{\rho}\ &(m_{A_3}^2)_{66}
	\end{array}\right), \label{mCPodd}
	\eea
	with some matrix elements of $M_{A_3}^2$ 
	defined as follows
	\bea
	&& (m_{A_3}^2)_{11}=-\fr{f_2v_1v_2v_S+9f_4v_{\eta}v_{\sigma}^2+4\sqrt{2}B_5v_{\sigma}v_{\chi}+v_{\eta}v_{\chi}(\sqrt{2}B_1+f_8v_{\ka}+f_5v_{\chi})}{2v_{\sigma}}, \crn
	&&  (m_{A_3}^2)_{12}=\fr{-3f_4v_{\sigma}^2+v_{\chi}(-\sqrt{2}B_1+f_8v_{\ka}+f_5v_{\chi})}{2},\crn 
	&&  (m_{A_3}^2)_{13}=\sqrt{2}B_5v_{\sigma}-\fr{v_{\eta}(\sqrt{2}B_1+f_8v_{\ka}-2f_5v_{\chi})}{2},\crn 
	&&  (m_{A_3}^2)_{22}=-2\sqrt{2}B_4v_{\ka}-\fr{v_{\sigma}[f_4v_{\sigma}^2+v_{\chi}(\sqrt{2}B_1+f_8v_{\ka}+f_5v_{\chi})]}{2v_{\eta}},\crn 
	&&  (m_{A_3}^2)_{23}=-\fr{v_{\sigma}(\sqrt{2}B_1-f_8v_{\ka}+2f_5v_{\chi})}{2},\crn 
	&&  (m_{A_3}^2)_{26}=-\sqrt{2}B_4v_{\eta}-\fr{f_8v_{\sigma}v_{\chi}}{2},\crn
	&&  (m_{A_3}^2)_{33}=-\fr{\sqrt{2}B_1v_{\eta}v_{\sigma}+f_8v_{\eta}v_{\sigma}v_{\ka}+\sqrt{2}B_5v_{\sigma}^2+4f_5v_{\eta}v_{\sigma}v_{\chi}+9f_{12}v_{\ka}v_{\chi}^2}{2v_{\chi}},\crn
	&& (m_{A_3}^2)_{44}=-2f_{11}v_{\ka}^2-\fr{(f_{10}+f_{9})v_S^2v_{\ka}}{2v_{\rho}},\crn
	&& (m_{A_3}^2)_{46}=2f_{11}v_{\ka}v_{\rho}+\fr{(f_{10}+f_{9})v_{S}^2}{2},\crn
	&& (m_{A_3}^2)_{55}=-8f_7v_S^2-2(f_{10}+f_{9})v_{\ka}v_{\rho}-\fr{f_2v_1v_2v_{\sigma}}{2v_S},\crn
	&&(m_{A_3}^2)_{66}=-\fr{\sqrt{2}B_4v_{\eta}^2+v_{\rho}[(f_{10}+f_{9})v_S^2+4f_{11}v_{\ka}v_{\rho}]+f_8v_{\eta}v_{\sigma}v_{\ka}+f_{12}v_{\chi}^3}{2v_{\ka}}.
	\eea 
	The matrix $M_{A}^2$ will be perturbatively diagonalized by a 
	 similar method used 
	in the diagonalization of $M_{\text{c}}^2$, i.e we split this matrix into leading and perturbative parts as $M_{A}^2=M_{A_0}^2+M_{A_p}^2$, where $M_{A_0}^2$ is obtained in the limit $B,v\to 0$. It should be noted that we have obtained all values of VEVs from the fermion mass and mixing spectrum studies, namely $v_{S}=v_{\sigma}=6$ TeV, $v_{\eta}=v_{\chi}=5$ TeV, $v_{\rho}=v_{\ka}=1$ TeV. Replacing all values of VEVs and at the leading order, the CP odd squared mass matrix 
	can be diagonalized as follows  
	\bea U_{A^0}M_{A^0}^2U^T_{A^0}=(M_{A^0}^{\text{diag}})^2\simeq f\times\text{diag}( 0,0, -30.76,-86,-148.7,-162.5,-206,-293) \ [\text{TeV}^2] \label{CPodd_LOmass},\eea 
	where the mixing matrix $U_{A^0}$ is defined by 
	\bea
	U_{A^0} &\simeq& \left(
	\begin{array}{cccccccc}
		0.993 & 0.122 & 0. & 0. & 0. & 0. & 0. & 0. \\
		0. & 0. & -0.288 & 0.721 & -0.481 & -0.288 & 0. & -0.288 \\
		0. & 0. & -0.11 & 0.354 & -0.053 & 0.911 & 0. & 0.174 \\
		0. & 0. & -0.002 & 0.535 & 0.771 & -0.215 & 0. & 0.271 \\
		-0.122 & 0.993 & 0. & 0. & 0. & 0. & 0. & 0. \\
		0. & 0. & 0.95 & 0.261 & -0.166 & 0.009 & 0. & -0.03 \\
		0. & 0. & -0.039 & 0.01 & -0.38 & -0.203 & 0. & 0.901 \\
		0. & 0. & 0. & 0. & 0. & 0. & -1. & 0. \\
	\end{array}
	\right). \label{U_A}
	\eea 
	This matrix shows the relations between physical and gauge eigenstates as follows 
	\bea 
(G_Z\ G_{Z'}\ \mathcal{A}_3\ \mathcal{A}_4 \ \mathcal{A}_5 \ \mathcal{A}_6 \ \mathcal{A}_7 \ \mathcal{A}_8 )^T=U_{A^0}(\phi_{1I} \ \phi_{2I} \ \sigma_{I} \ \eta_I \ \chi_I \ \rho_I \ S_I \ \ka_I)^T. \label{phys_gauge_CPodd} 
	\eea 
	It should be noted that here the squared masses for the 
	CP odd scalars are expressed in units of 
	TeV$^2$. 
	For instance, if $f\sim -10^{-2}$, $B\sim 10^{-3}$ TeV, we obtain the range of masses for the CP odd scalars as $M_A \sim (0,0, 0.554,0.927,1.219,1.27 1.428,1.705)$ TeV. The mass spectrum of CP odd scalars shows that two massless states correspond to two Goldstone bosons $G_Z$ and $G_{Z'}$ associated with the longitudinal components of the SM $Z$ and new heavy neutral $Z'$ gauge bosons, respectively. \footnote{We want to emphasize that the full matrix $M_A^2$ has exactly two vanishing eigenvalues corresponding to two Goldstone bosons $G_{Z, Z'}$. These two massless states will not receive masses even when we consider corrections arising from higher order terms in the perturbative expansion.}.  When considering next-to-leading order via the matrix $M_{A_p}^2\sim \mathcal{O}(Bv)$, the physical squared masses and states of CP odd scalars are slightly modified as follows
	\bea
	 \delta M_{A}^2\simeq B[\text{TeV}]\times \text{diag}(0,0,-2.027,-14.245,0,-19.027,-15.692,0) \ [\text{TeV}].
\eea 
For $B\sim 10^{-3}$ TeV, these corrections for CP odd Higgs squared masses are of the order $\sim \mathcal{O}(10^{-3}-10^{-2})$ TeV$^2$, which is significantly 2 to 3 orders of magnitude lower compared with the leading order results in Eq. (\ref{CPodd_LOmass}). Similarly, the physical states are also slightly shifted 
when considering the first order. Thus, we can safely ignore the very subleading second-order correction 
and use only the leading first-order result.
	
	\subsection{CP even sector \label{CPeven_scalar}}
	The squared mass matrix for CP even scalar fields in the basis $(\phi^0_{1R}\ \phi^0_{2R} \ \sigma_R \ \eta_R \ \chi_R \ \rho_R\ S_R\ \ka_R)^T$ have the following form: 
	\bea 
	M_{\text{S}}^2&=&\left( 
	\begin{array}{cc }
		M^2_{S_1} & (M^2_{S_2})^T \\ 
		M^2_{S_2}& M^2_{S_3}\\ 
	\end{array}\right),\eea 
	where the submatrices $M_{S_{1,2,3}}^2$ are given by:
	\bea 
	M_{S_2}^2&=&\left( 
	\begin{array}{cc}
		\fr{f_2v_2v_S}{2}+\la_{1\sigma}v_1v_{\sigma}& 	\fr{f_2v_1v_S}{2}+\la_{2\sigma}v_2v_{\sigma} \\ 
		\la_{1\eta}v_1v_{\eta}& 	\la_{2\eta}v_2v_{\eta}  \\ 
		\la_{1\chi}v_1v_{\chi} &\la_{2\chi}v_2v_{\chi}\\ 
		\la_{1\rho}v_1v_{\rho}  & \la_{2\rho}v_2v_{\rho} \\
		\fr{f_2v_2v_{\sigma}}{2}+\la_{1S}v_1v_S  & 	\fr{f_2v_1v_{\sigma}}{2}+\la_{2S}v_2v_S   \\
		\la_{1\ka}v_1v_{\ka}  &\la_{2\ka}v_2v_{\ka}  \\
		
	\end{array}\right),\crn 
	M_{S_{1}}^2&=&\left( 
	\begin{array}{cc}
		-\fr{f_2v_2v_Sv_{\sigma}}{2v_1}+2\la_1v_1^2& \fr{f_2v_Sv_{\sigma}}{2}+v_1v_2(\la_{12}+\tilde{\la}_{12}) \\ 
		\fr{f_2v_Sv_{\sigma}}{2}+v_1v_2(\la_{12}+\tilde{\la}_{12})& -\fr{f_2v_1v_Sv_{\sigma}}{2v_2}+2\la_2v_2^2 \\ 
		
	\end{array}\right),\\
	M_{S_3}^2&=&\left( 
	\begin{array}{cccccc}
		(m_{S_3}^2)_{11}	&  (m_{S_3}^2)_{12}& (m_{S_3}^2)_{13}&
		\la_{\sigma\rho}v_{\sigma}v_{\rho} & \la_{\sigma S}v_{\sigma}v_S+\fr{f_2v_1v_2}{2}& \la_{\sigma \ka}v_{\sigma}v_{\ka}+\fr{f_8v_{\eta}v_{\chi}}{2}\\ 
		(m_{S_3}^2)_{12}& 	 (m_{S_3}^2)_{22}  &  (m_{S_3}^2)_{23} & 	\la_{\eta\rho}v_{\eta}v_{\rho} & \la_{\eta S}v_{\eta}v_S&  (m_{S_3}^2)_{26}\\ 
		(m_{S_3}^2)_{13}& (m_{S_3}^2)_{23} & 	(m_{S_3}^2)_{33}& \la_{\chi\rho}v_{\chi}v_{\rho} & \la_{\chi S}v_{\chi}v_S& (m^2_{S_3})_{36}\\ 
		\la_{\sigma\rho}v_{\sigma}v_{\rho} & 	\la_{\eta\rho}v_{\eta}v_{\rho}  & 	\la_{\chi\rho}v_{\chi}v_{\rho} & (m_{S_3}^2)_{44}&  (m_{S_3}^2)_{45}&  (m_{S_3}^2)_{46}\\ 
		\la_{\sigma S}v_{\sigma}v_{S}+\fr{f_2v_{1}v_{2}}{2}  & \la_{\eta S}v_{\eta}v_S  & \la_{\chi S}v_{\chi}v_S &\la_{\rho S}v_{\rho}v_S+(f_9+f_{10})v_Sv_{\ka}& (m_{S_3}^2)_{55}& (m_{S_3}^2)_{56}\\
		\la_{\sigma \ka}v_{\sigma}v_{\ka}+\fr{f_8v_{\eta}v_{\chi}}{2} & (m_{S_3}^2)_{26} &  (m_{S_3}^2)_{36} &  (m_{S_3}^2)_{46} &  (m_{S_3}^2)_{56} &  (m_{S_3}^2)_{66}
	\end{array}\right),\nonumber
	\label{mCPeven}
	\eea
	with some matrix elements of $M_{S_3}^2$ defined as: 
	\bea 
	&& (m_{S_3}^2)_{11}=-\fr{f_2v_1v_2v_S-3f_4v_{\eta}v_{\sigma}^2+v_{\eta}v_{\chi}(\sqrt{2}B_1+f_8v_{\ka}+f_5v_{\chi})-4\la_{\sigma}v_{\sigma}^3}{2v_{\sigma}},\crn 
	&& (m_{S_3}^2)_{12}=-\fr{3f_4v_{\sigma}^2}{2}+\fr{v_{\chi}(\sqrt{2}B_1+f_8v_{\ka}+f_5v_{\chi})}{2}+\la_{\sigma\eta}v_{\sigma}v_{\eta},\crn 
	&& (m_{S_3}^2)_{13}=\fr{B_1v_{\eta}}{\sqrt{2}}+\fr{f_8v_{\eta}v_{\ka}}{2}+\sqrt{2}B_5v_{\sigma}+f_5v_{\eta}v_{\chi}+\la_{\sigma\chi}v_{\sigma}v_{\chi},\crn 
	&& (m_{S_3}^2)_{26}=\sqrt{2}B_4v_{\eta}+\la_{\eta\ka}v_{\eta}v_{\ka}+\fr{f_8v_{\sigma}v_{\chi}}{2},\crn 
	&& (m_{S_3}^2)_{55}=4f_7v_{S}^2+2\la_Sv_S^2-\fr{f_2v_1v_2v_{\sigma}}{2v_S},\crn
	&& (m^{2}_{S_3})_{56}=\la_{\ka S}v_{\ka}v_S+(f_9+f_{10})v_Sv_{\rho},\crn
	&&  (m^{2}_{S_3})_{66}=-\fr{\sqrt{2}B_4v_{\eta}^2+(f_9+f_{10})v_S^2v_{\rho}+f_8v_{\eta}v_{\sigma}v_{\chi}+f_{12}v_{\chi}^3-4\la_{\ka}v_{\ka}^3}{2v_{\ka}}.
	\eea 
	We continue to approximately diagonalize the given CP even squared mass matrix under 
	the assumption that $B\ll v \ll v_{\sigma,...,\ka}$, using a similar procedure employed in the diagonalization of 
	 $M_A^2$ and $M_c^2$. However, this matrix contains more parameters, such as $\la_{ij},\varkappa_{1i,2i}$, 
 compared to $M_A^2$, $M_c^2$, thus we will made 
	 several assumptions regarding these parameters. Our first assumption concerns the $\la_i$ 
	 couplings of four new singlets $i$, $(i=\sigma,...,\ka)$. We consider a benchmark scenario where 
	 $v_{\sigma}=v_S=6$ TeV, $v_{\rho}=v_{\ka}=1$ TeV, $v_{\eta}=v_{\chi}=5$ TeV, 
	 with the following conditions
	\bea  
	\la_{\sigma}=\la_S=\la, \hs \la_{\rho}=\la_{\ka}\sim \la\fr{v_{\rho}}{v_{\sigma}},\hs \la_{\eta}=\la_{\chi}\sim \la\fr{v_{\chi}}{v_{\sigma}}. 
	\eea 
	Furthermore, we set the couplings $\la_{ij}$ between new singlets $i$ and $j$ relating to $\la$ as 
	\bea 
	\la_{ij}\sim \la \mathcal{O}\left(\fr{v_iv_j}{v_{\sigma}^2}\right),\eea
	and for $\varkappa_{1,2i}$ are couplings between two doublets $\phi_{1,2}$ with new singlets $i$, we assume for the sake of simplicity that 
	\bea
	\varkappa_{1,2i}\sim \la_{1,2} \mathcal{O}\left(\fr{v_{1,2}}{v_i}\right) \ll  1.
	\eea  
	Due to these assumptions, the matrix $M_S^2$ can be decomposed as 
	\bea 
	M_S^2=M_{S_0}^2+M_{S_p}^2.
	\eea 
	At the leading order, we obtain the physical masses 
	\bea 
	U_{S^0}M_{S_0}^2U_{S^0}^T&\simeq& f\times\text{diag}(0,-20.95,-44.46,-144.12,-148.73,-221.55,-228.5,-548.54)\ [\text{TeV}^2],\label{CPeven_LOmass}
	\eea 
	where the matrix $U_{S^0}$ is given by 
	\bea 
	U_{S^0}&\simeq &\left(
	\begin{array}{cccccccc}
		0.993 & 0.122 & 0. & 0. & 0. & 0. & 0. & 0. \\
		0. & 0. & -0.259 & -0.192 & 0.204 & 0.805 & 0.378 & 0.253 \\
		0. & 0. & -0.405 & -0.326 & 0.267 & -0.536 & 0.605 & -0.075 \\
		0. & 0. & -0.038 & 0.155 & 0.605 & -0.202 & -0.303 & 0.689 \\
		-0.122 & 0.993 & 0. & 0. & 0. & 0. & 0. & 0. \\
		0. & 0. & 0.545 & -0.739 & -0.18 & -0.074 & 0.023 & 0.343 \\
		0. & 0. & -0.129 & 0.35 & -0.632 & -0.135 & 0.333 & 0.576 \\
		0. & 0. & 0.674 & 0.405 & 0.297 & 0.01 & 0.537 & -0.076 \\
	\end{array}
	\right), \label{U_S}
	\eea 
	and the physical states are determined by
	\bea 
	(H_1\ H_2\ H_3\ H_4 \ H_5 \ H_6\ H_7 \ H_8 )^T=U_{S^0}(\phi_{1R} \ \phi_{2R} \ \sigma_{R} \ \eta_R \ \chi_R \ \rho_R \ S_R \ \ka_R)^T.\label{phys_gauge_CPeven}
	\eea 
	We see that at the leading order, there is a physical state $H_1$ with zero mass which is composed of real components of doublets $\phi_{1R}$ and $\phi_{2R}$. This state will be identified as an SM-like Higgs boson $h$ discovered by the LHC. Furthermore, there are seven heavy CP even scalars with masses that are proportional to coupling $f$ and are on the TeV scale. When considering the next-to-leading order, the SM-like Higgs boson gains a small mass through the perturbative expansion 
	\bea
	\delta m_{H_1\equiv h}^2&\simeq& 0.1175\la_1+2.675\times10^{-5}\la_2+1.773\times10^{-3}(\la_{12}+\bar{\la}_{12}) \ [\text{TeV}^2]\simeq 0.1175\la_1 \ [\text{TeV}^2]. 
	\eea 
	
	We comment that the SM-like Higgs boson mass is at the electroweak scale, and to assign its mass $ 125$ GeV, as measured by the ATLAS and CMS experiments \cite{Workman:2022ynf}, the coupling $\la_1\sim 0.133$. As for the remaining seven heavy scalars $H_{2,..,8}$, their masses are shifted by 
	 \bea \delta m_{H}^2\simeq B [\text{TeV}]\times (-3.803,-3.763,-10.21,0,-13.25,-7.55,4.366) [\text{TeV}].
	 \eea  
	 For $B\sim \mathcal{O}(10^{-3})$ TeV, we obtain $ \delta m_{H}^2\sim \mathcal{O}(10^{-3}-10^{-2})$ TeV$^2$, which much smaller compared with leading order ones given in Eq. (\ref{CPeven_LOmass}). Similarly, their eigenstates 
	  are also slightly changed. 
	 Consequently, we will adopt the leading order results, as in the CP odd scalar sector.

	\section{Flavor phenomenology\label{Sec4}}
With the Yukawa terms shown in Eqs. (\ref{Lyq},\ref{Lyl}), we find that the model under consideration contains several couplings of both charged and neutral CP even(odd) Higgs bosons with SM and new exotic fermions. Besides that, 
the different $U(1)_X$ charges of the third and two first quark generations 
causes quark flavor-changing $\bar{q}_iq_jZ'$ interactions mediated by new neutral gauge boson $Z'$ at tree-level. Consequently, these couplings give rise to 
quark flavor-violating observables at tree or one-loop level.  In this section, we study these processes in detail. 
	
	\subsection{Lepton flavor phenomenology \label{lepton_flavor}}
\begin{figure}[H]
		\centering
		\begin{tabular}{c}
			\includegraphics[width=11cm]{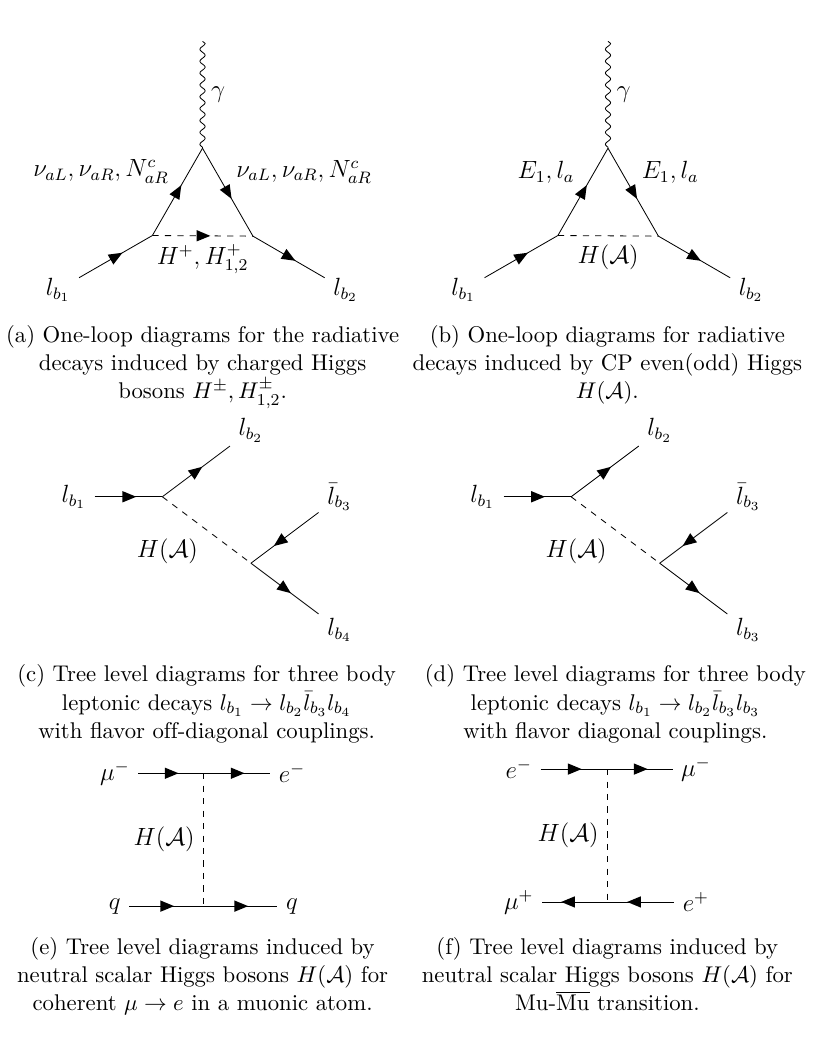}
		\end{tabular}
		\caption{\label{lepton_diagrams}Feynman diagrams for leptonic observables.}
	\end{figure}

	From the leptonic Yukawa interactions 
	of Eq. (\ref{Lyl}), we get the following terms describing the lepton flavor violating (LFV) processes 
	\bea 
	-\mathcal{L}_{\text{lepton}}&=&\sum_{i=1}^{3}y_{i}^{\left( l\right) }%
	\overline{l}_{iL}\fr{v_2+\phi^0_{2R}+i\phi^0_{2I}}{\sqrt{2}}l_{3R}+\sum_{i=1}^{3}y_{i}^{\left( l\right) }%
	\overline{\nu}_{iL}\phi_2^+l_{3R}%
	\notag \\ &&+
	\sum_{i=1}^{3}y_{i}^{\left( E\right)}\overline{l}_{iL}\fr{v_2+\phi^0_{2R}+i\phi^0_{2I}}{\sqrt{2}}E_{1R}+%
	\sum_{n=1}^2x_{n}^{(E)}\overline{E}_{1L}\fr{v_S+S_R-iS_I}{\sqrt{2}}l_{nR} \notag \\
	&&-\sum_{i=1}^{3}\sum_{j=1}^{3}y_{ij}^{\left( \nu \right) }\overline{l}_{iL}\phi_2^-\nu _{jR}+\sum_{i=1}^{3}\sum_{n=1}^{2}z_{in}^{\left(
		l\right) }\overline{N_{iR}^{C}}\zeta
	_{2}^{+}l_{nR}+H.c..
	\eea 
	We work based on physical states of charged leptons $e'_{aL(R)}$, which are related to the flavor states $e_{iL(R)}$ via the following transformations: 
	\bea 
	e_{iL(R)}=(V_{e_{L(R)}})_{ia}e'_{aL(R)},\eea
	where $V_{e_{L(R)}}$ are mixing matrix of the left (right)-handed lepton fields. We want to emphasize that there is barely any mixing between mass states of active neutrinos $\nu_{L}$ and heavy neutrinos $\nu_R, N_R$ via the inverse seesaw mechanism. For simplicity, we assume this mixing is suppressed and ignored, and only active neutrinos are mixed between themselves via the $3\times 3$ matrix $V_{\nu_L}$ which is defined by the relation $V_{\nu_L}^{\dagger}V_{e_L}=V_{\text{PMNS}}$, and the flavor states are related with the physical states 
	as follows
	\bea
	\nu_{iL}=(V_{\nu_L})_{ia}\nu'_{aL}.
	\eea 
	Furthermore, we assume the flavor states of right-handed neutrino $\nu_{iR}$,$N_{iR}$ are related to physical states via matrices $V_{\nu_R}$ and $V_{N_R}$ 
	\bea
	\nu_{iR}=(V_{\nu_R})_{ia}\nu'_{aR}, \hs N_{iR}=(V_{N_R})_{ia}N^{'}_{aR}. 
	\eea 
	Besides that, using the relations between physical and gauge eigenstates of charged, CP odd and CP even scalars shown in Eqs. (\ref{phys_gauge_Mc},\ref{phys_gauge_CPodd},\ref{phys_gauge_CPeven}), 
	the leptonic Yukawa interactions can be rewritten as follows: 
	\bea 
	-\mathcal{L}_{\text{lepton}}&=&\overline{\nu'}_{aL}g_R^{H^{+}\bar{\nu}_{aL}l_b}P_Rl'_bH^+
	+\overline{N^{'C}_{aR}}g_R^{H^+\overline{N}_al_b}P_Rl'_{b}H^{+}+\overline{\nu'}_{aR}g_R^{H^{+}\bar{\nu}_{R}l}P_Ll'_{b}H^{+}\crn 
	&& +\overline{\nu'}_{aL}g_R^{H_{1,2}^{+}\bar{\nu}_{aL}l_b}P_Rl'_bH_{1,2}^+
	+\overline{N^{'C}_{aR}}g_R^{H_{1,2}^+\overline{N}_al_b}P_Rl'_{b}H_{1,2}^{+}+\overline{\nu'}_{aR}g_R^{H_{1,2}^{+}\bar{\nu}_{R}l}P_Ll'_{b}H_{1,2}^{+}\crn 
	&&+\sum_{p=1}^8\overline{E'}_{1}(g_L^{H_p\bar{E}_1l_b}P_L+g_R^{H_p\bar{E}_1l_b}P_R)l'_bH_p+i\sum_{p=3}^8\overline{E'}_{1}(g_L^{\mathcal{A}_p\bar{E}_1l_b}P_L+g_R^{\mathcal{A}_p\bar{E}_1l_b}P_R)l'_b\mathcal{A}_p\crn 
	&&+ \sum_{p=1}^8\overline{l'}_ag_R^{H_p\bar{l}_al_b}P_Rl'_bH_p+i\sum_{p=3}^8\overline{l'}_ag_R^{\mathcal{A}_p\bar{l}_al_b}P_Rl'_b\mathcal{A}_p+ H.c., \label{lag_lepton}
	\eea 
	with the coefficients 
	given explicitly in Appendix \ref{Coefficients_lep_quark}. Here the index $p$ is taken from 3 to 8 for interactions of CP odd Higgs $\mathcal{A}$ since we only care about the physical fields. 
	Let's discuss the roles of each term in the equation provided above. The terms in first line of Eq. (\ref{lag_lepton}) contribute to the radiative decays $l_{b_1}\to l_{b_2}\gamma $ 
	with the one loop level exchange of charged Higgs $H^{+}, H_{1,2}^{+}$ and active neutrinos $\nu_{aL}$, right-handed neutrinos $N^c_{aR},\nu_{aR}$ running in the internal lines, respectively. Furthermore, the second and third lines contain terms describing radiative decays mediated by the neutral CP even (odd) Higgs $H(\mathcal{A})$ and new exotic lepton $E_1$ or ordinary charged leptons $l_a$. The Feynman diagrams for such observables are shown in subfigures (a,b) in Fig . (\ref{lepton_diagrams}). Additionally, the last line provides sources for tree-level lepton flavor violating decays of the SM like Higgs boson $H_1\equiv h$ if index $p=1$,  $h\to \bar{l}_al_b$, and potentially the tree-level three leptonic body decays $l\to 3l'$ via the CP even (odd) Higgs bosons $H(\mathcal{A})$, demonstrates in subfigures (c,d) in Fig. (\ref{lepton_diagrams}). These LFV terms also trigger other processes, namely the coherent $\mu\to e$ conversion in a muonic atom and the transition between muonium Mu to antimuonium $\overline{\text{Mu}}$ state, see in subfigures (e) and (f) of Fig. (\ref{lepton_diagrams}). 
	\subsubsection{LFV decays}
	The radiative cLFV observables $l_{b_1} \to l_{b_2}\gamma$ with $l_{b_1}=\{\mu,\tau\},l_{b_2}=\{e,\mu\} $ and  $l_{b_1}\neq l_{b_2}$ can be described via the effective Hamiltonian contributing by new charged Higgs and neutral CP even (odd) Higgs at the one-loop levels 
	\bea
	\mathcal{H}^{\text{lepton}}_{\text{eff}}=\bar{e}_{b_2}\sigma_{\mu\nu}[(C_L)_{b_2b_1}P_L+(C_R)_{b_2b_1}P_R] e_{b_1}F^{\mu\nu}, \label{leptonfcnc}\eea 
	where the coefficients $(C_{L,R})_{b_2b_1}$ are determined by evaluating one-loop diagrams involving charged Higgs $H^{\pm},H_{1,2}^{\pm}$, neutral CP even (odd) Higgs $H(\mathcal{A})$ , and the SM charged gauge boson $W_{\mu}^{\pm}$ as follows 
	\bea
	(C_{L})_{b_2b_1}&=&(C^{H^{+}\bar{\nu}_{L}l}_{L})_{b_2b_1}+(C^{H^+\bar{N}l}_{L,R})_{b_2b_1}+(C^{H_{1,2}^{+}\bar{\nu}_{L}l}_{L,R})_{b_2b_1}+(C^{H_{1,2}^+\bar{N}l}_{L,R})_{b_2b_1}+(C^{H_{p}(\mathcal{A}_p){\bar{E}_1}l}_{L})_{b_2b_1}\crn &&+(C^{H_{p}(\mathcal{A}_p)\bar{l}l}_{L})_{b_2b_1},\crn 
	(C_{R})_{b_2b_1}&=&(C^{W^+\bar{\nu}_{L}l}_{R})_{b_2b_1}+(C^{H^{+}\bar{\nu}_Rl}_{R})_{b_2b_1}+(C^{H_{1,2}^{+}\bar{\nu}_Rl}_{R})_{b_2b_1}+(C^{H_{p}(\mathcal{A}_p){\bar{E}_1}l}_{R})_{b_2b_1}.  
	\eea  
	Besides, these coefficients are obtained in the limit that the external mass of daughter lepton $m_{b_2}$ is very small in compared with the mass of decaying lepton $m_{b_1}$, i.e $m_{b_2}\ll m_{b_1}$, $m_{b_2}\sim 0$. Therefore, we have the following expressions for $(C_{L,R})_{b_2b_1}$
	\bea 
		(C^{W^+\bar{\nu}_{L}l}_R)_{b_2b_1}&=&\fr{-em_{b_1}}{32\pi^2m^2_{W^{\pm}}}\sum_{a=1}^3 (g_L^{W^+\bar{\nu}_{aL}l_{b_2}})^*g_L^{W^+\bar{\nu}_{aL}l_{b_1}}f^W_{\ga}\left(\fr{m_{\nu_{aL}}^2}{m_{W^+}^2}\right), \crn 
	(C^{H^+\bar{\nu}_{L}l}_L)_{b_2b_1}&=&\fr{-em_{b_1}}{32\pi^2m^2_{H^{\pm}_{1,2}}}\sum_{a=1}^3 (g_R^{H^+\bar{\nu}_{aL}l_{b_2}})^*g_R^{H^+\bar{\nu}_{aL}l_{b_1}}f_{\ga}\left(\fr{m_{\nu_{aL}}^2}{m_{H^+}^2}\right),\crn 
	(C^{H^+\overline{N}l}_L)_{b_2b_1}&=&\fr{-em_{b_1}}{32\pi^2m^2_{H^{\pm}_{1,2}}}\sum_{a=1}^3 (g_R^{H^+\overline{N}_al_{b_2}})^*g_R^{H^+\overline{N}_al_{b_1}}f_{\ga}\left(\fr{m_{N_a}^2}{m_{H^+}^2}\right),\crn 
	(C^{H^+\overline{\nu}_{R}l}_R)_{b_2b_1}&=&\fr{-em_{b_1}}{32\pi^2m^2_{H^{\pm}_{1,2}}}\sum_{a=1}^3 (g_L^{H^+\overline{\nu}_{aR}l_{b_2}})^*g_L^{H^+\overline{\nu}_{aR}l_{b_1}}f_{\ga}\left(\fr{m_{\nu_{aR}}^2}{m_{H^+}^2}\right), \crn
	(C^{H_{1,2}^+\bar{\nu}_{L}l}_L)_{b_2b_1}&=&\fr{-em_{b_1}}{32\pi^2m^2_{H^{\pm}_{1,2}}}\sum_{a=1}^3 (g_R^{H_{1,2}^+\bar{\nu}_{aL}l_{b_2}})^*g_R^{H_{1,2}^+\bar{\nu}_{aL}l_{b_1}}f_{\ga}\left(\fr{m_{\nu_{aL}}^2}{m_{H_{1,2}^+}^2}\right),\crn 
	(C^{H_{1,2}^+\overline{N}l}_L)_{b_2b_1}&=&\fr{-em_{b_1}}{32\pi^2m^2_{H^{\pm}_{1,2}}}\sum_{a=1}^3 (g_R^{H_{1,2}^+\overline{N}_al_{b_2}})^*g_R^{H_{1,2}^+\overline{N}_al_{b_1}}f_{\ga}\left(\fr{m_{N_a}^2}{m_{H_{1,2}^+}^2}\right),\crn 
	(C^{H_{1,2}^+\overline{\nu}_{R}l}_R)_{b_2b_1}&=&\fr{-em_{b_1}}{32\pi^2m^2_{H^{\pm}_{1,2}}}\sum_{a=1}^3 (g_L^{H_{1,2}^+\overline{\nu}_{aR}l_{b_2}})^*g_L^{H_{1,2}^+\overline{\nu}_{aR}l_{b_1}}f_{\ga}\left(\fr{m_{\nu_{aR}}^2}{m_{H_{1,2}^+}^2}\right), \crn (C^{H_{p}\overline{E}_1l}_R)_{b_2b_1}&=&\sum_{p=1}^8\fr{em_{b_1}}{32\pi^2m^2_{H_p}} \left[(g_L^{H_{p}\overline{E}_1 l_{b_2}})^*g_L^{H_{p}\overline{E}_1l_{b_1}}f'_{\ga}\left(\fr{m_{E_1}^2}{m_{H_{p}}^2}\right)+(g_L^{H_p\overline{E}_1l_{b_2}})^*g_R^{H_{p}\overline{E}_1l_{b_1}}\fr{m_{E_1}}{m_{b_1}}h'_{\ga}\left(\fr{m_{E_1}^2}{m_{H_{p}}^2}\right)\right],\crn 
	(C^{H_{p}\overline{E}_1l}_L)_{ab}&=&\sum_{p=1}^8\fr{em_{b_1}}{32\pi^2m^2_{H_p}} \left[(g_R^{H_{p}\overline{E}_1 l_{b_2}})^*g_R^{H_{p}\overline{E}_1l_{b_1}}f'_{\ga}\left(\fr{m_{E_1}^2}{m_{H_{p}}^2}\right)+(g_R^{H_p\overline{E}_1l_{b_2}})^*g_L^{H_{p}\overline{E}_1l_{b_1}}\fr{m_{E_1}}{m_{b_1}}h'_{\ga}\left(\fr{m_{E_1}^2}{m_{H_{p}}^2}\right)\right],\crn 
	(C^{\mathcal{A}_{p}\overline{E}_1l}_R)_{b_2b_1}&=&\sum_{p=3}^8\fr{em_{b_1}}{32\pi^2m^2_{\mathcal{A}_p}} \left[(g_L^{\mathcal{A}_{p}\overline{E}_1 l_{b_2}})^*g_L^{\mathcal{A}_{p}\overline{E}_1l_{b_1}}f'_{\ga}\left(\fr{m_{E_1}^2}{m_{\mathcal{A}_{p}}^2}\right)+(g_L^{\mathcal{A}_{p}\overline{E}_1l_{b_2}})^*g_R^{\mathcal{A}_{p}\overline{E}_1l_{b_1}}\fr{m_{E_1}}{m_{b_1}}h'_{\ga}\left(\fr{m_{E_1}^2}{m_{\mathcal{A}_{p}}^2}\right)\right],\crn 
	(C^{\mathcal{A}_{p}\overline{E}_1l}_L)_{b_2b_1}&=&\sum_{p=1}^8\fr{em_{b_1}}{32\pi^2m^2_{\mathcal{A}_{p}}} \left[(g_R^{\mathcal{A}_{p}\overline{E}_1 l_{b_2}})^*g_R^{\mathcal{A}_{p}\overline{E}_1l_{b_1}}f'_{\ga}\left(\fr{m_{E_1}^2}{m_{\mathcal{A}_{p}}^2}\right)+(g_R^{\mathcal{A}_{p}\overline{E}_1l_{b_2}})^*g_L^{\mathcal{A}_{p}\overline{E}_1l_{b_1}}\fr{m_{E_1}}{m_{b_1}}h'_{\ga}\left(\fr{m_{E_1}^2}{m_{\mathcal{A}_{p}}^2}\right)\right],\crn 
	(C^{H_{p}\overline{l}l}_L)_{b_2b_1}&=&\sum_{a=1}^3\sum_{p=1}^8\fr{em_{b_1}}{32\pi^2m^2_{H_p}} (g_R^{H_{p}\overline{l}_{a} l_{b_2}})^*(g_R^{H_{p}\overline{l}_{a}l_{b_1}})f'_{\ga}\left(\fr{m_{l_a}^2}{m_{H_{p}}^2}\right), \crn 
	(C^{\mathcal{A}_{p}\overline{l}l}_L)_{b_2b_1}&=&\sum_{a=1}^3\sum_{p=3}^8\fr{em_{b_1}}{32\pi^2m^2_{\mathcal{A}_p}} (g_R^{\mathcal{A}_{p}\overline{l}_{a} l_{b_2}})^*(g_R^{\mathcal{A}_{p}\overline{l}_{a}l_{b_1}})f'_{\ga}\left(\fr{m_{l_a}^2}{m_{\mathcal{A}_{p}}^2}\right), 
	\eea 
	where the couplings of SM gauge boson $W $ are given as : $g_L^{W^+\bar{\nu}_{aL}l_{b_{1(2)}}}=\sum_{i=1}^3 (V_{\nu_L}^*)_{ia}(V_{e_L})_{ib_{1(2)}}$. Besides, the loop functions $f^W_{\ga}, h^{'}_{\ga },f^{(')}_{\ga}$ are defined in Appendix \ref{Loopfuncs}. The formulas of the branching ratio of cLFV decays are expressed by  
	\bea
	\text{BR}(l_{b_1}\to l_{b_2} \gamma)&=&\fr{m_{l_{b_1}}^3}{4\pi \Ga_{l_{b_1}}}(|(C_L)_{b_2b_1}|^2+|(C_R)_{b_2b_1}|^2), \eea 
	where $\Ga_{l_{b_1}}$ is the total decay width of decaying lepton $l_{b_1}$. 
	For LFV decays of SM like Higgs boson  $h\to \bar{l}_{a}l_{b}$ , their branching ratio are given as follows 
	\bea 
	\text{BR}(h\to \bar{l}_{a}l_b)=\fr{m_h}{8\pi\Ga^{\text{SM}}_h}(|g_R^{h\bar{l}_al_b}|^2+|(g_R^{h\bar{l}_bl_a})^*|^2),
	\eea 
	where $\Ga^{\text{SM}}_h\simeq 4.1$ MeV is the total width of SM Higgs boson $h$. 
	
	Besides, the LFV couplings in the last line of Eq . (\ref{lag_lepton}) can also cause the three body leptonic decays. The expressions for their branching ratios read 
	\bea 
	\text{BR}(\tau\to3\mu)&=&R\left\{\sum_{p=1}^8\left[2\left|\fr{g_R^{H_p\bar{l}_2l_3}g_R^{H_p\bar{l}_2l_2}}{m_{H_p}^2}\right|^2+2\left|\fr{(g_R^{H_p\bar{l}_3l_2})^*(g_R^{H_p\bar{l}_2l_2})^*}{m_{H_p}^2}\right|^2+\left|\fr{g_R^{H_p\bar{l}_2l_3}(g_R^{H_p\bar{l}_2l_2})^*}{m_{H_p}^2}\right|^2+\left|\fr{(g_R^{H_p\bar{l}_3l_2})^*g_R^{H_p\bar{l}_2l_2}}{m_{H_p}^2}\right|^2\right]\right.\crn && +\left . \sum_{p=3}^8\left[2\left|\fr{g_R^{\mathcal{A}_p\bar{l}_2l_3}g_R^{\mathcal{A}_p\bar{l}_2l_2}}{m_{\mathcal{A}_p}^2}\right|^2+2\left|\fr{(g_R^{\mathcal{A}_p\bar{l}_3l_2})^*(g_R^{\mathcal{A}_p\bar{l}_2l_2})^*}{m_{\mathcal{A}_p}^2}\right|^2+\left|\fr{g_R^{\mathcal{A}_p\bar{l}_2l_3}(g_R^{\mathcal{A}_p\bar{l}_2l_2})^*}{m_{\mathcal{A}_p}^2}\right|^2+\left|\fr{(g_R^{\mathcal{A}_p\bar{l}_3l_2})^*g_R^{\mathcal{A}_p\bar{l}_2l_2}}{m_{\mathcal{A}_p}^2}\right|^2\right]\right\}, \crn 
	\text{BR}(\tau\to\mu e\mu)&=&R\left\{\sum_{p=1}^8\left[2\left|\fr{g_R^{H_p\bar{l}_2l_3}g_R^{H_p\bar{l}_2l_1}}{m_{H_p}^2}\right|^2+2\left|\fr{(g_R^{H_p\bar{l}_3l_2})^*(g_R^{H_p\bar{l}_1l_2})^*}{m_{H_p}^2}\right|^2+\left|\fr{g_R^{H_p\bar{l}_2l_3}(g_R^{H_p\bar{l}_1l_2})^*}{m_{H_p}^2}\right|^2+\left|\fr{(g_R^{H_p\bar{l}_3l_2})^*g_R^{H_p\bar{l}_2l_1}}{m_{H_p}^2}\right|^2\right]\right.\crn && +\left . \sum_{p=3}^8\left[2\left|\fr{g_R^{\mathcal{A}_p\bar{l}_2l_3}g_R^{\mathcal{A}_p\bar{l}_2l_1}}{m_{\mathcal{A}_p}^2}\right|^2+2\left|\fr{(g_R^{\mathcal{A}_p\bar{l}_3l_2})^*(g_R^{\mathcal{A}_p\bar{l}_1l_2})^*}{m_{\mathcal{A}_p}^2}\right|^2+\left|\fr{g_R^{\mathcal{A}_p\bar{l}_2l_3}(g_R^{\mathcal{A}_p\bar{l}_1l_2})^*}{m_{\mathcal{A}_p}^2}\right|^2+\left|\fr{(g_R^{\mathcal{A}_p\bar{l}_3l_2})^*g_R^{\mathcal{A}_p\bar{l}_2l_1}}{m_{\mathcal{A}_p}^2}\right|^2\right]\right\},\crn 
	\text{BR}(\tau\to e\mu\mu)&=&R\sum_{p=1}^8\left[\left|\fr{g_R^{H_p\bar{l}_1l_3}g_R^{H_p\bar{l}_2l_2}}{m_{H_p}^2}+\fr{g_R^{H_p\bar{l}_2l_3}g_R^{H_p\bar{l}_1l_2}}{m_{H_p}^2}\right|^2+\left|\fr{(g_R^{H_p\bar{l}_3l_1})^*(g_R^{H_p\bar{l}_2l_2})^*}{m_{H_p}^2}+\fr{(g_R^{H_p\bar{l}_3l_2})^*(g_R^{H_p\bar{l}_2l_1})^*}{m_{H_p}^2}\right|^2\right .\crn  && \left .+  \left|\fr{(g_R^{H_p\bar{l}_3l_1})^*g_R^{H_p\bar{l}_2l_2}}{m_{H_p}^2}\right|^2+  \left|\fr{(g_R^{H_p\bar{l}_3l_2})^*g_R^{H_p\bar{l}_1l_2}}{m_{H_p}^2}\right|^2\right] \crn &&+ R\sum_{p=3}^8\left[\left|\fr{g_R^{\mathcal{A}_p\bar{l}_1l_3}g_R^{\mathcal{A}_p\bar{l}_2l_2}}{m_{\mathcal{A}_p}^2}+\fr{g_R^{\mathcal{A}_p\bar{l}_2l_3}g_R^{\mathcal{A}_p\bar{l}_1l_2}}{m_{\mathcal{A}_p}^2}\right|^2+\left|\fr{(g_R^{\mathcal{A}_p\bar{l}_3l_1})^*(g_R^{\mathcal{A}_p\bar{l}_2l_2})^*}{m_{\mathcal{A}_p}^2}+\fr{(g_R^{\mathcal{A}_p\bar{l}_3l_2})^*(g_R^{\mathcal{A}_p\bar{l}_2l_1})^*}{m_{\mathcal{A}_p}^2}\right|^2\right .\crn  && \left .+\left|\fr{(g_R^{\mathcal{A}_p\bar{l}_3l_1})^*g_R^{\mathcal{A}_p\bar{l}_2l_2}}{m_{\mathcal{A}_p}^2}\right|^2+  \left|\fr{(g_R^{\mathcal{A}_p\bar{l}_3l_2})^*g_R^{\mathcal{A}_p\bar{l}_1l_2}}{m_{\mathcal{A}_p}^2}\right|^2 \right],
	\eea  
   where $R=\fr{m_{\tau}^5}{1536\pi^3\Ga_{\tau}}$ is the overall factor. Similarly, we can obtain the branching ratios of the other three body leptonic decays such as $\mu\to3e$, $\tau\to 3e$, and $\tau\to e\mu e$ by replacing appropriate indices. 
	
	We would like to emphasize that the model includes 
	contributions to the anomalous magnetic moments for electron or muon, denoted as $\Delta a_{e,\mu}$ respectively, as studied in previous work \cite{Hernandez:2021xet}. However, as aforementioned, the paper considered the observables within a framework of a very simplifying benchmark scenario containing few scalar contributions. Therefore, in this work, we revisit $\Delta a_{e,\mu}$ in a more comprehensive manner, considering all possible contributions. The expressions for $\Delta a_{e,\mu}$ are written as follows 
	\bea 
	\Delta a_{e(\mu)}&=&-\fr{4m_{e\mu}}{e}\text{Re}[(C_R)_{b_1b_1(b_2b_2)}], \crn 
	 (C_R)_{b_1b_1(b_2b_2)}&=&(C^{H^{+}\bar{\nu}_Rl}_R)_{b_1b_1(b_2b_2)}+(C^{H_{1,2}^+\bar{\nu}_Rl}_R)_{b_1b_1(b_2b_2)}+(C^{H(\mathcal{A})ll}_R)_{b_1b_1(b_2b_2)}.
	 \eea 
	
	To close this section, we offer a concise qualitative discussion on how our model impacts the electric dipole moment of the neutron. Following the considerations given in \cite{Jung:2013hka,Logan:2020mdz}, the neutron's electric dipole moment in multi-Higgs doublet models like the one considered in this work originates from various factors. Firstly, there's the tree-level exchange of CP-violating scalars, resulting in four-fermion operators involving both up- and down-type quarks. Secondly, there's the CP-violating three-gluon operator, known as the Weinberg operator, along with Barr-Zee type two-loop diagrams. These contribute to the electric dipole moment and chromoelectric dipole moments of both up- and down-type quarks. Notably, the contributions from the first and third sources are diminished due to the small masses of the light quarks \cite{Jung:2013hka, Logan:2020mdz}. 
	
	Consequently, it is expected that the main contribution to the electric dipole moment of the neutron in the extended 2HDM theory considered in this work will arise from the CP violating two loop level self-gluon trilinear interaction involving the exchange of charged scalars along with top and bottom quarks through virtual processes as in Refs. \cite{Jung:2013hka,Logan:2020mdz}. Thus, the upper limit on the electric dipole moment of the neutron $|d_e|\leqslant 1.1\times 10^{-29}$\textit{e cm} \cite{ACME:2018yjb} can set constraints on the ratio between CP-violating parameter combinations and squared charged scalar masses, as discussed in detail in Ref. \cite{Jung:2013hka}. Some recent detailed studies of the consequences of multi Higgs doublet models in the electric dipole moment of the neutron are done in Refs. \cite{Logan:2020mdz,Eeg:2019eei,Eeg:2016fsy}. A detailed numerical analysis of the constraints arising from the upper experimental bound on the electric dipole moment of the neutron in our model under consideration goes beyond the scope of the present work and is deferred for future work.

	\subsubsection{$\text{Mu}-\overline{\text{Mu}}$ transition}
	The LFV couplings of CP odd(even) Higgs $H(\mathcal{A})$ in the last line of Eq. (\ref{lag_lepton}) also cause another process,  called the muonium (Mu: $\mu^+e^-$) to antimuonium ($\overline{\text{Mu}}$: $\mu^-e^+$) transition. This process can be induced via the following effective Lagrangian 
		\bea 
		\mathcal{L}^{\text{Mu}-\overline{\text{Mu}}}_{\text{eff}}=-\sum_{i=1,2}\fr{\mathcal{G}_i}{\sqrt{2}}\mathcal{Q}_i, 
		\eea   
		where the coefficients and corresponding operators are defined as follows
		\bea 
		\mathcal{Q}_1 &=&[\bar{\mu}(1-\ga_{5})e][\bar{\mu}(1-\ga_{5})e],\hs \mathcal{G}_1=\sum_{p=1}^8\fr{[(g_R^{H_p\bar{l}_1l_2})^*]2}{4\sqrt{2}m_{H_p}^2}+\sum_{p=3}^8\fr{[(g_R^{A_p\bar{l}_1l_2})^*]^2}{4\sqrt{2}m_{\mathcal{A}_p}^2}\crn
		\mathcal{Q}_2 &=&[\bar{\mu}(1+\ga_{5})e][\bar{\mu}(1+\ga_{5})e], \hs \mathcal{G}_2=\sum_{p=1}^8\fr{(g_R^{H_p\bar{l}_2l_1})^2}{4\sqrt{2}m_{H_p}^2}+\sum_{p=3}^8\fr{(g_R^{\mathcal{A}_p\bar{l}_2l_1})^2}{4\sqrt{2}m_{\mathcal{A}_p}^2}.
		\eea 
		The $\text{Mu}-\overline{\text{Mu}}$ transition probability in the presence of external magnetic field $B$ is given by \cite{Fukuyama:2021iyw}
		\bea
		P(\text{Mu}\to \overline{\text{Mu}}) =2\tau^2\left(|c_{0,0}|^2|\mathcal{M}^B_{0,0}|^2+|c_{1,0}|^2|\mathcal{M}^B_{1,0}|^2+\sum_{m=\pm 1}|c_{1,m}|^2\fr{|M_{1,m}|^2}{1+(\tau\Delta E)^2}\right),
		\eea 
		where $\tau\sim 2.2\times 10^{-6} s$  is the expected lifetime of the Mu system. Additionally, $|c_{0,0}|^2=0.32, |c_{1,0}|^2=0.18$ are the population of Mu states. $\Delta E$ is the energy splitting between the $(1,1)$ and $(1,-1)$ states caused by external magnetic field $B$. The factor $X=0.631$ is for a magnetic field $B=0.1$ Tesla. It should be noted that the transition probability for $(1,\pm1)$ states is suppressed for the case $B\geq \mathcal{O}(10^{-6})$ Tesla. The transition probability in this case reads 
		\bea
		P(\text{Mu}\to \overline{\text{Mu}}) &\simeq & 5.74\times 10^{-7}\fr{|\mathcal{G}_1+\mathcal{G}_2|^2}{G_F^2}.
		\eea 
		The PSI experiment for Mu-$\overline{\text{Mu}}$ transition reported that P$(\text{Mu}\to \overline{\text{Mu}}) <8.3\times 10^{-11}$ \cite{Willmann:1998gd}, which implies 
		\bea 
		|\mathcal{G}_1+\mathcal{G}_2|<1.2\times 10^{-2}G_F \label{Mu-AntiMu}.  
		\eea 
		
	\subsubsection{$\mu \to e$ coherent conversion}
	On the one hand, the LFV couplings of neutral Higgs boson $H(\mathcal{A})$ can also induce the coherent $\mu \to e$ conversion in a muonic atom. Specifically, $\mu^-$ is captured by atomic nuclei target and subsequently converts into  $e^-$ without emitting a neutrino due to the influence of the nuclear field, as described in the Feynman diagram (e) of Fig. (\ref{lepton_diagrams}). In this work, the $\mu \to e$ conversion arises from the non-photonic contribution and can be described via the effective Lagrangian at the quark level as follows 
	\bea
	\mathcal{L}^{\mu \to e}_{\text{eff}}=-G_Fm_{\mu}m_N\sum_{q=u,d,s}[C^{q}_{SL}(\bar{e}P_L\mu)+C^{q}_{SR}(\bar{e}P_R\mu)](\bar{q}q)+H.c.,
	\eea 
	where the operators $\bar{q}\ga_5q$ do not contribute to the coherent conversion process \cite{Kitano:2002mt}, therefore we do not include them. $m_{\mu}$ and $m_N$ are the masses of muon and nuclei $N$, respectively. The coefficients $C^{q}_{SL},C^{q}_{SR}$ are given by
	\bea 
	C^{q}_{SL}&=&\sum_{p=1}^8\fr{(g_R^{H_p\bar{l}_2l_1})^*\text{Re}(g_R^{H_p\bar{q}q})}{m_{H_p}^2}+\sum_{p=3}^8\fr{(g_R^{\mathcal{A}_p\bar{l}_2l_1})^*\text{Im}(g_R^{\mathcal{A}_p\bar{q}q})}{m_{\mathcal{A}_p}^2}, \crn  C^{q}_{SR}&=&\sum_{p=1}^8\fr{g_R^{H_p\bar{l}_1l_2}\text{Re}(g_R^{H_p\bar{q}q})}{m_{H_p}^2}+\sum_{p=3}^8\fr{g_R^{\mathcal{A}_p\bar{l}_1l_2}\text{Im}(g_R^{\mathcal{A}_p\bar{q}q})}{m_{\mathcal{A}_p}^2},
	\eea  
	where coefficients $g^{H_p(\mathcal{A}_p)\bar{q}q}, (q=u,d,s)$ are defined below in the Section. (\ref{quark_flavor}). To evaluate the rate of $\mu\to e $ transition in a nuclei, we should transform the above Lagrangian from the quark to nucleon level 
	\bea 
		\mathcal{L}^{\mu N\to e N}_{\text{eff}}=-\sum_{\mathcal{N}=p,n}[C^{\mathcal{N}}_{SL}(\bar{e}P_L\mu)+C^{\mathcal{N}}_{SR}(\bar{e}P_R\mu)](\bar{\psi}_N\psi_N)+H.c.,
	\eea  
	with $\psi_N$ is defined as the nucleon field, whereas the coefficients are rewritten as  
	\bea C^{p(n)}_{SL(R)}&=&\sum_{q=u,d,s}C^q_{SL(R)}f^q_{S_{p(n)}}, \crn 
	f^u_{S_{p(n)}}&=&\fr{m_u}{m_u+m_d}\fr{\sigma_{\pi N}}{m_{p}}(1\pm \xi), \hs 	f^d_{S_{p(n)}}=\fr{m_d}{m_u+m_d}\fr{\sigma_{\pi N}}{m_{p}}(1\mp \xi),\hs 	f^s_{S_{p}}=\fr{m_s}{m_u+m_d}\fr{\sigma_{\pi N}}{m_{p}}y,
	\eea   
	where the nucleon matrix elements $\sigma_{\pi N}=39.8$ MeV, $\xi=0.18$ and $y=0.09$. $m_{u,d,s}$ are the quark masses evaluated at the scale of 2 GeV \cite{Cheng:2012qr, Workman:2022ynf}. Then the branching ratio of $\mu\to e$ conversion in a target of atomic nuclei $N$ is given by 
	\bea
	\text{BR}(\mu N\to e N)&=&\fr{4G_F^2m_{\mu}^7}{\Ga^{N}_{\text{capt}}}\left[|m_pC^{p}_{SR}S^p_N+m_nC^{n}_{SR}S^n_N|^2+ |m_pC^{p}_{SL}S^p_N+m_nC^{n}_{SL}S^n_N|^2\right], \label{mu-e-N}
	\eea 
	where $\Ga_{\text{capt}}^N$ is the total capture rate, $S^{p,n}_N$ are the overlap integrals of atomic nuclei $N$.
	For instance, we consider the $\mu \to e$ transition captured by Gold (Au) nuclei, we have $S^{p}_{\text{Au}}=0.0614, S^{n}_{\text{Au}}=0.0918$ \cite{Kitano:2002mt}, $\Ga_{\text{capt}}^{\text{Au}}\simeq 8.7\times 10^{-18}$ GeV \cite{Suzuki:1987jf}. The predicted branching ratio in Eq. (\ref{mu-e-N}) will be compared to the experimental limit reported by SINDRUM-II \cite{SINDRUMII:2006dvw}. 
	
	All the above-mentioned leptonic observables should be compared with the corresponding upper experimental limits shown in Table. \ref{lepton_exp}. Here we 
	require the consistency with the $3\sigma$ experimentally allowed range 
	for the 
	observable $\Delta a_{e(\mu)}$ by the following reason. The first and second charged fermion families of the model, i.e., the electron and muon receive tree and loop-level masses, respectively, from the inverse seesaw mechanism. This makes the interactions of the first and second generations of fermions with other particles quite suppressed, therefore we will compare the predictions with the $3\sigma$ experimentally allowed ranges of $\Delta a_{e(\mu)}$. For other 
	observables related to electron or muon, we apply this setup. 
	\begin{table}[H]
		\begin{centering}
			\begin{tabular}{|c|c|c|c|}
				\hline
				LFV Observables & Experimental limits  & LFV Observables & Experimental constraints  
				\tabularnewline
				\hline 
				BR$ (\mu \to e\ga)$ & $\leq 4.2\times 10^{-13}$  \cite{SINDRUMII:2006dvw,BaBar:2009hkt,MEG:2016leq} &  BR$(h\to e\mu)$ & $<6.1\times 10^{-5} $ \cite{Workman:2022ynf}
				\tabularnewline
				BR$(\tau\to e\ga)$ & $\leq 3.3\times 10^{-8}$ \cite{SINDRUMII:2006dvw,BaBar:2009hkt,MEG:2016leq} 	 &   BR$(h\to e\tau)$ & $<2.2\times 10^{-3} $ \cite{Workman:2022ynf}
				\tabularnewline
				BR$ (\tau\to \mu\ga)$ & $\leq 4.4\times 10^{-8}$  \cite{SINDRUMII:2006dvw,BaBar:2009hkt,MEG:2016leq}  &  BR$(h\to \mu\tau)$ & $<1.5\times 10^{-3} $ \cite{Workman:2022ynf}
				\tabularnewline
				BR$ (\mu^-\to e^-e^+e^-)$ & $\leq 1.0\times 10^{-12}$  \cite{Workman:2022ynf} &  BR$(\tau^-\to \mu^-e^+\mu^-)$ & $\leq 9.8\times 10^{-9} $ \cite{Workman:2022ynf} \tabularnewline
				
				BR$(\tau^-\to e^-e^+e^-)$ & $\leq 1.4\times 10^{-8} $ \cite{Workman:2022ynf} & BR$(\tau^-\to e^-\mu^+e^-)$ & $\leq 8.4\times 10^{-9} $ \cite{Workman:2022ynf}
				\tabularnewline
				BR$(\tau^-\to e^-\mu^+\mu^-)$ & $\leq 1.6\times 10^{-8} $\cite{Workman:2022ynf}  & BR$(\tau^-\to \mu^-\mu^+\mu^-)$ & $\leq 1.1\times 10^{-8} $ \cite{Workman:2022ynf} \tabularnewline
					BR$(\tau^-\to \mu^+\mu^-)$ & $\leq 1.6\times 10^{-8} $\cite{Workman:2022ynf}  & P$(\text{Mu}-\overline{\text{Mu}})$ & $<8.3\times 10^{-11} $\cite{Willmann:1998gd}   \tabularnewline
			BR$(\mu^- \text{Au}\to e^-\text{Au})$ & $\leq 7.0\times 10^{-13} $ \cite{SINDRUMII:2006dvw} & $\Delta a_{e}^{\text{Rb}}$ & $0.48(30)\times 10^{-12} $\cite{Workman:2022ynf}  \tabularnewline
			$\Delta a_{\mu}$ & $249(48)\times 10^{-11}$ \cite{Muong-2:2023cdq}  & & \tabularnewline
				\hline 
			\end{tabular}
			\par
			\protect\caption{\label{lepton_exp} Experimental constraints for leptonic flavor observables.}
		\end{centering}
	\end{table} 
	\subsection{Quark flavor phenomenology \label{quark_flavor}}
\begin{figure}[H]
		\centering
		\begin{tabular}{c}
			\includegraphics[width=11cm]{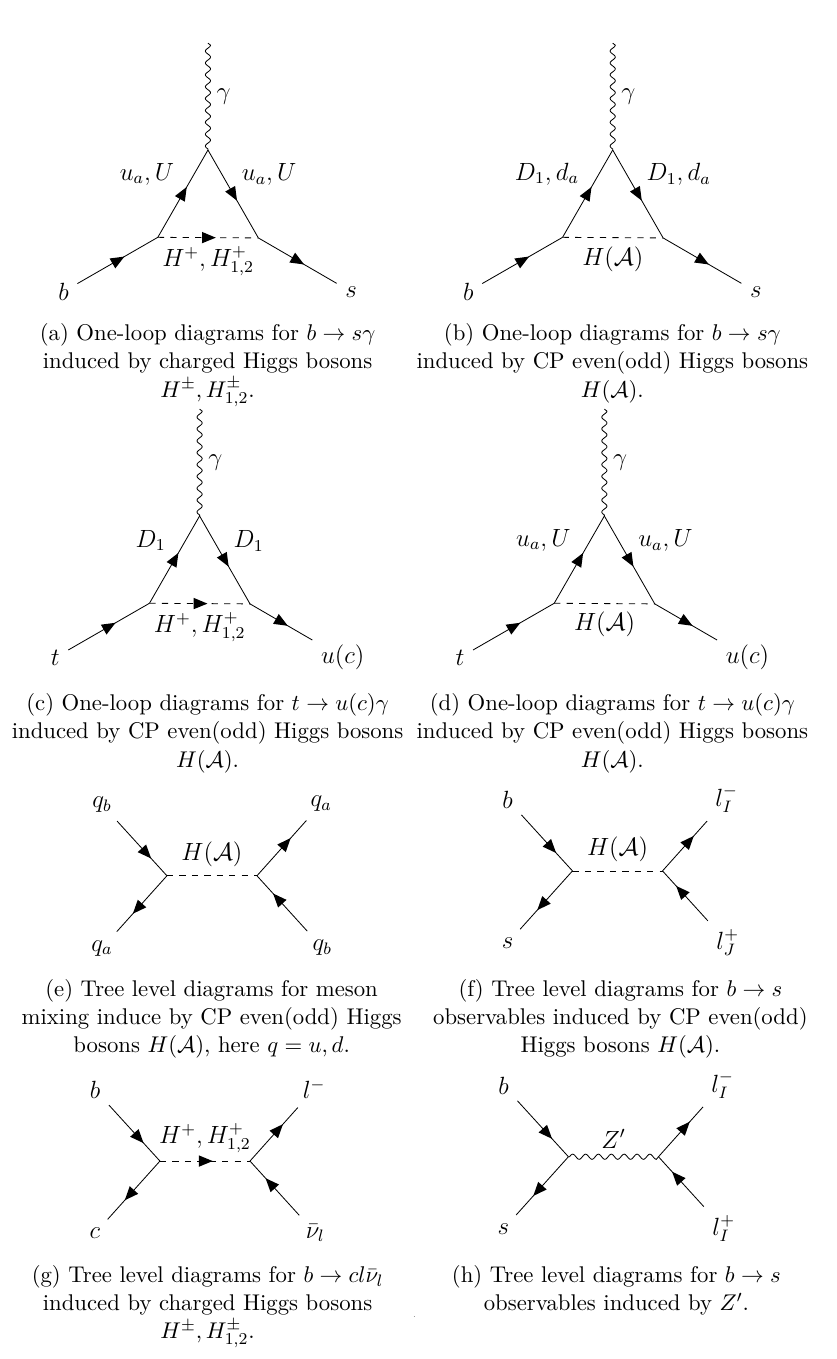}
		\end{tabular}
		\caption{\label{quark_diagrams}Feynman diagrams for $b\to s$, $b\to c$ and $t\to u(c)$ quark transitions.}
	\end{figure}

	The 
	Yukawa terms contributing to the down type quark transitions, 
	such as $b\to s (d)$, are obtained from Eq. (\ref{Lyq}) as follows
	\bea 
	-\mathcal{L}^{d_a\to d_b}_{\text{quark}}&=& \sum_{i=1}^3y_i^{(d)}\bar{u}_{3L}\phi_2^+d_{iR} +\sum_{i=1}^3y_i^{(d)}\bar{d}_{3L}\fr{v_2+\phi^0_{2R}+i\phi^0_{2I}}{\sqrt{2}}d_{iR} +\sum_{i=1}^3w_1^{(d)}\bar{U}_L\zeta_1^+d_{iR} +\sum_{i=1}^3x_i^{(d)}\bar{D}_{1L}\fr{v_{\sigma}+\sigma_R+i\sigma_I}{\sqrt{2}} d_{iR}\crn   && -\sum_{i=1}^3y_i^{(u)}\bar{d}_{3L}\phi_1^-u_{iR}-\sum_{n=1}^2x_n^{(U)}\bar{d}_{nL}\phi_2^{-}U_R+\sum_{n=1}^2x_n^{(D)}\bar{d}_{nL}\fr{v_1+\phi^0_{1R}+i\phi^0_{1I}}{\sqrt{2}}D_{1R} +H.c.. \label{Lqd}
	\eea
	For the observables related to the up-type quark transitions such as $t\to (u,c)$, we have the following Yukawa interactions: 
	-\bea \mathcal{L}^{u_a\to u_b}_{\text{quark}}&=& \sum_{i=1}^3y_i^{(u)}\bar{u}_{3L}\fr{v_1+\phi^0_{1R}+i\phi^0_{1I}}{\sqrt{2}}u_{iR}+\sum_{n=1}^2x_n^{(U)}\bar{u}_{nL}\fr{v_2+\phi^0_{2R}+i\phi^0_{2I}}{\sqrt{2}}U_R\crn &&
	+\sum_{n=1}^2x_n^{(D)}\bar{u}_{nL}\phi_1^+D_{1R}+\sum_{i=1}^3w_i^{(u)}\bar{D}_{1L}\zeta_1^-u_{iR}+\sum_{i=1}^3x_{i}^{(u)}\bar{U}_L\fr{v_{\sigma}+\sigma_R-i\sigma_I}{\sqrt{2}}u_{iR}+H.c.. \label{Lqu}\eea 
	We now rewrite the above-given quark Yukawa interactions in a physical basis taking into account that 
	the physical quark eigenstates $u'_L,d'_L$ are related with the quark interaction eigenstates by the following transformations \bea u_{iL(R)}=(V_{u_{L(R)}})_{ia}u'_{aL(R)},\hs  d_{iL(R)}=(V_{d_{L(R)}})_{ia}d'_{aL(R)}, \eea
	where $V_{u(d)_{L,R}}$ are the mixing matrices of left(right) of up(type) quarks, respectively. Furthermore, the physical states of scalar fields are similar as pointed out in the lepton flavor sector. The quark Yukawa terms of   
	Eqs. (\ref{Lqd},\ref{Lqu}) are then be rewritten as follows 
	\bea 
	-\mathcal{L}^{d_a\to d_b}&=&\bar{u}'_a(g_L^{H^{+}\bar{u}_ad_b}P_L+g_R^{H^{+}\bar{u}_ad_b}P_R)d'_bH^+ +\bar{u}'_a(g_L^{H_{1,2}^{+}\bar{u}_ad_b}P_L+g_R^{H_{1,2}^{+}\bar{u}_ad_b}P_R)d'_bH_{1,2}^+\crn && +\bar{U}'(g_L^{H^{+}\bar{U}d_b}P_L+g_R^{H^+\bar{U}d_b}P_)d'_bH^+
	+\bar{U}'(g_L^{H_{1,2}^{+}\bar{U}d_b}P_L+g_R^{H_{1,2}^+\bar{U}d_b}P_)d'_bH_{1,2}^+ \crn &&  
	+\sum_{p=1}^8\bar{D}'_{1}(g_L^{H_p\bar{D}_1d_b}P_L+g_R^{H_p\bar{D}_1d_b}P_R)d'_{b}H_p+i\sum_{p=3}^8\bar{D}'_{1}(g_L^{\mathcal{A}_p\bar{D}_1d_b}P_L+g_R^{H_p\bar{D}_1d_b}P_R)d'_{b}\mathcal{A}_p \crn &&
	+\sum_{p=1}^8\bar{d}'_{a}(g_R^{H_p\bar{d}_ad_b}P_R)d'_{b}H_p+i\sum_{p=3}^8\bar{d}'_{i}(g_L^{\mathcal{A}_p\bar{d}_ad_b}P_R)d'_{b}\mathcal{A}_p+H.c.
	,  \label{lag_di_dj}
	\eea 
	\bea 
	-\mathcal{L}^{u_a\to u_b}&=&\bar{D}_1'(g_L^{H^{-}\bar{D}_1u_b}P_L+g_R^{H^-\bar{D}_1u_b}P_R)u'_bH^- + \bar{D}_1'(g_L^{H_{1,2}^{-}\bar{D}_1u_b}P_L+g_R^{H_{1,2}^-\bar{D}_1u_b}P_R)u'_bH_{1,2}^- \crn && 
	+\sum_{p=1}^8\bar{U}'(g_L^{H_p\bar{U}u_b}P_L+g_R^{H_p\bar{U}u_b}P_R)u'_{b}H_p+i\sum_{p=3}^8\bar{U}'(g_L^{\mathcal{A}_p\bar{U}u_b}P_L+g_R^{H_p\bar{U}u_b}P_R)u'_{b}\mathcal{A}_p \crn && 
	+\sum_{p=1}^8\bar{u}'_{a}(g_R^{H_p\bar{u}_au_b}P_R)u'_{b}H_p+i\sum_{p=3}^8\bar{u}'_{a}(g_R^{\mathcal{A}_p\bar{u}_au_b}P_R)u'_{b}\mathcal{A}_p+H.c.
	,  \label{lag_ui_uj}
	\eea 
	where the coefficients in Eqs. (\ref{lag_di_dj},\ref{lag_ui_uj}) are defined in Appendix \ref{Coefficients_lep_quark}.
	Here, we will clarify in detail the roles of each term in the above given Yukawa interactions. The terms in the first and second lines of Eq. (\ref{lag_di_dj}) contribute to flavor-changing neutral current processes (FCNC), such as the inclusive decay branching ratio BR$(\bar{B}\to X_s\gamma)$ at one-loop level. This process involves the virtual exchange of charged Higgs $H^{\pm}, H_{1,2}^{\pm}$ and up type quarks (both SM $u$ or new exotic ones $U$) in the internal lines of the loops, as illustrated 
	in subfigure (a) in Fig. (\ref{quark_diagrams}). The terms of the third line similarly contribute to such processes, but through the one-loop level exchange of neutral CP even (odd) Higgs bosons $H(\mathcal{A})$ and new exotic down type quark $D_1$, as shown in (subfigure (b) in Fig. (\ref{quark_diagrams})). The observables are also influenced by the one-loop exchange of the same neutral Higgs, but with the internal quarks in the loop being the SM ones $d$, rather than the new quark $D_1$, shown in the last line of Eq. (\ref{lag_di_dj}). Otherwise, the terms in Eq. (\ref{lag_ui_uj}) contribute to observables in up type quark transition $u_a\to u_b$, such as the branching ratios of FCNC top quark decays, i.e BR$(t\to u(c)\ga)$, BR$(t\to h(u,c))$ (see the subfigure (c) and (d) in Fig. (\ref{quark_diagrams}).) Furthermore, we want to emphasize that the terms in the last line of Eq. (\ref{lag_di_dj}) and Eq. (\ref{lag_ui_uj}) trigger the meson mixing $K^{0}-\bar{K}^{0}, B_{s,d}^{0}-\bar{B}^{0}_{s,d}$ and $D^{0}-\bar{D}^{0}$ at the tree-level via the exchange of CP even(odd) Higgs bosons $H(\mathcal{A})$ (see the subfigure (e)). Particularly, for the index $p=1$, the terms also describe the tree-level flavor violating decays of  SM-like Higgs boson $h\to \bar{d}_a d_b, h\to \bar{u}_a u_b $.
	
	It is important to note that when combining the lepton flavor conserving (violating) interactions by $H(\mathcal{A})$ in Eq. (\ref{lag_lepton}), the model provides the tree-level contributions to several observables, namely the branching ratio of leptonic decays BR$(B_s\to l^+l^-)$, BR$(B_s\to \tau^+\mu^-)$; semileptonic decays BR$(B\to K\tau^+\tau^-)$, BR$(B^+\to K^+\tau^+\mu^-)$, the lepton flavor universality violating (LFUV) ratios $R_{K^{(*)}}=\fr{\text{BR}(B\to K^{(*)}\mu^+\mu^-)}{\text{BR}(B\to K^{(*)}e^+e^-)}$  (shown in subfigure (f)).
	 Additionally, the terms in the first line also contribute to the observables related to flavor-changing charged currents (FCCCs) at the tree-level ($b\to c\bar{\tau}\nu_{\tau}$), such as LFUV ratios $R_{D^{(*)}}=\fr{\text{BR}(\bar{B}\to D^{(*)}\tau\bar{\nu}_{\tau})}{\text{BR}(\bar{B}\to D^{(*)}l\bar{\nu}_l)}, \ l=e,\mu$, shown in subfigure (g) in Fig. (\ref{quark_diagrams}). 
	
	Besides the above-mentioned contributions, the model also provides other contributions arising from the new neutral gauge boson $Z'$ with quark flavor violating couplings $Z'\bar{q}_{iL}q_{jL}$ at tree-level, due to the different $U(1)_X$ charges of left-handed third quark $q_{3L}$ generation compared to the first and second ones $q_{(1,2)L}$. This kind of contribution not only yields new physics contributions to  
	meson mass splittings 
	$\Delta m_{K, B_s, B_d}$ at tree-level as pointed out in the previous work \cite{Hernandez:2021xet} but also give rise to other FCNC observables such as BR$(\bar{B}\to X_s\gamma)$ at one-loop level (with $Z'$ and SM down type quark $d_i$ are internal lines), BR$(B_s\to l^+l^-)$ at the tree-level (subfigure (h) in Fig. (\ref{quark_diagrams})). However, the contribution of this $Z'$ to the inclusive decay BR$(\bar{B}\to X_s\gamma)$ is negligible, in comparison with contributions of charged Higgs bosons \cite{Buras:2012dp}.  All of these contributions will be analyzed 
	in detail in the section on numerical studies.

	\subsubsection{FCNC $d_a\to d_b$ observables}
		Firstly, we study the FCNC $d_a\to d_b$ observables which can described via the general effective Hamiltonian 
	\bea 
	\mathcal{H}_{\text{eff}}&=&-\fr{4G_F}{\sqrt{2}}V^{*}_{ts}V_{tb}\left\{\sum_{i=7,8,9,10}[C_i(\mu)\mathcal{O}_i(\mu)+C'_i(\mu)\mathcal{O}'_i(\mu)]+\sum_{i=S,P}[C^{IJ}_i(\mu)\mathcal{O}^{IJ}_i(\mu)+C^{'IJ}_i(\mu)\mathcal{O}^{'IJ}_i(\mu)]\right\},
	\eea
	where $C^{(')}_{i}(\mu)$ are the Wilson coefficients (WCs) corresponding to the effective operators $\mathcal{O}^{(')}_i$ for $b\to s$ observables at the scale $\mu=\mathcal{O}(m_b)$. The indices $I, J$ denote different lepton flavors in WCs $C^{(')}_{S, P}$ since the model contains lepton flavor violating coupling related to CP even(odd) Higgs bosons \footnote{However, the WCs $C_{9,10}$ are only generated by new neutral gauge boson $Z'$ which have the same couplings with three lepton generations, thus these are no lepton flavor universality violations (LFUV) caused by these WCs. }. All WCs are defined as follows
	\bea 
	\mathcal{O}^{(')}_7 &=&\fr{e}{16\pi^2}m_b(\bar{s}\sigma^{\mu\nu}P_{R(L)} b)F_{\mu\nu}, \hs \mathcal{O}^{(')}_8=\fr{g_s}{16\pi^2}m_b(\bar{s}\sigma^{\mu\nu}T^a P_{R(L)} b)G^a_{\mu\nu}, \\ 
	\mathcal{O}^{(')}_{9}&=&\fr{e^2}{16\pi^2}(\bar{s}\ga^{\mu}P_{L(R)}b)(\bar{l}\ga_{\mu}l),\hs \mathcal{O}^{(')}_{10}=\fr{e^2}{16\pi^2}(\bar{s}\ga^{\mu}P_{L(R)} b)(\bar{l}\ga_{\mu}\ga_5 l), \\
	\mathcal{O}^{IJ(')}_{S}&=&\fr{e^2}{16\pi^2}(\bar{s}P_{L(R)}b)(\bar{l}_Il_J),\hs 
	\mathcal{O}^{IJ(')}_{P}=\fr{e^2}{16\pi^2}(\bar{s}P_{L(R)}b)(\bar{l}_I\ga_5l_J).
	\eea 
	 The primed WCs $C'_i,\tilde{C'}_S$ corresponding to the operators $\mathcal{O}'_i,\mathcal{\tilde{O}'}_S$ are obtained by flipping the chirality. It is important to emphasize that  
	 the quark mixing matrix $V_q=V_{u_L}^{\dagger}V_{d_L}$ obtained from the diagonalization of the 
	 low energy up and down type quark mass matrices is found to be unitary. 
 Its magnitude $|V|$ is consistent with the absolute entry value constraints of the CKM matrix given in \cite{Workman:2022ynf}, as shown in Sec (\ref{Sec5_input_par}). However, the matrix itself $V$ has the different entries $(V_q)_{ij}$ compared to corresponding CKM ones defined 
 by the "standard" parametrization with three mixing angles and one CP violation phase \cite{Workman:2022ynf}. 
 Therefore, the SM WCs in our model are modified by $C^{\text{SM}}_{7,8,9,10}\to C^{\text{SM}}_{7,8,9,10}\fr{(V_q^*)_{32}(V_q)_{33}}{V^{*}_{ts}V_{tb}}$. We can split the WCs as combination of both SM and NP, i.e   $C^{(')}_{7,8,9,10}=C^{\text{SM}}_{7,8,9,10}\fr{(V_q^*)_{32}(V_q)_{33}}{V^{*}_{ts}V^{}_{tb}}+C^{(')\text{NP}}_i$, where $C^{\text{SM}}_{7,8,9,10}$ have been calculated in SM \cite{Misiak:2006ab, Misiak:2020vlo, Beneke:2019slt}, whereas $C^{\text{SM}}_{S, P}$ are absent. For WCs of NP $C_{7,8,9,10, S, P}^{(')\text{NP}}$, we can write them as
  the summation of different NP contributions as follows 
	\bea
	C_{7,8}^{\text{NP}}(\mu)&=& C^{H^+\bar{u}d}_{7,8}(\mu)+C^{H_{1(2)}^+\bar{u}d}_{7,8}(\mu)+C^{H^+\bar{U}d}_{7,8}(\mu)+C^{H_{1(2)}^+\bar{U}d}_{7,8}(\mu)+C^{H_{p}\bar{D}_1d}_{7,8}(\mu)+C^{\mathcal{A}_{p}\bar{D}_1d}_{7,8}(\mu),\crn 
	C_{7,8}^{'\text{NP}}(\mu)&=& C^{'H^+\bar{u}d}_{7,8}(\mu)+C^{'H_{1(2)}^+\bar{u}d}_{7,8}(\mu)+C^{'H^+\bar{U}d}_{7,8}(\mu)+C^{'H_{1(2)}^+\bar{U}d}_{7,8}(\mu)+C^{'H_{p}\bar{D}_1d}_{7,8}(\mu)+C^{'\mathcal{A}_{1p}\bar{D}_1d}_{7,8}(\mu)\crn &&+ C^{'H_{p}\bar{d}d}_{7,8}(\mu)+C^{'\mathcal{A}_{p}\bar{d}d}_{7,8}(\mu),\crn
	C_{9}^{\text{NP}}(\mu)&=&C_{9}^{Z'}(\mu),\hs  C_{10}^{\text{NP}}(\mu)=C_{10}^{Z'}(\mu),\hs C_{9,10}^{'\text{NP}}(\mu)=0, \crn 
	C_{S}^{(')IJ}(\mu)&=&\sum_{p=1}^8(C^{(')H_{p}\bar{d}_2d_3}_S)^{IJ}(\mu),\hs C_{P}^{(')IJ}(\mu)=\sum_{p=3}^8(C^{(')\mathcal{A}_{p}\bar{d}_2d_3}_P)^{IJ}(\mu). \label{WCs}
	\eea 
	
	It should be noted that the WCs depend on the energy scale. The model contains several energy scales namely 
	masses of the new Higgses as well as the $Z'$ gauge boson mass. Because the QCD running effect is negligible 
at high energy, we can assume that these scales are approximately the same $\mu_H\sim\mathcal{O}(m_H)$, thus implying that 
the WCs only depend on a single energy scale $\mu=\mu_H$.  For simplicity, we calculate the loop contributions at on-shell, i.e., $q^2=0$, $p_s^2=m_s^2$, and $p_b^2=m_b^2$. Because $m_s\ll m_b$, we set the $s$ quark mass to be zero, $m_s=0$, and keep the mass of $b$ quark at the linear order, i.e. $m_b^2=0$. Thus, we have the following expression of WCs at scale $\mu=\mu_H$ for $b\to s$ transitions as follows
	\bea 
	C^{H^+\bar{u}d}_7(\mu_H)&=&\fr{-\sqrt{2}}{8G_FV^*_{ts}V_{tb}m^2_{H^{\pm}}}\sum_{a=1}^3\left[(g_L^{H^+\bar{u}_ad_2})^*g_L^{H^+\bar{u}_ad_3}f''_{\ga}\left(\fr{m_{u_a}^2}{m_{H^+}^2}\right)+(g_L^{H^+\bar{u}_ad_2})^*g_R^{H^+\bar{u}_ad_3}\fr{m_{u_a}}{m_b}h''_{\ga}\left(\fr{m_{u_a}^2}{m_{H^+}^2}\right)\right],\crn 
	C^{'H^+\bar{u}d}_7(\mu_H)&=&\fr{-\sqrt{2}}{8G_FV^*_{ts}V_{tb}m^2_{H^{\pm}}}\sum_{a=1}^3\left[(g_R^{H^+\bar{u}_ad_2})^*g_R^{H^+\bar{u}_ad_3}f''_{\ga}\left(\fr{m_{u_a}^2}{m_{H^+}^2}\right)+(g_R^{H^+\bar{u}_ad_2})^*g_L^{H^+\bar{u}_ad_3}\fr{m_{u_a}}{m_b}h''_{\ga}\left(\fr{m_{u_a}^2}{m_{H^+}^2}\right)\right],\crn 
	C^{H_{1(2)}^+\bar{u}d}_7(\mu_H)&=&\fr{-\sqrt{2}}{8G_FV^*_{ts}V_{tb}m^2_{H^{\pm}_{1(2)}}}\sum_{a=1}^3\left[(g_L^{H_{1(2)}^+\bar{u}_ad_2})^*g_L^{H_{1(2)}^+\bar{u}_ad_3}f''_{\ga}\left(\fr{m_{u_a}^2}{m_{H_{1(2)}^+}^2}\right)+(g_L^{H_{1(2)}^+\bar{u}_ad_2})^*g_R^{H_{1(2)}^+\bar{u}_ad_3}\fr{m_{u_a}}{m_b}h''_{\ga}\left(\fr{m_{u_a}^2}{m_{H_{1(2)}^+}^2}\right)\right],\crn 
	C^{'H_{1(2)}^+\bar{u}d}_7(\mu_H)&=&\fr{-\sqrt{2}}{8G_FV^*_{ts}V_{tb}m^2_{H^{\pm}_{1(2)}}}\sum_{a=1}^3\left[(g_R^{H_{1(2)}^+\bar{u}_ad_2})^*g_R^{H_{1(2)}^+\bar{u}_ad_3}f''_{\ga}\left(\fr{m_{u_a}^2}{m_{H_{1(2)}^+}^2}\right)+(g_R^{H_{1(2)}^+\bar{u}_ad_2})^*g_L^{H^+\bar{u}_ad_3}\fr{m_{u_a}}{m_b}h''_{\ga}\left(\fr{m_{u_a}^2}{m_{H_{1(2)}^+}^2}\right)\right],\crn 
	C^{H^+\bar{U}d}_7(\mu_H)&=&\fr{-\sqrt{2}}{8G_FV^*_{ts}V_{tb}m^2_{H^{\pm}}}\left[(g_L^{H^+\bar{U}d_2})^*g_L^{H^+\bar{U}d_3}f''_{\ga}\left(\fr{m_{U}^2}{m_{H^+}^2}\right)+(g_L^{H^+\bar{U}d_2})^*g_R^{H^+\bar{U}d_3}\fr{m_{U}}{m_b}h''_{\ga}\left(\fr{m_{U}^2}{m_{H^+}^2}\right)\right],\crn 
	C^{'H^+\bar{U}d}_7(\mu_H)&=&\fr{-\sqrt{2}}{8G_FV^*_{ts}V_{tb}m^2_{H^{\pm}}}\left[(g_R^{H^+\bar{U}d_2})^*g_R^{H^+\bar{U}d_3}f''_{\ga}\left(\fr{m_{U}^2}{m_{H^+}^2}\right)+(g_R^{H^+\bar{U}d_2})^*g_L^{H^+\bar{U}d_3}\fr{m_{U}}{m_b}h''_{\ga}\left(\fr{m_{U}^2}{m_{H^+}^2}\right)\right],\crn 
	C^{H_{1(2)}^+\bar{U}d}_7(\mu_H)&=&\fr{-\sqrt{2}}{8G_FV^*_{ts}V_{tb}m^2_{H^{\pm}_{1(2)}}}\left[(g_L^{H_{1(2)}^+\bar{U}d_2})^*g_L^{H_{1(2)}^+\bar{U}d_3}f''_{\ga}\left(\fr{m_{U}^2}{m_{H_{1(2)}^+}^2}\right)+(g_L^{H_{1(2)}^+\bar{U}d_2})^*g_R^{H_{1(2)}^+\bar{U}d_3}\fr{m_{U}}{m_b}h''_{\ga}\left(\fr{m_{U}^2}{m_{H_{1(2)}^+}^2}\right)\right] , \crn
	C^{'H_{1(2)}^+\bar{U}d}_7(\mu_H)&=&\fr{-\sqrt{2}}{8G_FV^*_{ts}V_{tb}m^2_{H^{\pm}_{1(2)}}}\left[(g_R^{H_{1(2)}^+\bar{U}d_2})^*g_R^{H_{1(2)}^+\bar{U}d_3}f''_{\ga}\left(\fr{m_{U}^2}{m_{H_{1(2)}^+}^2}\right)+(g_R^{H_{1(2)}^+\bar{U}d_2})^*g_L^{H_{1(2)}^+\bar{U}d_3}\fr{m_{U}}{m_b}h''_{\ga}\left(\fr{m_{U}^2}{m_{H_{1(2)}^+}^2}\right)\right],\crn  
	C^{H_p\bar{D}_1d}_7(\mu_H)&=&\fr{\sqrt{2}}{24G_FV^*_{ts}V_{tb}}\sum_{p=1}^8\fr{1}{m^2_{H_p}} \left[(g_L^{H_p\bar{D}_1d_2})^*g_L^{H_p\bar{D}_1d_3}f'_{\ga}\left(\fr{m_{D_1}^2}{m_{H_p}^2}\right)+(g_L^{H_p\bar{D}_1d_2})^*g_R^{H_p\bar{D}_1d_3}\fr{m_{D_1}}{m_b}h'_{\ga}\left(\fr{m_{D_1}^2}{m_{H_p}^2}\right)\right],\crn 
	C^{'H_p\bar{D}_1d}_7(\mu_H)&=&\fr{\sqrt{2}}{24G_FV^*_{ts}V_{tb}}\sum_{p=1}^8\fr{1}{m^2_{H_p}} \left[(g_R^{H_p\bar{D}_1d_2})^*g_R^{H_p\bar{D}_1d_3}f'_{\ga}\left(\fr{m_{D_1}^2}{m_{H_p}^2}\right)+(g_R^{H_p\bar{D}_1d_2})^*g_L^{H_p\bar{D}_1d_3}\fr{m_{D_1}}{m_b}h'_{\ga}\left(\fr{m_{D_1}^2}{m_{H_p}^2}\right)\right],\crn 
	C^{\mathcal{A}_p\bar{D}_1d}_7(\mu_H)&=&\fr{\sqrt{2}}{24G_FV^*_{ts}V_{tb}}\sum_{p=3}^8\fr{1}{m^2_{\mathcal{A}_p}} \left[(g_L^{\mathcal{A}_p\bar{D}_1d_2})^*g_L^{\mathcal{A}_p\bar{D}_1d_3}f'_{\ga}\left(\fr{m_{D_1}^2}{m_{\mathcal{A}_p}^2}\right)+(g_L^{\mathcal{A}_p\bar{D}_1d_2})^*g_R^{\mathcal{A}_p\bar{D}_1d_3}\fr{m_{D_1}}{m_b}h'_{\ga}\left(\fr{m_{D_1}^2}{m_{H_p}^2}\right)\right], \crn 
	C^{'\mathcal{A}_p\bar{D}_1d}_7(\mu_H)&=&\fr{\sqrt{2}}{24G_FV^*_{ts}V_{tb}}\sum_{p=3}^8\fr{1}{m^2_{\mathcal{A}_p}}\left[(g_R^{\mathcal{A}_p\bar{D}_1d_2})^*g_R^{\mathcal{A}_p\bar{D}_1d_3}f'_{\ga}\left(\fr{m_{D_1}^2}{m_{\mathcal{A}_p}^2}\right)+(g_R^{\mathcal{A}_p\bar{D}_1d_2})^*g_L^{\mathcal{A}_p\bar{D}_1d_3}\fr{m_{D_1}}{m_b}h'_{\ga}\left(\fr{m_{D_1}^2}{m_{H_p}^2}\right)\right],\crn 
	C^{'H_p\bar{d}d}_7(\mu_H)&=&\fr{\sqrt{2}}{24G_FV^*_{ts}V_{tb}}\sum_{p=1}^8\fr{1}{m_{H_p}^2}\sum_{a=1}^3 \left[(g_R^{H_{p}\bar{d}_ad_2})^*g_R^{H_p\bar{d}_ad_3}f'_{\ga}\left(\fr{m_{d_a}^2}{m_{H_p}^2}\right)\right], \crn 
	C^{'\mathcal{A}_p\bar{d}d}_7(\mu_H)&=&\fr{\sqrt{2}}{24G_FV^*_{ts}V_{tb}}\sum_{p=3}^8\fr{1}{m_{\mathcal{A}_p}^2}\sum_{a=1}^3\left[(g_R^{\mathcal{A}_p\bar{d}_ad_2})^*g_R^{\mathcal{A}_p\bar{d}_ad_3}f'_{\ga}\left(\fr{m_{d_a}^2}{m_{\mathcal{A}_p}^2}\right)\right], \label{WCs1}
	\eea 
	where the functions $f^{'(")}_{\ga},h^{'(")}_{\ga}$ are given in Appendix \ref{Loopfuncs}. The WCs $C_8^{(')}$ have similar form as $C_7$ but replacing functions $f^{',('')}_{\ga},h^{'('')}_{\ga}$ by $f^{'(")}_{g},h^{'(")}_{g}$. For the WCs associated $Z'$ boson, there are WCs $C_{9,10}$ as follows 
	\bea
	C^{(Z')}_{9}(\mu_H) &=& \fr{8\sqrt{2}\pi^2g^2_X}{9e^2G_FV^*_{ts}V_{tb}m_{Z'}^2}(V^*_{d_L})_{32}(V_{d_L})_{33}, \hs  C^{(Z')}_{10}(\mu_H) = \fr{4\sqrt{2}\pi^2g^2_X}{9e^2G_FV^*_{ts}V_{tb}m_{Z'}^2}(V^*_{d_L})_{32}(V_{d_L})_{33}. \label{WCs_quark1}
	\eea
	Here we denote $g_X$ is the coupling of $Z'$ gauge boson, which is given in \cite{Hernandez:2021xet}. We want to emphasize that WCs $C_{9,10}$ in our model are blinded with lepton flavor, i.e $C_{9,10}^{e}=C_{9,10}^{\mu}=C_{9,10}^{\tau}$ since all three lepton generations are identical under the gauge symmetry group. Furthermore, the LFUV ratios $R_{K^{(*)}}$ mostly depend on $C_{9,10}$, thus making these observable to be nearly identity, which satisfies the current experimental results \cite{LHCb:2020pcv}. 
	On the other hand, the CP even (odd) Higgs cause scalar and pseudoscalar WCs $C_{S, P}^{(')IJ}$, which have the following expressions 
	\bea 
	(C^{H_p\bar{d}_2d_3}_S)^{IJ}(\mu_H)&=&\fr{16\pi^2}{e^2}\fr{\sqrt{2}}{4G_FV^*_{ts}V_{tb}} \sum_{p=1}^8\fr{(g_R^{H_p\bar{d}_3d_2})^*(g_R^{H_p\bar{l}_Il_J}+g_R^{H_p\bar{l}_Jl_I})}{2m_{H_p}^2}, \\  (C^{'H_p\bar{d}_2d_3}_S)^{IJ}(\mu_H)&=&\fr{16\pi^2}{e^2}\fr{\sqrt{2}}{4G_FV^*_{ts}V_{tb}} \sum_{p=1}^8\fr{g_R^{H_p\bar{d}_2d_3}(g_R^{H_p\bar{l}_Il_J}+g_R^{H_p\bar{l}_Jl_I})}{2m_{H_p}^2}, \\  (C^{\mathcal{A}_p\bar{d}_2d_3}_P)^{IJ}(\mu_H)&=&-\fr{16\pi^2}{e^2}\fr{\sqrt{2}}{4G_FV^*_{ts}V_{tb}} \sum_{p=3}^8\fr{(g_R^{\mathcal{A}_p\bar{d}_3d_2})^*(g_R^{\mathcal{A}_p\bar{l}_Il_J}+g_R^{\mathcal{A}_p\bar{l}_Jl_I})}{2m_{\mathcal{A}_p}^2}, \\ (C^{'\mathcal{A}_p\bar{d}_2d_3}_P)^{IJ}(\mu_H)&=&-\fr{16\pi^2}{e^2}\fr{\sqrt{2}}{4G_FV^*_{ts}V_{tb}} \sum_{p=3}^8\fr{g_R^{\mathcal{A}_p\bar{d}_2d_3}(g_R^{\mathcal{A}_p\bar{l}_Il_J}+g_R^{\mathcal{A}_p\bar{l}_Jl_I})}{2m_{\mathcal{A}_p}^2}.\eea
	With the definitions of WCs, we have the following formula for the branching ratio BR$(B_s\to \mu^+\mu^-)$ \cite{Mohanta:2005gm},
	\bea
	\mathrm{BR}(B_s \to \mu^+\mu^-)&=&  \frac{\tau_{B_s}}{64 \pi^3}\al^2 G_F^2 f_{B_s}^2|V_{tb}V^*_{ts}|^2 m_{B_s}
	\sqrt{1-\fr{4m_{\mu}^2}{m_{B_s}^2}}  \left\{ \left(1-\frac{4m_{\mu}^2}{m_{B_s}^2}\right)\left |\fr{m_{B_s}^2}{m_b+m_s} \left(C^{\mu\mu}_S-C^{'\mu\mu}_S \right)\right |^2 \right.  \nonumber \\  
	&&+\left.\left|2m_{\mu}C_{10}  +\fr{m_{B_s}^2}{m_b+m_s}\left(C^{\mu\mu}_P-C^{'\mu\mu}_P \right) \right|^2\right\},
	\eea
	where $\tau_{B_s}$ is the lifetime of $B_s$ meson, $\al_{\text{em}}$ is the fine-structure constant. Furthermore, we have to take into account the effect of $B_s$--$\bar{B}_s$ oscillations, therefore theoretical prediction is related to the experimental value by \cite{DeBruyn:2012wj}
	\be
	\mathrm{BR}(B_s \to \mu^+\mu^-)_{\text{exp}}\simeq \fr{1}{1-y_s}\mathrm{BR}(B_s \to \mu^+\mu^-),
	\ee
	where $y_s=\frac{\Delta \Ga_{B_s}}{2 \Ga_{B_s}}$ which has numerical values is given Table \ref{input-par}.
	
	For the inclusive decay $\bar{B}\to X_s\gamma$, we have its branching ratio given by  \cite{Gambino:2001ew,Buras:2011zb}
	\be
	\text{BR}(\bar{B}\to X_s\ga)=\fr{6\al_{\text{em}}}{\pi C}\left|\fr{V_{ts}^*V_{tb}}{V_{cb}}\right|^2 \left[|C_7(\mu_b)|^2+|C^{'}_7(\mu_b)|^2+N(E_{\gamma})\right]\text{BR}(\bar{B}\rightarrow X_c e\bar{\nu}) , \label{bra1}
	\ee
	where $N(E_{\gamma})$ is a non-perturbative contribution which amounts around $4\%$ of the branching ratio. We compute the leading order contribution to $N(E_{\gamma})$ followed the Eq. (3.8) in Ref. \cite{Misiak:2020vlo} and then obtain $N(E_{\ga})\simeq 3.3\times 10^{-3}$. Additionally, $C$ is the semileptonic phase-space factor, $C=|V_{ub}/V_{cb}|^2\Ga(\bar{B}\rightarrow X_c e\bar{\nu}_e)/\Ga(\bar{B}\rightarrow X_u e\bar{\nu}_e)$, and BR$(\bar{B}\rightarrow X_c e\bar{\nu})$ is the branching ratio for semileptonic decay. It is necessary to consider the QCD corrections to complete the calculation for such decay. The WCs $C_{7}^{(')}(\mu_b)$ are evaluated at the matching scale $\mu_b=2$ GeV by running down from the higher scale $\mu_{H}$ via the renormalization group equations. Its expression can be split as follows
	\bea C_{7}(\mu_b)=\fr{(V_q^*)_{32}(V_q)_{33}}{V^{*}_{ts}V^{}_{tb}}C_{7}^{\text{SM}}(\mu_b)+C_{7}^{\text{NP}}(\mu_b),\hs  C^{'}_{7}(\mu_b)=C_{7}^{'\text{NP}}(\mu_b), \eea 
	where $C_{7}^{\text{SM}}(\mu_b)$ is the SM WC and have been calculated up to next-to-next-leading order of QCD corrections at the scale $\mu_b=2.0$ GeV \cite{Misiak:2006ab,Misiak:2020vlo}. Furthermore, for the NP contributions at the matching scale $\mu_b$, WCs $C_{7,8}$, we have \cite{Buras:2011zb}
	\bea 
	&& C_{7}^{(')H^{\pm}}(\mu_b)  =\ka_7C_7^{(')H^{\pm}}(m_{H^{\pm}})+\ka_8 C_8^{(')H^{\pm}}(m_{H^{\pm}}),\label{WCs2}\eea 
	where $\ka_{7,8}$ are so called "magic numbers" and given in  \cite{Buras:2011zb}.

	Besides, we also are interested in other observables which are affected by short-distance effects, such as the branching ratios of decays $B^+\to K^+\tau^+\tau^-$, $B^+\to K^+\tau^+\mu^-$, $B_s\to \tau^+\mu^-$. 
	These branching ratios are given by  \cite{Cornella:2021sby}
	\bea 
	10^9\times \text{BR}(B^+\to K^+\tau^+\tau^-)&=&2.2|C_9|^2+6|C_{10}|^2+8.3|C^{\tau\tau}_{S}-C_S^{'\tau\tau}|^2+8.9|C^{\tau\tau}_P-C_P^{'\tau\tau}|^2\crn && +4.8\text{Re}[(C^{\tau\tau}_S-C_S^{'\tau\tau})C^*_9]+5.9\text{Re}[(C_P^{\tau\tau}-C_P^{'\tau\tau})C^*_{10}], \\ 
	10^9\times \text{BR}(B^+\to K^+\tau^+\mu^-)&=&13.58|C^{\tau\mu}_{S}-C_{S}^{\tau\mu}|^2+14.54|C^{\tau\mu}_P-C_P^{'\tau\mu}|^2, \\
	\text{BR}(B_s\to \tau^-\mu^+)&=&\fr{\tau_{B_s}}{64\pi^3}\al^2G_F^2f_{B_s}^2|V_{ts}^*V_{tb}|^2\fr{m_{B_s}^5}{(m_b+m_s)^2}\left(1-\fr{m_{\tau}^2}{m_{B_s}^2}\right)^2\crn &&\times (|C_S^{\tau\mu}-C_S^{'\tau\mu}|^2+|C_{P}^{\tau\mu}-C_P^{'\tau\mu}|^2). \eea 
	
	We would like to note that the meson mass splittings 
	such as $\Delta m_{K}, \Delta m_{B_s}$ and $\Delta m_{B_d}$ were investigated in \cite{Hernandez:2021xet}. However, the authors have just considered these observables in simplified scenarios in which there were only a few Higgs contributions and assumptions real down quark Yukawa couplings. Therefore, in this work, we reconsider the meson mixing in more detail with all Higgs contributions and general Yukawa couplings. The new physics contributions to meson masses differences involve neutral gauge $Z'$ and CP even (odd) Higgs $H(\mathcal{A})$ bosons as given by 
	\bea
	&& \Delta m_K 
	\simeq \fr{2}{27}\fr{g_X^2}{m_{Z'}^2} \mathrm{Re}\left\{[(V_{d_L})^*_{31}(V_{d_L})_{32}]^2\right\}  m_K f^2_K \nonumber \crn && +\frac{5}{48}\mathrm{Re} \left\{\sum_{p=1}^8\left[ \left(g_R^{H_p\bar{d}_1d_2}\right)^2+ \left[\left(g_R^{H_p\bar{d}_2d_1}\right)^*\right]^2\right]\fr{1}{m_{H_p}^2}\right . \crn  && \left .-\sum_{p=3}^8\left[ \left(g_R^{\mathcal{A}_p\bar{d}_1d_2}\right)^2+ \left[\left(g_R^{\mathcal{A}_p\bar{d}_2d_1}\right)^{*}\right]^2\right]\fr{1}{m_{\mathcal{A}_p}^2}\right\}\left(\frac{m_K}{m_s+m_d} \right)^2 m_K f_K^2    \crn 
	&& - \frac{1}{4}\mathrm{Re}\left\{\sum_{p=1}^8 \left(g_R^{H_p\bar{d}_2d_1}\right)^*\left(g_R^{H_p\bar{d}_1d_2}\right)\fr{1}{m_{H_p}^2}+\sum_{p=3}^8 \left(g_R^{\mathcal{A}_p\bar{d}_2d_1}\right)^*\left(g_R^{\mathcal{A}_p\bar{d}_1d_2}\right)\fr{1}{m_{\mathcal{A}_p}^2}\right\}\left(\frac{1}{6}+\frac{m_K^2}{\left(m_s+m_d\right)^2} \right)m_K f_K^2,\label{ptdt16}
	\eea 
	\bea && \Delta m_{B_s}
\simeq \fr{2}{27}\fr{g_X^2}{m_{Z'}^2} \mathrm{Re}\left\{[(V_{d_L})^*_{32}(V_{d_L})_{33}]^2\right\} m_{B_s} f^2_{B_s} \nonumber \crn && +\frac{5}{48}\mathrm{Re} \left\{\sum_{p=1}^8\left[ \left(g_R^{H_p\bar{d}_2d_3}\right)^2+ \left[\left(g_R^{H_p\bar{d}_3d_2}\right)^{*}\right]^2\right]\fr{1}{m_{H_p}^2}\right . \crn  &&- \left .\sum_{p=3}^8\left[ \left(g_R^{\mathcal{A}_p\bar{d}_2d_3}\right)^2+ \left[\left(g_R^{\mathcal{A}_p\bar{d}_3d_2}\right)^{*}\right]^2\right]\fr{1}{m_{\mathcal{A}_p}^2}\right\}\left(\frac{m_{B_s}}{m_s+m_b} \right)^2 m_{B_s} f_{B_s}^2 \nonumber \crn && - \frac{1}{4}\mathrm{Re}\left\{\sum_{p=1}^8 \left(g_R^{H_p\bar{d}_2d_3}\right)^*\left(g_R^{H_p\bar{d}_3d_2}\right)\fr{1}{m_{H_p}^2}+\sum_{p=3}^8 \left(g_R^{\mathcal{A}_p\bar{d}_2d_3}\right)^*\left(g_R^{\mathcal{A}_p\bar{d}_3d_3}\right)\fr{1}{m_{\mathcal{A}_p}^2}\right\}\left(\frac{1}{6}+\frac{m_{B_s}^2}{\left(m_s+m_b\right)^2} \right)m_{B_s} f_{B_s}^2, 
	\label{ptdt17}\eea
	
	\bea
	&& \Delta m_{B_d}
	\simeq \fr{2}{27}\fr{g_X^2}{m_{Z'}^2} \mathrm{Re}\left\{[(V_{d_L})^*_{31}(V_{d_L})_{33}]^2\right\}  m_{B_d} f^2_{B_d} \nonumber \crn && +\frac{5}{48}\mathrm{Re} \left\{\sum_{p=1}^8\left[ \left(g_R^{H_p\bar{d}_1d_3}\right)^2+ \left[\left(g_R^{H_p\bar{d}_3d_1}\right)^{*}\right]^2\right]\fr{1}{m_{H_p}^2} \right . \crn  && - \left .\sum_{p=3}^8\left[ \left(g_R^{\mathcal{A}_p\bar{d}_1d_3}\right)^2+ \left[\left(g_R^{\mathcal{A}_p\bar{d}_3d_1}\right)^{*}\right]^2\right]\fr{1}{m_{\mathcal{A}_p}^2}\right\}\left(\frac{m_{B_d}}{m_b+m_d} \right)^2 m_{B_d} f_{B_d}^2     \crn &&  - \frac{1}{4}\mathrm{Re}\left\{\sum_{p=1}^8 \left(g_R^{H_p\bar{d}_3d_1}\right)^*\left(g_R^{H_p\bar{d}_1d_3}\right)\fr{1}{m_{H_p}^2}+\sum_{p=3}^8 \left(g_R^{\mathcal{A}_p\bar{d}_3d_1}\right)^*\left(g_R^{\mathcal{A}_p\bar{d}_1d_3}\right)\fr{1}{m_{\mathcal{A}_p}^2}\right\}\left(\frac{1}{6}+\frac{m_{B_d}^2}{\left(m_b+m_d\right)^2} \right)m_{B_d} f_{B_d}^2 . 
	\label{ptdt18}\eea
	We want to remark that the new interactions in the considering model also generate a contribution to another meson mixing parameter of the Kaon system, named CP-violating parameters $\ep_K$.  By using the formula in Ref. \cite{Aebischer:2023mbz} 
	, we find that the NP contribution to the $\ep_K$ parameter in the model can naively be expressed as $\ep_K^{H_p}\sim  \text{Re}[g_R^{H_p\bar{s}d}]\text{Im}[g_R^{H_p\bar{s}d}]$ for the neutral Higgs contribution, and as  $\ep_K^{Z'}\sim  \text{Re}[g_L^{Z'\bar{s}d}]\text{Im}[g_L^{Z'\bar{s}d}]$ for the $Z'$ contribution, with $g_L^{Z'\bar{s}d}=\fr{g_X}{3}(V^*_{u_L})_{31}(V_{u_L})_{32}$. Using the numerical inputs shown below, we approximately estimate these contributions as $\ep_K^{H_p} \sim \mathcal{O}(10^{-19})$ whereas $\ep_K^{Z'}\sim \mathcal{O}(10^{-11})$. These values are remarkably tiny compared with the SM predictions \cite{Brod:2019rzc} and the experimental values of $\ep_K^{\text{SM,exp}}\sim \mathcal{O}(10^{-3})$ \cite{Workman:2022ynf}, thus we ignore the NP contribution for this observable. On the other hand, we also do not consider NP contributions to the meson mixing observables in the up-type quark sector $\bar{D}^0-D^0$ such as mass difference $\Delta m_D$. This is justified because $\Delta m_D\sim \text{Re}[(g_R^{H_p\bar{u}c})^2+((g_R^{H_p\bar{c}u})^{*})^2]\sim \mathcal{O}(10^{-26})$ GeV for neutral Higgs contributions $H_p$, and $\Delta m_D^{Z'}\sim \text{Re}[(g_L^{Z'\bar{u}c})^2] \sim \mathcal{O}(10^{-19})$ GeV for $Z'$, which are significantly smaller compared with the current experimental constraints $\Delta m_D^{\text{exp}}=x\Ga_D\in[5.78,7.38]\times 10^{-16} $ GeV \cite{HFLAV:2022pwe}. Besides, for the theoretical calculation in SM, the double Cabibbo and GIM suppression, makes the short-distance box diagrams contributing to $\Delta m_D$ much smaller by several orders compared with long-distance diagrams \cite{Isidori:2011qw, Franco:2012ck}, responsible for the large theoretical uncertainty for this observable. We cannot get the SM prediction with high precision for $\Delta m_D$. This is not like the case of $\Delta m_{B_s, B_d}$ which is mostly affected by short-distance box diagrams. For instance, we currently have that the magnitude $\Delta m_D$ in the SM is much large than the corresponding NP contribution, i.e., $\Delta m_D^{\text{SM}}\sim \mathcal{O}(10^{-16}-10^{-15})\gg \Delta m_D^{\text{NP}}$. Consequently, the NP contributions of the model are significantly suppressed, and thus we do not consider the $\bar{D}^0-D^0$ meson mixing in our analysis.
	\subsubsection{FCCC $b\to c$  observables}
	The effective Hamiltonian that induces the $b\to c$ transition reads   
	\bea 
	\mathcal{H}^{b\to c}_{\text{eff}}&=& \fr{4G_F}{\sqrt{2}}V_{cb}[\tilde{C}^{I}_V\mathcal{\tilde{O}}^{I}_V(\mu)+\tilde{C}^{I}_S(\mu)\mathcal{\tilde{O}}^{I}_S(\mu)+\tilde{C}^{'I}_S(\mu)\mathcal{\tilde{O}}^{'I}_S(\mu)], 
	\eea
	where index $I$ denotes the lepton flavor, and the operators are given as follows
	\bea 
	\mathcal{\tilde{O}}^{I}_{V}&=&(\bar{c}\ga_{\mu}P_{L}b)(\bar{l}_I\ga^{\mu}P_L\nu_{I}),\hs 
	\mathcal{\tilde{O}}^{(')I}_{S}=(\bar{c}P_{L(R)}b)(\bar{l}_I P_L\nu_{I}).	
	\eea 
		Here the operator $\mathcal{\tilde{O}}_{V}$ is generated via the exchange of the SM charged gauge boson $W_{\mu}^{\pm}$. In the SM, the WC of this operator is $\tilde{C}^{\text{SM}}_V=1$ for all lepton flavor.  However, as pointed out in the $b\to s$ observables sector and there is mixing in lepton sector, the WCs 
		$\tilde{C}_{V}$ will be modified as $\mathcal{\tilde{C}}^{I}_{V}=\fr{(V_q)_{23}}{V_{cb}}(V^*_l)_{II}$. Here $V_l$ is the leptonic mixing matrix defined by $V_l=V^{\dagger}_{e_L}V_{\nu_L}$ (shown below). Furthermore, the new charged Higgs bosons $H^{\pm},H^{\pm}_{1,2}$ cause the operators  $\mathcal{\tilde{O}}^{(')}_{S}$. Their corresponding WCs are given as follows 
		\bea 
	\mathcal{\tilde{C}}^{(')I}_{S}&=&\tilde{C}^{(')H^{+}\bar{u}_2d_3}_{S}(\mu_H)+\tilde{C}^{(')H_{1,2}^{+}\bar{u}_2d_3}_{S}(\mu_H), 
	\eea
	with  
	\bea  
\tilde{C}^{(')H^{+}\bar{u}_2d_3}_{S}(\mu_H)&=&\fr{\sqrt{2}}{4G_FV_{cb}}\fr{g_{L(R)}^{H^+\bar{u}_2d_3}(g_R^{H^+\bar{\nu}_Il_I})^*}{m_{H^{\pm}}^2},\hs 	\tilde{C}^{(')H_{1,2}^{+}\bar{u}_2d_3}_{S}(\mu_H)=\fr{\sqrt{2}}{4G_FV_{cb}}\fr{g_{L(R)}^{H_{1,2}^+\bar{u}_2d_3}(g_R^{H_{1,2}^+\bar{\nu}_Il_I})^*}{m_{H_{1,2}^{\pm}}^2}. 
	\eea  
	Regarding the observables related to the $b\to c$ transition, we consider LFUV ratios $R_{D^{(*)}}$ which are defined by 
	\bea R_{D^{(*)}}=\fr{\int_{m_{\tau}^2}^{(m_B-m_D^{(*)})^2}\fr{d\Ga(\bar{B}\to D^{(*)}\tau\bar{\nu}_{\tau})}{dq^2} dq^2}{\int_{m_{l}^2}^{(m_B-m_D^{(*)})^2}\fr{d\Ga(\bar{B}\to D^{(*)}l\bar{\nu}_{l})}{dq^2} dq^2}, 
	\eea  
	with $q^2$ is the squared transfer momentum, and $l$ denotes either to $e$ or $\mu$. The differential decay widths of $\bar{B}\to Dl\bar{\nu}_l$ and $\bar{B}\to D^*l\bar{\nu}_l$ are given as follows 
	\bea 
	\fr{d\Ga(\bar{B}\to Dl\bar{\nu}_l)}{dq^2}&=&\fr{\eta^2_{\text{EW}}G_F^2|V_{cb}|^2m_B\sqrt{\la}}{192\pi^3}\left(1-\fr{m_l^2}{q^2}\right)^2\left\{\left[\fr{\la}{m_B^4}\left(1+\fr{m_l^2}{2q^2}\right)(f_{+}(q^2))^2+\fr{3m_l^2}{2q^2}\left(1-\fr{m_D^2}{m_B^2}\right)^2(f_0(q^2))^2\right]|C_{V}|^2\right  .\crn && + \left. \fr{3q^2}{2(m_b-m_c)^2}\left(1-\fr{m_D^2}{m_B^2}\right)^2(f_0(q^2))^2|\tilde{C}_{S}+\tilde{C}'_S|^2\right  .\crn &&  \left.+\fr{3m_l}{m_b-m_c}\text{Re}[
	\tilde{C}_{V}(\tilde{C}*_S+\tilde{C}^{'*}_S)]\left(1-\fr{m_D^2}{m_B^2}\right)^2(f_0(q^2))^2\right\}, \\
		\fr{d\Ga(\bar{B}\to D^*l\bar{\nu}_l)}{dq^2}&=&\fr{\eta^2_{\text{EW}}G_F^2|V_{cb}|^2m_B\sqrt{\la^3}}{192\pi^3}\left(1-\fr{m_l^2}{q^2}\right)^2\left\{
	\left(1+\fr{m_l^2}{5q^2}\right)\fr{5q^2(m_B+m_{D^*})^2}{2\la^*}(A_{1}(q^2))^2+	\fr{64m_B^2m_{D^*}^2}{\la^*}(A_{12}(q^2))^2\right  .\crn && + \left. \fr{3m_l^2}{2q^2}(A_0(q^2))^2|C_{V}|^2+ \fr{3q^2}{2(m_b+m_c)^2}(A_0(q^2))^2|\tilde{C}_{S}-\tilde{C}'_S|^2+\fr{m_l}{\sqrt{q^2}}\text{Re}[
	\tilde{C}_{V}(\tilde{C}^*_S-\tilde{C}^{'*}_S)](A_0(q^2))^2\right\}, 
	\eea
	where $\eta_{\text{EW}}\simeq 1.0066$ is the QED correction \cite{Sirlin:1981ie}. $f_{+,0}(q^2)$ are the vector and scalar form factors (FFs) of $B\to D$ transition, whereas $A_{12}(q^2),A_0(q^2),A_1(q^2)$ are FFs for $B\to D^*$ transition. All FFs are depend on the squared momentum transfer $q^2=(m_B-m_D)^2$, and defined explicitly in Refs. \cite{Bordone:2019vic,Bernlochner:2017jka}. Otherwise, $\la^{(*)}$ is the Kallen function $\la=m_B^4+m_{D^{(*)}}^4+q^4-2(m_B^2m_{D^{(*)}}^2+m_B^2q^2+m_{D^{(*)}}^2q^2)$, $m_l$ is the mass of daughter lepton $l$. Taking integral, we obtain the following expressions for 
	$R_{D^{(*)}}$ ratios as functions of WCs $\tilde{C}_{V_L},\tilde{C}^{(')}_S$ 
	\bea 
	R_{D}&\simeq &R_{D}^{\text{SM}}\fr{|\tilde{C}^{\tau}_{V_L}|^2+1.46\text{Re}[\tilde{C}^{\tau}_{V_L}(\tilde{C}^{'*\tau}_{S}+\tilde{C}^{*\tau}_{S})]+0.98|\tilde{C}^{'\tau}_{S}+\tilde{C}^{\tau}_S|^2}{|\tilde{C}^{\mu}_{V_L}|^2+0.14\text{Re}(\tilde{C}^{\mu}_{V_L}(\tilde{C}^{'*\mu}_{S}+\tilde{C}^{*\mu}_{S})+0.95|\tilde{C}^{'\mu}_{S}+\tilde{C}^{\mu}_S|^2}, \crn R_{D^*}&\simeq &R_{D^*}^{\text{SM}}\fr{|\tilde{C}^{\tau}_{V_L}|^2+0.127\text{Re}[\tilde{C}^{\tau}_{V_L}(\tilde{C}^{'*\tau}_{S}-\tilde{C}^{*\tau}_{S})]+0.035|\tilde{C}^{'\tau}_{S}-\tilde{C}^{\tau}_S|^2}{|\tilde{C}^{\mu}_{V_L}|^2+0.037\text{Re}(\tilde{C}^{\mu}_{V_L}(\tilde{C}^{'*\mu}_{S}-\tilde{C}^{*\mu}_{S})+0.05|\tilde{C}^{'\mu}_{S}-\tilde{C}^{\mu}_S|^2}.
	\eea 
It is important to note that the above expression of LFU ratios $R_{D^{(*)}}$ are written in the most general form and different than Refs \cite{Iguro:2018fni,Blanke:2018yud,Iguro:2022uzz} for the following reasons. Firstly, in the considering model, the charged Higgs bosons $H^{\pm}, H_{1,2}^{\pm}$ couplings to different lepton flavors are distinguishable, as given explicitly in Sec.  \ref{Coefficients_lep_quark}. As a result, this enables our predicted ratios $R_{D^{(*)}}$ to depend explicitly on both couplings of NP with $\tau$ and $\mu$ flavor. Secondly, the SM WC $\tilde{C}_V$ is modified since the CKM matrix in the model has different entries $(V_q)_{ij}$ compared to CKM ones defined by the standard parameterization \cite{Workman:2022ynf} as well as there are mixing between charged lepton and neutrino flavors. Thirdly, there are  several charged Higgs contributions $H^{\pm},H_{1,2}^{\pm}$ to these observables. Combining above mentions, $R_{D^{(*)}}$ in our model are differences compared to Refs. \cite{Iguro:2018fni,Blanke:2018yud,Iguro:2022uzz} which assume one charged Higgs contribution coupling to only $\tau$ flavor. Therefore, the below numerical studies about these observables will demonstrate differences from suck works. 

	\subsubsection{$t \to u(c)$ transitions}
	 In terms of the observables related to the up type quark transitions $t\to u(c)$, we consider the branching ratios for the FCNC top quark decays 
		$t\to u(c)h$, $t\to u(c)\ga$. 
	The branching ratios for the $t\to u(c)h$ decay are given by: 
	\bea
	\text{BR}(t\to u(c)h)=\fr{(|g_R^{h\bar{u}_{1(2)}u_3}|^2+|(g_R^{h\bar{u}_3u_{1(2)}})^*|^2)}{16\pi\Ga_t}\fr{(m_t^2-m_h^2)^2}{m_t^3},
	\eea 
where $\Ga_{t}=1.42^{+0.19}_{-0.15}$ GeV is the total decay width of top quark \cite{Workman:2022ynf}. On the other hand, the branching ratios for the radiative decays $t\to u(c)\ga i$ have the following expressions 
	\bea
	\text{BR}(t\to u_i\ga)&=&\fr{\Ga(t\to u(c)\ga)}{\Ga^{\text{total}}_{t}}=\fr{m_t^3(|C^{tu_i\ga}_L|^2+|C_R^{tu_i\ga}|^2)}{16\pi\Ga^{\text{total}}_t},\hs i=1,2,
	\eea 
	with $C^t_{L,R}$ are the coefficients combining by different contributions 
	\bea 
	C_R^{tu_i\ga}&=&\fr{iem_t}{16\pi^2m_{H^{\pm}}^2}\left[(g_{L}^{H^{-}\bar{D}_1u_i})^*g_L^{H^{-}\bar{D}_1u_3}f'''_{\ga}\left(\fr{m_{D_1}^2}{m_{H^{\pm}}^2}\right)+(g_{L}^{H^{-}\bar{D}_1u_i})^*g_R^{H^{-}\bar{D}_1u_3}\fr{m_{D_1}}{m_t}h'''_{\ga}\left(\fr{m_{D_1}^2}{m_{H^{\pm}}^2}\right)\right] \crn &&+\fr{iem_t}{16\pi^2m_{H_{1(2)}^{\pm}}^2}\left[(g_{L}^{H_{1(2)}^{-}\bar{D}_1u_i})^*g_L^{H_{1(2)}^{-}\bar{D}_1u_3}f'''_{\ga}\left(\fr{m_{D_1}^2}{m_{H_{1(2)}^{\pm}}^2}\right)+(g_{L}^{H_{1(2)}^{-}\bar{D}_1u_i})^*g_R^{H_{1(2)}^{-}\bar{D}_1u_3}\fr{m_{D_1}}{m_t}h'''_{\ga}\left(\fr{m_{D_1}^2}{m_{H_{1(2)}^{\pm}}^2}\right)\right] \crn &&+\fr{iem_t}{24\pi^2}\sum_{p=1}^8\fr{1}{m_{H_p}^2}\left[(g_{L}^{H_{p}\bar{U}u_i})^*g_L^{H_{p}\bar{U}u_3}f'_{\ga}\left(\fr{m_{U}^2}{m_{H_{p}}^2}\right)+(g_{L}^{H_{p}\bar{U}u_i})^*g_R^{H_{p}\bar{U}u_3}\fr{m_U}{m_t}h'_{\ga}\left(\fr{m_{U}^2}{m_{H_{p}}^2}\right)\right]\crn&& +\fr{iem_t}{24\pi^2}\sum_{p=3}^8\fr{1}{m_{\mathcal{A}_p}^2}\left[(g_{L}^{\mathcal{A}_{p}\bar{U}u_i})^*g_L^{\mathcal{A}_{p}\bar{U}u_3}f'_{\ga}\left(\fr{m_{U}^2}{m_{\mathcal{A}_{p}}^2}\right)+(g_{L}^{\mathcal{A}_{p}\bar{U}u_i})^*g_R^{\mathcal{A}_{p}\bar{U}u_3}\fr{m_U}{m_t}h'_{\ga}\left(\fr{m_{U}^2}{m_{\mathcal{A}_{p}}^2}\right)\right], \crn 
	C_L^{tu_i\ga}&=&\fr{iem_t}{16\pi^2m_{H^{\pm}}^2}\left[(g_{R}^{H^{-}\bar{D}_1u_i})^*g_R^{H^{-}\bar{D}_1u_3}f'''_{\ga}\left(\fr{m_{D_1}^2}{m_{H^{\pm}}^2}\right)+(g_{R}^{H^{-}\bar{D}_1u_i})^*g_L^{H^{-}\bar{D}_1u_3}\fr{m_{D_1}}{m_t}h'''_{\ga}\left(\fr{m_{D_1}^2}{m_{H^{\pm}}^2}\right)\right] \crn &&+\fr{iem_t}{16\pi^2m_{H_{1(2)}^{\pm}}^2}\left[(g_{R}^{H_{1(2)}^{-}\bar{D}_1u_i})^*g_R^{H_{1(2)}^{-}\bar{D}_1u_3}f'''_{\ga}\left(\fr{m_{D_1}^2}{m_{H_{1(2)}^{\pm}}^2}\right)+(g_{R}^{H_{1(2)}^{-}\bar{D}_1u_i})^*g_L^{H_{1(2)}^{-}\bar{D}_1u_3}\fr{m_{D_1}}{m_t}h'''_{\ga}\left(\fr{m_{D_1}^2}{m_{H_{1(2)}^{\pm}}^2}\right)\right] \crn &&+\fr{iem_t}{24\pi^2}\sum_{p=1}^8\fr{1}{m_{H_p}^2}\left[(g_{R}^{H_{p}\bar{U}u_i})^*g_R^{H_{p}\bar{U}u_3}f'_{\ga}\left(\fr{m_{U}^2}{m_{H_{p}}^2}\right)+(g_{R}^{H_{p}\bar{U}u_i})^*g_L^{H_{p}\bar{U}u_3}\fr{m_U}{m_t}h'_{\ga}\left(\fr{m_{U}^2}{m_{H_{p}}^2}\right)\right]\crn&& +\fr{iem_t}{24\pi^2}\sum_{p=3}^8\fr{1}{m_{\mathcal{A}_p}^2}\left[(g_{L}^{\mathcal{A}_{p}\bar{U}u_i})^*g_L^{\mathcal{A}_{p}\bar{U}u_3}f'_{\ga}\left(\fr{m_{U}^2}{m_{\mathcal{A}_{p}}^2}\right)+(g_{R}^{\mathcal{A}_{p}\bar{U}u_i})^*g_L^{\mathcal{A}_{p}\bar{U}u_3}\fr{m_U}{m_t}h'_{\ga}\left(\fr{m_{U}^2}{m_{\mathcal{A}_{p}}^2}\right)\right]\crn &&+\fr{iem_t}{24\pi^2}\left[\sum_{p=1}^8\sum_{a=1}^3\fr{1}{m_{H_p}^2}(g_{R}^{H_p\bar{u}_au_i})^*g_R^{H_p\bar{u}_au_i}f'_{\ga}\left(\fr{m_{u_a}^2}{m_{H_p}^2}\right)+ \sum_{p=3}^8\sum_{a=1}^3\fr{1}{m_{\mathcal{A}_p}^2}(g_{R}^{\mathcal{A}_p\bar{u}_au_i})^*g_R^{\mathcal{A}_p\bar{u}_au_i}f'_{\ga}\left(\fr{m_{u_a}^2}{m_{\mathcal{A}_p}^2}\right)\right], 
	\eea  
	where the loop functions are shown in Appendix \ref{Loopfuncs}. Similarly to the mentioned decays $b\to s\ga$, we should consider the QCD corrections in $t\to u(c)\ga$ decays. However, the QCD effects at next-to-leading order (NLO) to $t\to u(c)\ga$ are negligible and modify the branching ratios of such decays around 0.2\% compared to the LO contribution \cite{Zhang:2008yn}. Moreover, both the measurements and theoretical predictions for such decays are currently not precisely known, as compared to the $b\to s\ga$ decay \footnote{These decays currently just have the experimental upper limits for branching ratios, which are indicated in Table \ref{SM_exp data}}. Thus the role of QCD corrections is insignificant, and we can ignore the QCD effects in such observables. 
	\subsubsection{Constraints on quark flavor processes}
	All the predicted observables should be compared with the corresponding experimental values shown in the last column in Table \ref{SM_exp data}. It is worth mentioning that the central values of SM prediction and the measurement results of some "clean" observables are quite close, including BR$(B_s\to \mu^+\mu^-)$, BR$(\bar{B}\to X_s\ga)$, LFUV ratios $R_{D^{(*)}}$ and BR$(B^-_i\to \tau\bar{\nu}_{\tau})$. However, we should take into account the effect of both SM and experimental uncertainties. Therefore, it is better to consider the ratios between SM and respective experimental values on each clean observable since the uncertainties can be reduced via the numerator and denominator of these ratios. Moreover, considering ratios like that also causes the overall factors to be canceled.  For instance, with the branching ratios BR$(B_s\to \mu^+\mu^-)$ and BR$(\bar{B}\to X_s\gamma)$, we have the following constraints at $3\sigma$ range as follows 
	\bea 
	\fr{\text{BR}(B_s\to \mu^+\mu^-)_{\text{exp}}}{\text{BR}(B_s\to \mu^+\mu^-)_{\text{SM}}}&& =\fr{1}{1-y_s}\fr{\left(1-\fr{4m_{\mu}^2}{m_{B_s}^2}\right)|\tilde{S}|^2+|\tilde{P}|^2}{|C_{10}^{\text{SM}}|^2}=0.9426(1\pm 0.2772), \label{Bsmm_constraint}
	\eea 
	with 
	\bea 
	&& \tilde{P}=C_{10}+\fr{m_{B_s}^2}{2m_{\mu}(m_b+m_s)}(C_P-C_P'), \hs  \tilde{S}=\fr{m_{B_s}^2}{2m_{\mu}(m_b+m_s)}(C_S-C_S'), \eea 
	\bea
	\fr{\text{BR}(\bar{B}\to X_s\gamma)_{\text{exp}}}{\text{BR}(\bar{B}\to X_s\gamma)_{\text{SM}}} &=& 1+\fr{|C_7^{\text{NP}}|^2+|C_7^{'\text{NP}}|^2+2C_7^{\text{SM}}\text{Re}[C^{\text{NP}}_7]}{|C_7^{\text{SM}}|^2+N(E_{\ga})}\crn 
	&=& 1.0265(1\pm0.2217) \label{bsga_constraint}.
	\eea
	For constraint of LFV ratios $R_{D^{(*)}}$, we also obtain 
	\bea
\fr{R_{D}^{\text{exp}}}{R_D^{\text{SM}}}&=&1.106(1\pm0.109), \hs   \fr{R_{D}^{*\text{exp}}}{R_D^{*\text{SM}}}=1.48(1\pm0.203) .\label{RD_RDs}	
	\eea
	In addition, we obtain constraints for $B^0_{s,d}$--$\bar{B}^0_{s,d}$ meson systems as 
	\be
	\fr{(\Delta m_{B_d})_{\mathrm{SM}}}{(\Delta m_{B_d})_{\mathrm{exp}}}
	=1.0721(1\pm0.1605), \hs 
	\fr{(\Delta m_{B_s})_{\mathrm{SM}}}{(\Delta m_{B_s})_{\mathrm{exp}}} =1.0566(1\pm0.1374).  \label{DmBq} \ee 
	
	However, in $K^0$--$\bar{K}^0$ meson system, the lattice QCD calculations for long-distance effect are not well-controlled. Therefore, we assume the present theory contributes about 30\% to $\Delta m_K$, it reads    
	\bea 
	\fr{(\Delta m_{K})_{\text{SM}}}{(\Delta m_{K})_{\text{exp}}}=1(1\pm0.3), \eea 
	and then translates to the following constraint
	\bea  \fr{(\Delta m_{K})_{\text{NP}}}{(\Delta m_{K})_{\text{exp}}}\in [-0.3,0.3], 
	\label{DmK}\eea
	in agreement with \cite{Buras:2015kwd}. 
	
	For other observables such as branching ratios of FCNC top quark decays $t\to hu(c), t\to u(c)\gamma$, $B_s\to \tau^+\mu^-, B^+\to K^+\tau^+\mu^-$ and $B^+\to K^+\tau^+\tau^-$, we will compare their theory predictions with corresponding upper experimental limits.
	\begin{table}[H]
		\begin{centering}
			\begin{tabular}{|c|c|c|}
				\hline
				Observables & SM predictions  & Experimental constraints  \tabularnewline
					\hline 
				$\Delta m_K$ & $0.467\times 10^{-2}   \  \text{ps}^{-1}$ \cite{Workman:2022ynf}& $0.5293(9)\times 10^{-2}  \  \text{ps}^{-1}$ \cite{Workman:2022ynf} \tabularnewline
				$\Delta m_{B_s}$ & $18.77(86) \  \text{ps}^{-1}$ \cite{Lenz:2019lvd} &  $17.765(6)  \  \text{ps}^{-1}$ \cite{HFLAV:2022pwe} \tabularnewline
				$\Delta m_{B_d}$ & $0.543(29)\ \text{ps}^{-1} $ \cite{Lenz:2019lvd}&	$0.5065(19)  \  \text{ps}^{-1}$  \cite{HFLAV:2022pwe} \tabularnewline
				$\text{BR}(B_s\to \mu^+\mu^-)$ & $(3.66 \pm 0.14 ) \times 10^{-9} $ \cite{Beneke:2019slt} &	$(3.45\pm 0.29) \times 10^{-9}$  \cite{HFLAV:2022pwe} \tabularnewline 
				$\text{BR}(\bar{B}\to X_s \gamma)$ & $(3.40 \pm 0.17) \times 10^{-4}$ \cite{Misiak:2020vlo} &	$(3.49\pm 0.19) \times 10^{-4}$  \cite{HFLAV:2022pwe}  \tabularnewline
				$R_{D}$ & $0.298\pm0.004$ \cite{Bigi:2016mdz,Jaiswal:2017rve,Martinelli:2021onb}&	$0.441\pm0.060\pm0.066$ \cite{LHCb:2023zxo}	 
				\tabularnewline
				$R_{D^*}$ & $0.254\pm0.005$ \cite{BaBar:2019vpl,Gambino:2019sif}&	$0.281\pm0.018\pm0.024$ \cite{LHCb:2023zxo}
				\tabularnewline	
				BR$(t\to hu)$ & $2\times10^{-17}$ \cite{Eilam:1990zc,Aguilar-Saavedra:2004mfd}&	$<1.9\times 10^{-4}$ \cite{Workman:2022ynf}
				\tabularnewline
				BR$(t\to hc)$ & $3\times 10^{-15}$ \cite{Eilam:1990zc,Aguilar-Saavedra:2004mfd}&	$<7.3\times 10^{-4}$ \cite{Workman:2022ynf}
				\tabularnewline
				BR$(t\to u\ga)$ & $4\times10^{-16}$ \cite{Eilam:1990zc,Aguilar-Saavedra:2004mfd}&	$<0.85\times 10^{-5}$ \cite{ATLAS:2022per}
				\tabularnewline
				BR$(t\to c\ga)$ & $5\times10^{-14}$ \cite{Eilam:1990zc,Aguilar-Saavedra:2004mfd}&	$<4.2\times 10^{-5}$ \cite{ATLAS:2022per}
				\tabularnewline
				BR$(B_s\to \tau^+\mu^-)$ & $0$ \cite{Workman:2022ynf}&	$<4.2\times 10^{-5}$ \cite{LHCb:2019ujz}
				\tabularnewline
				BR$(B^+\to K^+\tau^+\mu^-)$ & $0$ \cite{Workman:2022ynf}&	$<3.3\times 10^{-5}$ \cite{BaBar:2012azg}
				\tabularnewline
				BR$(B^+\to K^+\tau^+\tau^-)$ & $(1.4\pm0.2)\times 10^{-7}$ \cite{Cornella:2019hct}&	$<2.25\times 10^{-3}$ \cite{Workman:2022ynf}
				\tabularnewline
				\hline 
			\end{tabular}
			\par
		\end{centering}	
		\caption{\label{SM_exp data} The SM predictions and experimental values for flavor-changing observables related to quark sectors.}
	\end{table}
	\section{Numerical results \label{Sec5}}
	\subsection{Set up input parameters \label{Sec5_input_par}}
	In this section, we perform a numerical analysis 
	of all observables above mentioned in the lepton and quark sectors. We first provide some comments about the input parameters. In the quark sector, we use the benchmark points satisfying the observed quark spectrum, given in the Ref. \cite{Hernandez:2021xet} 
and then we obtain the numerical forms of the mixing matrices $V_{(u,d)_{L, R}}$ for the left- and right-handed SM up (type) quarks, respectively, as follows 
	\bea 
	&& 	V_{u_L}\simeq \left( 
	\begin{array}{ccc}
		-0.691643 & -0.72224 & -3.4606\times 10^{-5} \\ 
		-0.03477-0.7214i & 0.03331+0.69084i & 5.4627\times 10^{-5}+2.987\times 10^{-5}i \\ 
		2.4033\times10^{-5}+2.99143\times10^{-5}i & -5.9984\times 10^{-5}+1.835\times10^{-5} & 0.77155-0.63617i 
	\end{array}%
	\right),  
	\crn  && 	V_{d_L}\simeq \left( 
	\begin{array}{ccc}
		0.59424 & -0.80365 & -0.03216 \\ 
		0.33464+0.73129i & 0.24752+0.53964i & -0.001978+0.02746i \\ 
		-0.00852-0.006394i & 0.01665-0.03739i& -0.57436+0.817511i
	\end{array}%
	\right) ,\label{VuL_VdL}
	\eea 
	
	\bea 
	&& 	V_{u_R}\simeq\left( 
	\begin{array}{ccc}
		0.062774+0.0789265i & -0.772558 + 0.217704 i & 0.45357 - 0.373982 i \\ 
		-0.324679 - 0.404588i& 0.444546 - 0.124976i& 0.555116 - 0.45771 i \\ 
		-0.531888 - 0.661678i& -0.363214 + 0.102927i& -0.285327 + 
		0.235261 i
	\end{array}%
	\right),  
	\crn  && 	V_{d_R}\simeq \left( 
	\begin{array}{ccc}
		-0.615248 - 0.345117 i& 0.152051 + 0.37215i& -0.33544 + 0.477733 i \\ 
		0.0635554 - 0.064921i& -0.158241 - 0.795923i& -0.332068 + 
		0.47217i\\ 
		-0.56473 - 0.418558i& -0.00509284 - 0.424059i&
		0.328169 - 0.467272 i
	\end{array}%
	\right) .\label{VuR_VdR}
	\eea 
	We would like to note that the obtained quark mixing matrix is defined as follows \bea V_q=V_{u_L}^{\dagger}V_{dL}=\left(
	\begin{array}{ccc}
		-0.950192+0.215979 i & 0.157931\, +0.159795 i & 0.00248254\, -0.00234475 i \\
		0.0871643\, -0.206829 i & 0.961472\, -0.153024 i & 0.0421355\, +0.00223271 i \\
		-0.00248469-0.0103215 i & 0.0366984\, -0.0182365 i & -0.96322+0.26536 i \\
	\end{array}
	\right), \eea 
	 is unitary and its magnitude $|V|$ is in agreement with the constraints of absolute values of the CKM matrix given in \cite{Workman:2022ynf}. However, the entries themselves $V_{ij}$ are different than the corresponding CKM ones. 
	 
	 Additionally, the 
	 exotic up (down) type quarks $U, T$ ($D_1, B$) are 
	 nearly degenerate with masses at the TeV scale, i.e  $m_U\simeq m_T \sim \mathcal{O}(1)$ TeV, $m_{D_1}\simeq m_B\sim \mathcal{O}(1)$ TeV. Moreover, they barely mix with ordinary quarks $u(d)$, as stated in \cite{Hernandez:2021xet}. Therefore, we ignore their mixing with the SM quarks and consider these new quarks as physical fields. 
	
	Similarly, in the lepton sector, we have found the numerical forms for the charged leptonic mixing matrices $V_{e_{L(R)}}$ obtained from the benchmark points in Eq. (\ref{benchmarkleptons}). These mixing matrices are given by 
	\bea 
	V_{e_L}\simeq \left(
	\begin{array}{ccc}
		0.120049 & 0.0789769 & -0.989622 \\
		-0.35756 & 0.933372 & 0.0311129 \\
		-0.926142 & -0.350114 & -0.140289 \\
	\end{array}
	\right) ,\hs 	V_{e_R}\simeq \left(
	\begin{array}{ccc}
		-0.472389 & 0.871878 & -0.129139 \\
		-0.88139 & -0.467177 & 0.0699824 \\
		-0.000685472 & -0.14688 & -0.989154 \\
	\end{array}
	\right).   \label{VeL_eR}
	\eea 
	The mixing matrix of active neutrinos $V_{\nu_L}$ is given by: 
		\bea 
	V_{\nu_L}
	\simeq\left(
	\begin{array}{ccc}
		-0.331797-0.0666808 i & 0.64392\, -0.0408052 i & -0.684818+0.0137822 i \\
		-0.598615+0.0635722 i & 0.420741\, +0.0389029 i & 0.676312\, -0.0410494 i \\
		-0.722483-0.0331869 i & -0.636202-0.0203087 i & -0.24586-0.106325 i \\
	\end{array}
	\right), \label{VnuL}
	\eea
	such the 
	leptonic mixing matrix $V_l=V_{e_L}^{\dagger}V_{\nu_L}$ is unitary and agrees with the constraints imposed by the allowed ranges of the absolute values of the PMNS mixing matrix entries, given in \cite{Gonzalez-Garcia:2021dve,Workman:2022ynf}.
	
	On the one hand, the masses and eigenstates of exotic charged leptons $E_{1,2}$ are supposed to be the same exotic quarks, i.e they do not mix with SM charged leptons and are nearly degenerate in mass ($m_{E_1}\sim m_{E_2}\sim \mathcal{O}(1)$ TeV). On the other hand, the mixing matrices of right-handed neutrinos $\nu_R$ and neutral leptons $N_R$ are assumed to be diagonal, i.e $V_{\nu_R}=I, V_{N_R}=I$, for simplicity. The active neutrino masses are chosen in the normal hierarchy as follows 
	\bea
	m_{\nu_{1L}}=0.5 \ \text{eV}, \hs m^2_{\nu_{2L}}= m_{\nu_{1L}}^2+\Delta m_{21}^2, \hs m^2_{\nu_{3L}}= m_{\nu_{1L}}^2+\Delta m_{31}^2, 
	\eea 
	where $\Delta m_{21}^2, \Delta m_{31}^2$ are given in \cite{Gonzalez-Garcia:2021dve}. Besides, the masses of heavy neutral lepton $N_R$ and right-handed neutrinos $\nu_R$ are obtained from the inverse seesaw mechanism as $m_{N_{aR}}\simeq  m_{\nu_{aR}}=\fr{v_{\sigma}}{\sqrt{2}}$ \cite{Hernandez:2021xet}. 
 With the remaining SM parameters used in our numerical study, we list them in Table \ref{input-par}. 
	\begin{table}[H]
		\protect\caption{\label{input-par} The numerical values of input parameters.}
		\begin{centering}
			\begin{tabular}{|c|c|c|c|}
				\hline
				Input parameters & Values  & Input parameters & Values \tabularnewline
				\hline 
			
				$f_{B_s}$ & $230.3(1.3) \ \text{MeV}$ \cite{FlavourLatticeAveragingGroupFLAG:2021npn}	 &  $ m_{B_s}$ & $5366.88(11) \ \text{MeV}$  \cite{Workman:2022ynf}\tabularnewline
				$ m_{u}$ & $1.24(22) \ \text{MeV}$  \cite{Xing:2020ijf} &  $m_{d}$ & $2.69(19)  \ \text{MeV} $ \cite{Xing:2020ijf}\tabularnewline
				
				$m_{c} $ & $0.63(2)  \ \text{GeV} $ \cite{Xing:2020ijf} & $ m_{s}$ & $53.5(4.6) \ \text{MeV}$  \cite{Xing:2020ijf} \tabularnewline
				$m_{t}$ & $172.9(4) \  \text{GeV} $ \cite{Xing:2020ijf} & $m_{b}$ & $2.86(3)  \ \text{GeV} $ \cite{Xing:2020ijf} \tabularnewline
				$N(E_{\ga})$ & $3.3\times 10^{-3} $ \cite{Misiak:2020vlo}& $C_7^{\text{SM}}(\mu_b=2.0 \ \text{GeV})$ & $-0.3636  $\cite{Misiak:2006ab,Czakon:2006ss,Misiak:2020vlo} \tabularnewline
				$C_9^{\text{SM}}(\mu_b=5.0 \ \text{GeV})$ & $4.344  $\ \cite{Beneke:2017vpq} &	$C_{10}^{\text{SM}}(\mu_b=5.0 \ \text{GeV})$ & $-4.198  $\  \cite{Beneke:2017vpq} \tabularnewline
				$y_s$ & $0.0645(3) $ \cite{HFLAV:2022pwe}& $\ka_7 $ &$0.408 $ \cite{Buras:2011zb} \tabularnewline
				$\ka_7 $ & $0.408 $ \cite{Buras:2011zb}& $\ka_8$ &$  0.129 $ \tabularnewline
					$m_W $ & $80.385\ \text{GeV}$  \cite{Czakon:2015exa} &
				$m_Z $ & $91.1876 \ \text{GeV}$ \cite{Czakon:2015exa} \tabularnewline
				$G_F$ & $1.166379 \times 10^{-5} \ \text{GeV}^{-2}$ \cite{Workman:2022ynf} &
				$s_W^2$ & $0.2312$ \cite{Workman:2022ynf} \tabularnewline
					$\la$ & $0.22519(83) $  \cite{UTfit:2022hsi} &
				$A$ & $0.828(11)$ \cite{UTfit:2022hsi} \tabularnewline
					$\bar{\rho}$ & $0.1609(95)$ \cite{UTfit:2022hsi} &
				$\bar{\eta}$ & $0.347(10)$ \cite{UTfit:2022hsi} \tabularnewline
				\hline 
			\end{tabular}
			\par
		\end{centering}

	\end{table} 

	With all specified setup of input parameters, the free parameters used in the analysis of the 
		observables are only the couplings $f, B$ appearing in the scalar potential, the masses of charged exotic fermions $m_{E_1},m_{D_1}, m_U$, the charged Higgs boson masses $m_{H^{\pm}}, m_{H_{1,2}^{\pm}}$, the mixing angle $\theta$ between $m_{H_{1,2}^{\pm}}$. For our numerical analysis, we will 
		randomly vary these unknown parameters in the following ranges 
	\bea 
	-f\in[10^{-3},10^{-1}],\hs \theta \in [0,\pi/2], \hs B\in [10^{-3},10^{-2}] \ \text{TeV}, \hs m_{E_1}\in [1,4] \ \text{TeV}, \hs m_{H^{\pm},H_{1,2}^{\pm}} \in [0.5, 1.5] \ \text{TeV} .
	\eea 
	
	Below, we provide a justification for the choice of these ranges.  
	The ranges of $f$ and $B$ are chosen based on the assumptions used in the diagonalization of the scalar mass matrices, whereas the mass ranges of exotic fermions and charged Higgs bosons are obtained from the experimental exclusion limits arising from collider searches \cite{Sanyal:2019xcp, CMS:2020osd}. Additionally, the mass range of charged Higgs bosons also satisfies the experimental constraints resulting from the experimentally allowed ranges of the anomalous magnetic moments of the electron and muon $\Delta a_{e,\mu}$, as shown in \cite{Hernandez:2021xet}. 
	\subsection{SM-like Higgs boson decays $h\to \bar{f}f$ and $h\to \bar{f}f'$ \label{a_hff}}
	Firstly, we are interesting in the deviations factors $a_{h\bar{f}f}$ from the couplings of SM like Higgs boson $h$ with fermions $f$ as follows 
	\bea a_{h\tau\tau}=\fr{(g_{h\tau\tau})_{\text{theory}}}{(g_{h\tau\tau})_{\text{SM}}},\hs a_{h\mu\mu}=\fr{(g_{h\mu\mu})_{\text{theory}}}{(g_{h\mu\mu})_{\text{SM}}}, \hs a_{h\bar{b}b}=\fr{(g_{h\bar{b}b})_{\text{theory}}}{(g_{h\bar{b}b})_{\text{SM}}},\hs a_{h\bar{t}t}=\fr{(g_{h\bar{t}t})_{\text{theory}}}{(g_{h\bar{t}t})_{\text{SM}}},
	\eea 
	where $(g_{h\bar{f}f})_{\text{theory}}$ are the couplings predicted by the model and are obtained from the last lines of Eq. (\ref{gLRd}) and Eq. (\ref{gLRu}. The $(g_{h\bar{f}f})_{\text{SM}}=m_f/v$ are the SM predictions. With the help of input parameters given in Sec. (\ref{Sec5_input_par}), we find the predicted values of the model for these factors and 
	are shown them in the Table (\ref{ahff}). 
	We see that factors $a_{h\tau^+\tau^-},a_{h\bar{b}b}$ and $a_{h\bar{t}t}$ are in agreement with 
	the $1\sigma$ experimentally allowed range reported by ALTAS and CMS \cite{ATLAS:2022per,CMS:2022dwd}, whereas the model estimates $a_{h\mu^+\mu^-}$ is in the $3\sigma$ experimentally allowed range of the ATLAS result. This can be understood because the second and first SM fermion families 
	receive tree and one loop level masses from the  
	inverse seesaw mechanism, whereas the masses for the third generation of SM fermions are generated at tree level by Yukawa interactions involving the Higgs doublets $\phi_{1,2}$ ($\phi_{1}$ for the top and $\phi_{2}$ for the bottom and tau lepton). Therefore, the couplings of the SM-like Higgs boson $h$ with the first and second SM fermion generations are smaller than the coupling of the $125$ GeV SM-like Higgs boson $h$ with the third family of the SM fermions. 
	\begin{table}[H]
		\protect\caption{\label{ahff} The comparison between predicted values and experimental limits of deviation factors. $a_{h\bar{f}f}$ }  
		\begin{centering}
			\begin{tabular}{|c|c|c|c|c|}
				\hline
				Observables $a_{h\bar{f}f}$ & $a_{h\mu\mu}$ & $a_{h\tau\tau}$ &$a_{hbb}$&$a_{htt}$ \tabularnewline
				\hline 
				Predicted values& $\simeq 0.143$&$\simeq 0.86$&$\simeq 1.01$&$\simeq 0.997$ \tabularnewline
				\hline 
			\end{tabular}
			\par
		\end{centering}	
	\end{table}
Besides, the model also predicts the LFV decays of SM like Higgs boson $h$, namely BR$(h\to \bar{l}_{b_1}l_{b_2})$, which are demonstrated in the below Table (\ref{h_fifj})
\begin{table}[H]
	\protect\caption{\label{h_fifj} The comparison between predicted values and experimental limits of BR$(h\to \bar{l}_{b_1}l_{b_2})$.}
	\begin{centering}
		\begin{tabular}{|c|c|c|}
			\hline
			Branching ratios & Predicted values & Experimental upper limits \tabularnewline
			\hline 
			$h\to e \mu $ & $ \simeq 2.7\times10^{-10}$ & $4.4\times 10^{-5}$ \cite{CMS:2023pte}\tabularnewline
			$h\to e\tau $ & $ \simeq 3.0\times 10^{-8}$& $2.0\times 10^{-3}$  \cite{ATLAS:2023mvd}\tabularnewline
			$h\to \mu\tau $ & $\simeq 1.2\times 10^{-3}$&  $1.8\times 10^{-3}$\cite{ATLAS:2023mvd} \tabularnewline
			\hline 
		\end{tabular}
		\par
	\end{centering}	
\end{table}
We see that all predicted branching ratios of LFV decays of SM like Higgs boson $h\to e\mu,e\tau,\mu\tau$ satisfy the current upper experimental bounds \cite{CMS:2023pte, ATLAS:2023mvd}. Specifically, the decays  $h\to e\mu,e\tau$ are significantly lower by several orders of magnitudes compared with corresponding experiment results, whereas $h\to \mu\tau$ channel is quite close to its measurement limit. This behavior of the branching ratio for the $h\to \mu\tau$ decay can be interpreted similarly as in Table \ref{ahff}, where the couplings of SM-like $125$ GeV Higgs boson $h$ with the third generation of SM charged fermions arise at tree-level, which are much larger than the couplings of $h$ with the remaining SM fermions resulting from the inverse seesaw mechanisms at tree and one-loop levels. It is worth mentioning that the flavor changing leptonic Yukawa interactions in our model are dynamically generated at one loop level below the scale of spontaneous breaking of the $U(1)_X\times Z_4$ symmetry, after we integrate out the heavy neutral leptonic seesaw messengers and the charged scalars running in the internal lines of the loop. Such interactions give rise to flavor changing Higgs decays $h\to e\mu$, $h\to e\tau$ and $h\to\mu\tau$ whose corresponding rates do depend on the corresponding flavor changing Yukawa couplings. As a consequence of our particular benchmark in the charged lepton sector, we find rates for the  $h\to e\mu$ and $h\to e\tau$ decays, very well below their current experimental limits, whereas for the $h\to\mu\tau$ decay, we find that its corresponding rate is below and very close to its upper experimental bound. This makes the $h\to\mu\tau$ decay one example of a smoking gun signature which can be used to assess the viability of our model. 

On the other hand, it is important to note that the model contains a $Z-Z'$ mixing, which is controlled by the small mixing angle $\tan{\theta_{ZZ'}}\sim\mathcal{O}\left(\frac{v^2}{\Lambda^2_{new}}\right)\sim\mathcal{O}\left(10^{-4}\right)$ (whose exact expression is given in Appendix D), then implying the existence of additional contributions to $Z$ boson decays including both flavor-conserving $Z\to \bar{f}f$ and flavor-violating ones $Z\to \bar{f}_af_b$ ($a\neq b$). The flavor-violating $Z$ decays are originated by the non-universal $U(1)_X$ gauge symmetry assignments of the SM left handed quark fields, as well as to the $Z-Z'$ mixing. The branching ratios for the flavor-violating $Z\to \bar{f_a}f_b$ decays read
\bea 
\text{BR}(Z\to \bar{f}_a f_b)=\fr{M_Z}{12\pi\Ga_{Z}}\left(\left|\left(g_{L}^{Z}\right)_{ab}\right|^2+\left|\left(g_{R}^{Z}\right)_{ab}\right|^2\right)=\fr{M_Z}{6\pi\Ga_Z}\left(\left|\left(g_{V}^{Z}\right)_{ab}\right|^2+\left|\left(g_{A}^{Z}\right)_{ab}\right|^2\right),
\eea
where the vector (axial) couplings of $Z$ gauge boson are provided in Appendix \ref{Z couplings}. As follows from the expression given above as well as from the expressions of the $Z$ couplings given in Appendix \ref{Z couplings}, the non-SM contributions to the $Z$ flavor-conserving couplings with SM fermions will be of the order of $\sim s_{\theta_{ZZ'}}\sim \mathcal{O}\left(10^{-4}\right)$, which implies that the branching ratios of $Z$ decays into SM fermions will be very close to the SM predictions and thus consistent with the current experimental constraints \cite{Workman:2022ynf}. For quark flavor-violating decays $Z\to bs,bd,cu$, the model predicts these processes can happen at the tree-level due to non-universal assignments of the quark fields under $U(1)_X$ gauge group and the $Z-Z'$ mixing. As seen from Appendix \ref{Z couplings}, we can estimate New Physics (NP) contributions to BR$(Z\to \bar{s}b)\sim s_{\theta_{ZZ'}}^2|(V^*_{d_L})_{32}(V_{d_L})_{33}|^2\sim (10^{-4})^2\times 10^{-3}\sim 10^{-11}$, BR$(Z\to \bar{d}b)\sim s_{\theta_{ZZ'}}^2|(V^*_{d_L})_{31}(V_{d_L})_{33}|^2\sim (10^{-4})^2\times 10^{-4}\sim 10^{-12}$, BR$(Z\to \bar{u}c)\sim s_{\theta_{ZZ'}}^2|(V^*_{u_L})_{31}(V_{u_L})_{32}|^2\sim (10^{-4})^2\times 10^{-18}\sim 10^{-26}$. These ones are much smaller than the upper experimental limits BR$(Z\to \left\{bs,bd,cu\right\})_{\text{exp}}<2.9\times 10^{-3}$ as well as SM prediction BR$(Z\to \left\{bs,bd,cu\right\})_{\text{SM}}=\left\{4.2(7)\times 10^{-8},1.8(3)\times 10^{-9},1.4(2)\times 10^{-18}\right\}$ \cite{Kamenik:2023hvi}, therefore the NP impact to quark flavor-violating decays of $Z$ boson are very weak and thus can be ignored.
 
On the other hand, we would also like to note that there are lepton flavor-violating $Z$ decays induced at one-loop level including the virtual exchange of charged Higgses $H^{\pm},H_{1,2}^{\pm}$, neutral CP even (odd) Higgses $H(\mathcal{A})$ running in the internal lines of the loop. For instance, the branching ratio for the $Z\to e\mu$ decay is approximately estimated as follows 
	\bea \text{BR}(Z\to e\mu)&\sim& \fr{1}{16\pi^2}\left\{|\sum_{F,\phi}(g^{\phi^+\bar{F}l_{1}}_R)^*g^{H^+\bar{F}l_{2}}_R+\sum_{p=1}^8[(g^{H_p\bar{E}_1l_{1}}_L+g^{H_p\bar{E}_1l_{1}}_R)^*(g^{H_p\bar{E}_1l_{2}}_L+g^{H_p\bar{E}_1l_{2}}_R)+(g^{H_p\bar{l}_al_{1}}_R)^*g^{H_p\bar{l}_al_{2}}_R]\right . \crn  && \left .+\sum_{p=3}^8[(g^{\mathcal{A}_p\bar{E}_1l_{1}}_L+g^{\mathcal{A}_p\bar{E}_1l_{1}}_R)^*(g^{\mathcal{A}_p\bar{E}_1l_{2}}_L+g^{\mathcal{A}_p\bar{E}_1l_{2}}_R)+(g^{\mathcal{A}_p\bar{l}_al_{1}}_R)^*g^{\mathcal{A}_p\bar{l}_al_{2}}_R]|^2 \right\}\sim \mathcal{O}(10^{-14}), \crn \eea 
	 where $1/(16\pi^2)$ is a loop factor, and $\phi=H^+,H_{1,2}^{+}$, $F=\nu_{aL},\nu_{aR}, N_a$.  Similarly, we find $ \text{BR}(Z\to \mu\tau)\sim \mathcal{O}(10^{-8})$ and $ \text{BR}(Z\to e\tau)\sim \mathcal{O}(10^{-13})$. These predicted values are very suppressed in comparison with the order of magnitude of the upper experimental limits $ \mathcal{O}(10^{-6}-10^{-7})$ \cite{Workman:2022ynf}. This shows the NP constraints for the lepton-flavor violating $Z$ decays are very weak and thus can also be skipped. 
	\subsection{Leptonic flavor observables}
\begin{figure}[H]
		\includegraphics[width=0.51\textwidth]{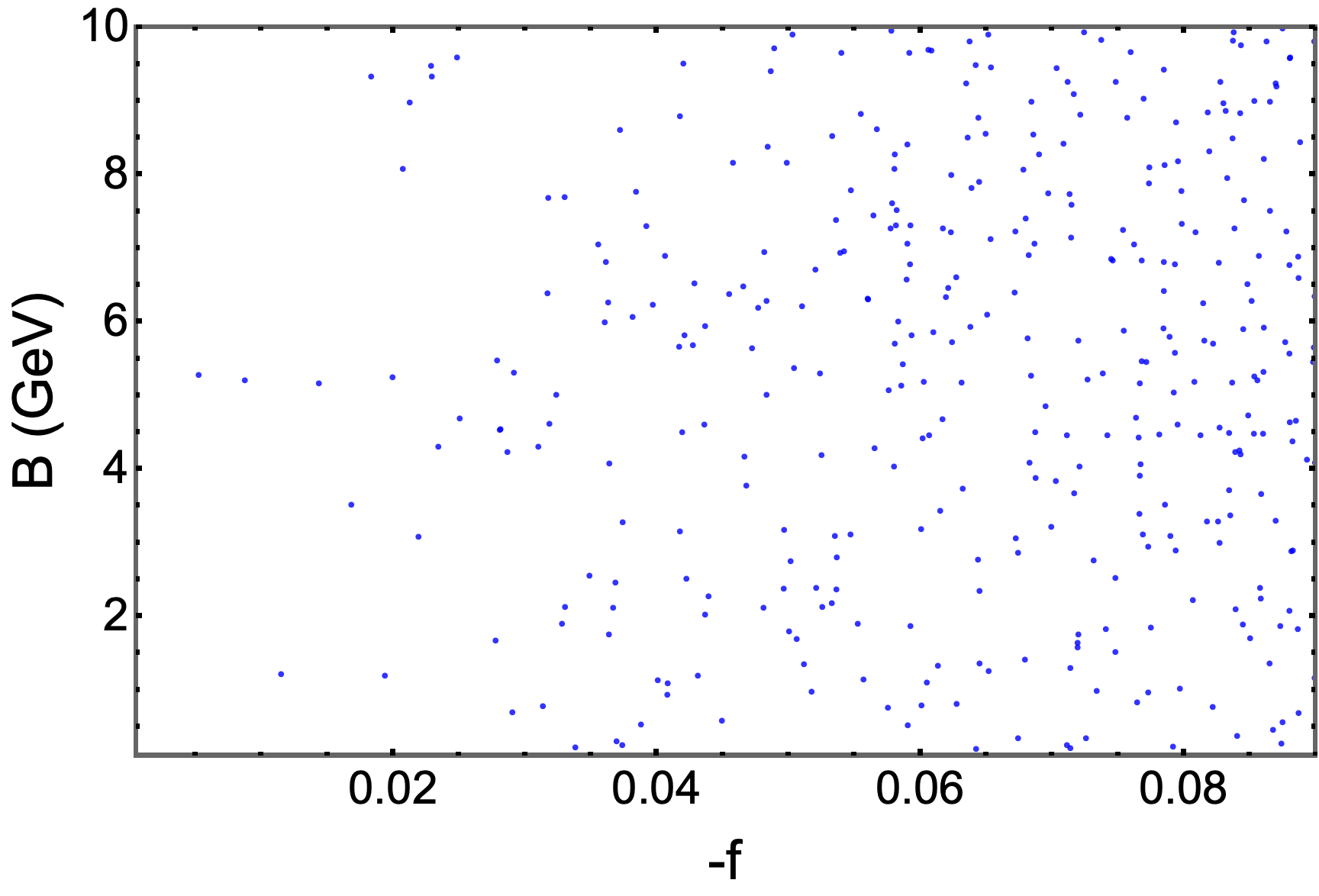}%
		\includegraphics[width=0.51\textwidth]{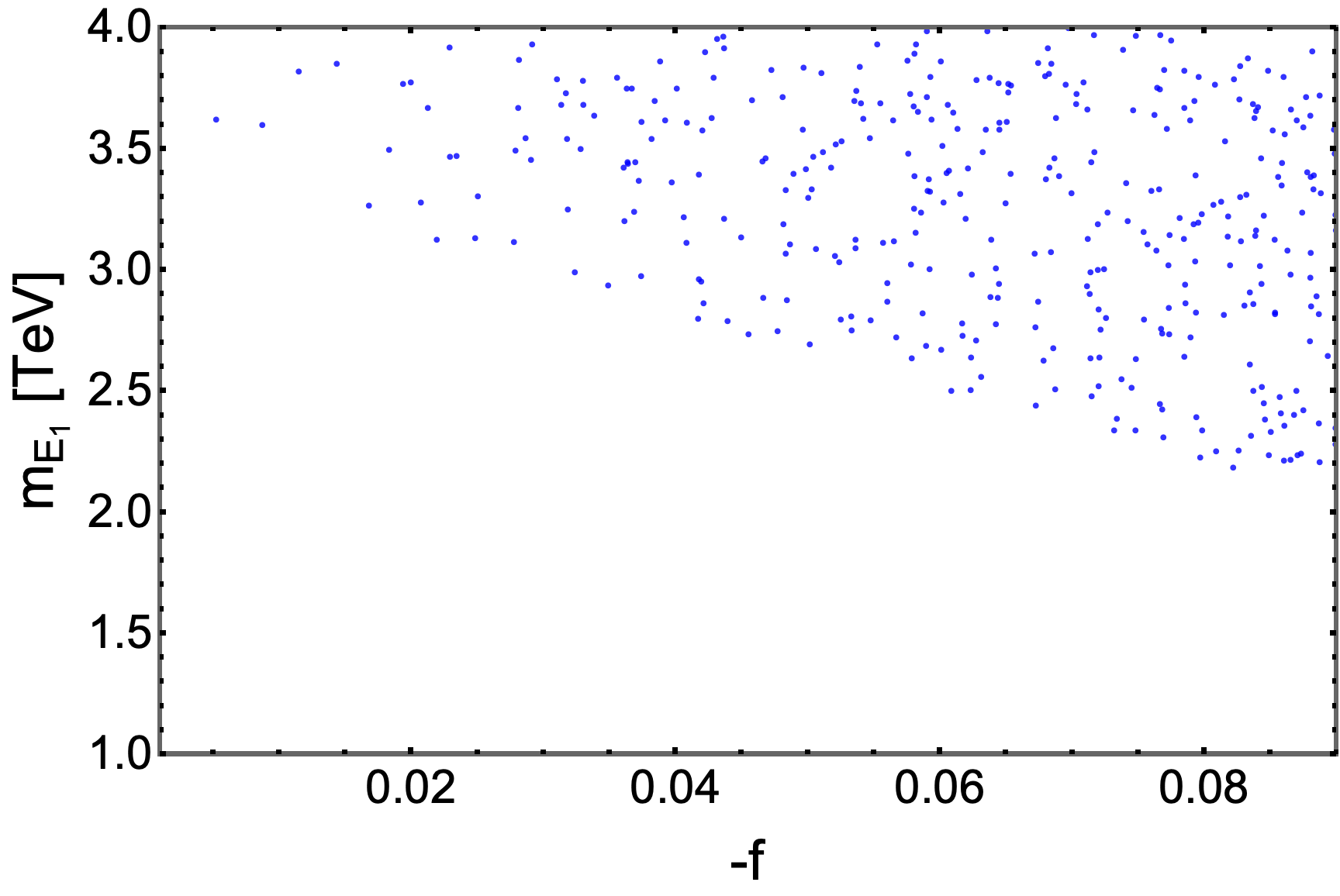}
		\caption{The left and right panels respectively show the correlations of couplings $B-f$ and $m_{E_1}-f$ satisfying the experimental limits of cLFV and anomalous magnetic moments $\Delta a_{e(\mu)}$.} \label{Fig_lep_1}
	\end{figure}

\begin{figure}[H]
		\includegraphics[width=0.51\textwidth]{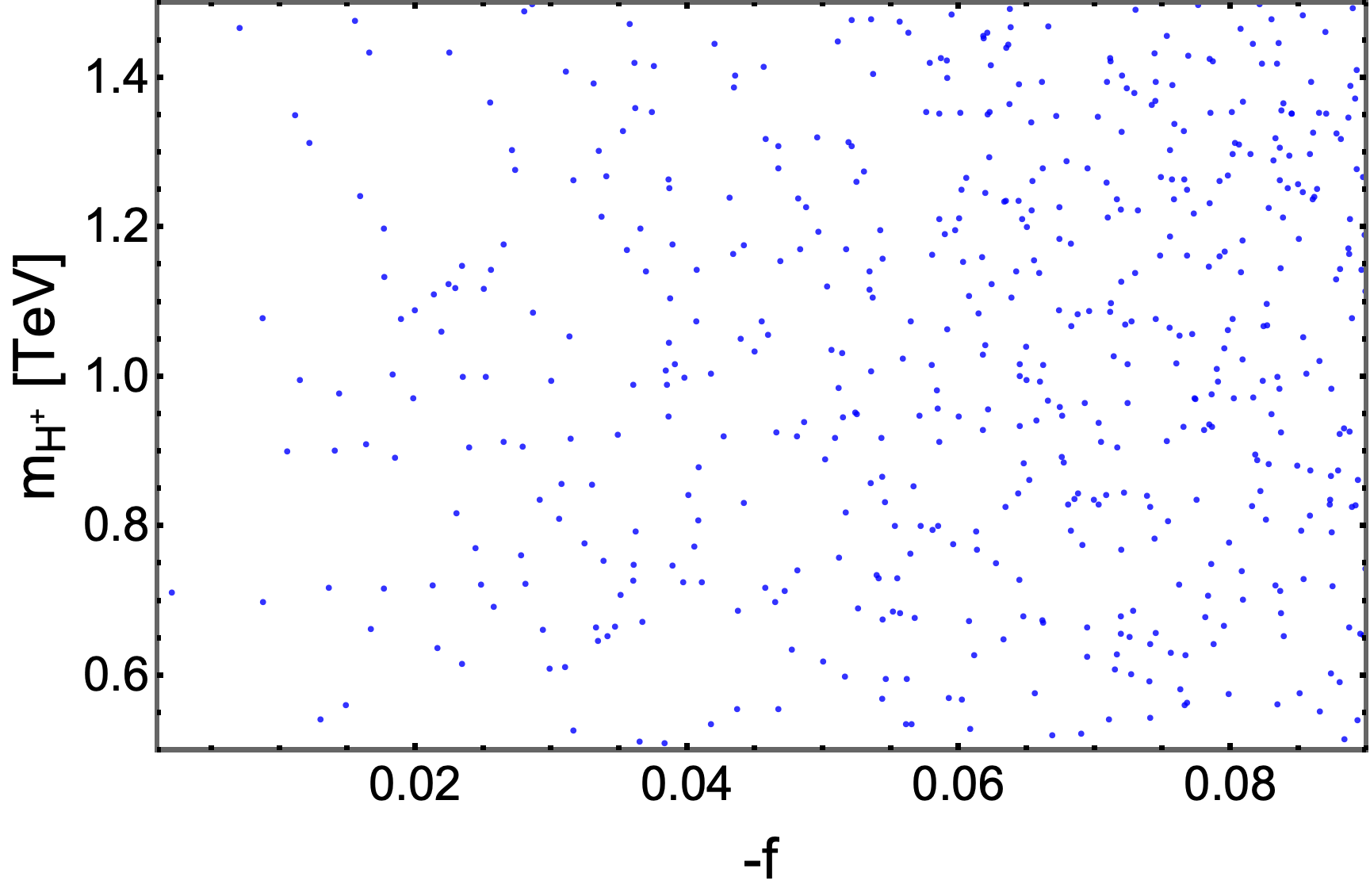}%
		\includegraphics[width=0.51\textwidth]{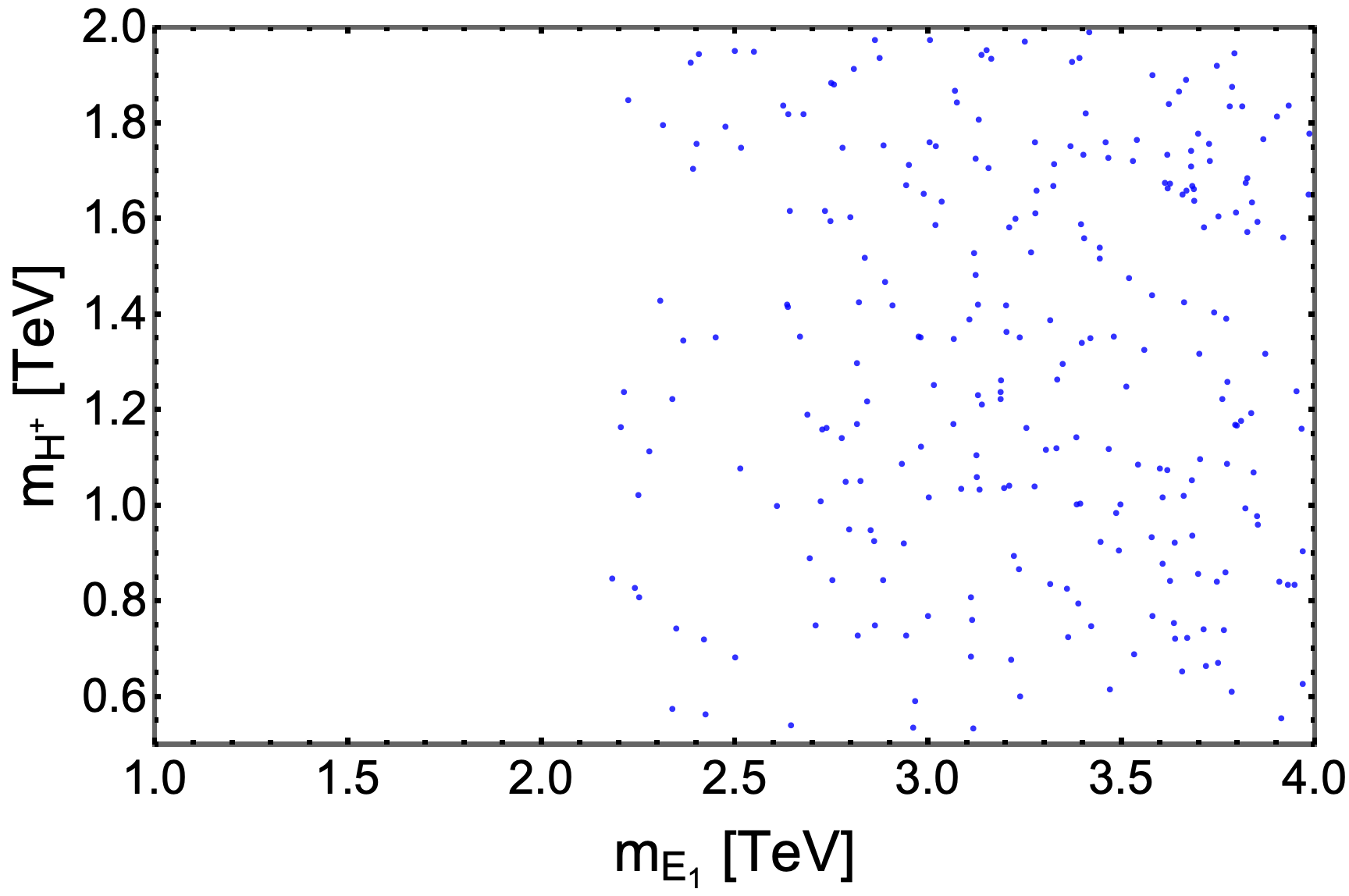}
		\caption{The left and right panels respectively show the correlations of couplings $-f$ vs $m_{H_1^{\pm}}$ and  $m_{E_1}$ vs$ m_{H_1^{\pm}}$ satisfying the experimental limits of cLFV and anomalous magnetic moments $\Delta a_{e(\mu)}$.}\label{Fig_lep_2}
	\end{figure}

	In what follows  
	we perform a numerical study of the constraints imposed by the upper limits of the 
	branching ratios of cLFV decays BR$(l_{b_1}\to l_{b_2}\ga)$ as well as by the allowed experimental ranges of the anomalous magnetic moments $\Delta a_{e(\mu)}$ via the Figs (\ref{Fig_lep_1}, \ref{Fig_lep_2}).  
	
	The left panel in Fig. (\ref{Fig_lep_1}) illustrates 
	the correlation between couplings of the scalar potential,  denoted as $-f$ and $B$. There is almost no dependence between these parameters; in other words, the change of $B$ does not affect $f$, especially for $-f\geq 0.04$. This indicates that the values of the $B$ coupling are insignificant to the leptonic observables. From the left panel, the values of $-f$ mainly range from $-f\geq 0.04$. On the other hand, the right panel demonstrates stronger correlations between the $-f$ parameter and the mass of exotic charged lepton $E_1$. We notice that the behavior of $m_{E_1}$ and $-f$ is reversed; in other words, an increase of the charged exotic lepton mass $m_{E_1}$ leads to a decrease of the 
	 $-f$ parameter 
	 and vice versa. From the right panel, we find the limit $m_{E_1}\geq 2.2$ TeV.
	
	Turning to Fig. (\ref{Fig_lep_2}), we illustrate the correlations between $-f$ or $m_{E_1}$ and mass of charged Higgs $m_{H_1^{\pm}}$ in the left and right panel, respectively. We observe that both panels indicate the irrelevance of $m_{H}^{\pm}$ to leptonic observables. Moreover, this behavior is similar if we consider other charged Higgs bosons $H_{1,2}^{\pm}$. Additionally, the Figs. (\ref{Fig_lep_1},\ref{Fig_lep_2}) demonstrate the remarkable impacts of the loop diagrams involving the exotic charged lepton $E_1$ and neutral Higgs bosons $H(\mathcal{A})$ on cLFV and anomalous magnetic moments $\Delta a_{e(\mu)}$. 
	
\begin{figure}[H]
		\centering 
		\includegraphics[width=0.75\textwidth]{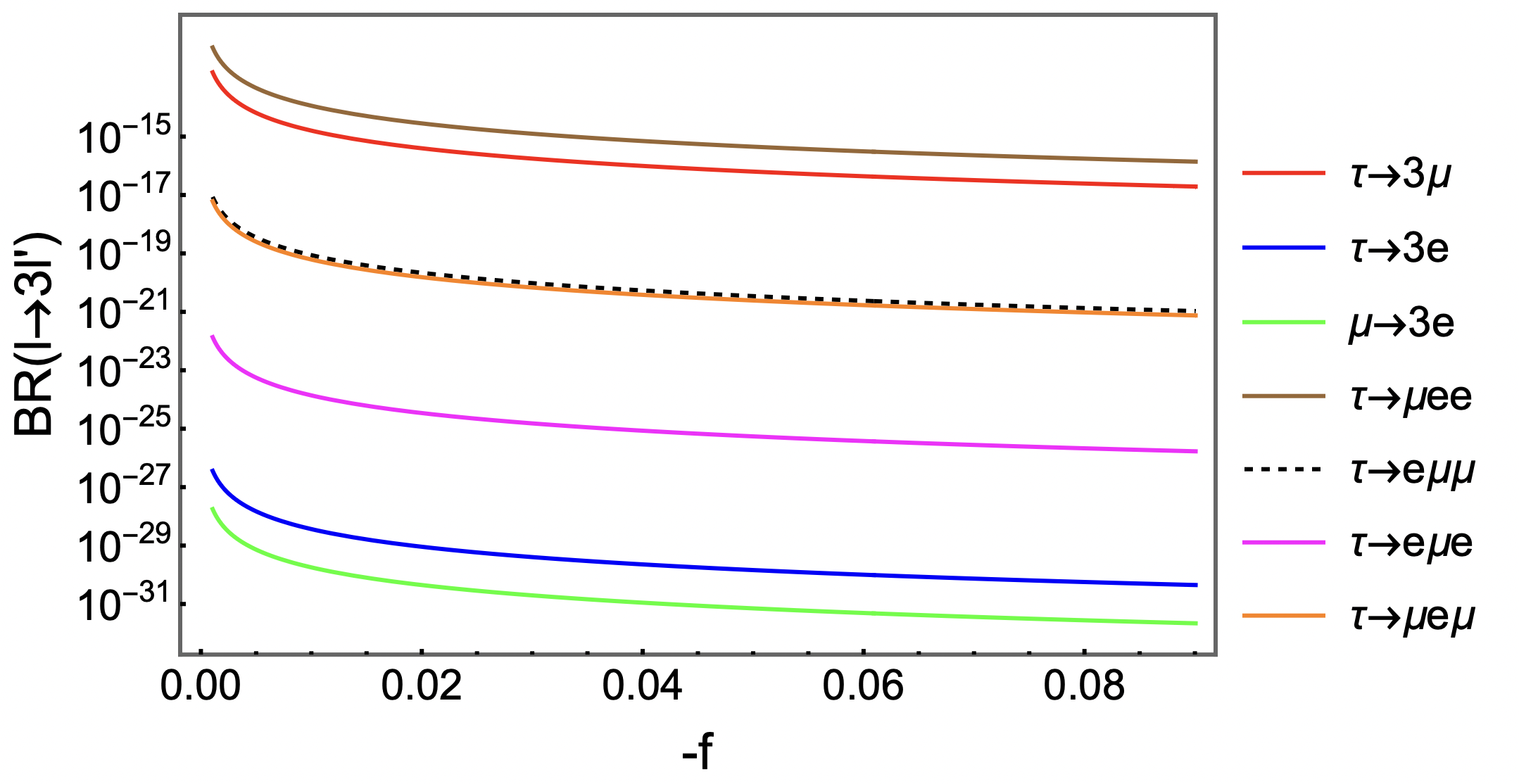}%
		\caption{The figure shows the dependence of branching ratios of three body leptonic decays as the function of coupling $-f$.} \label{Fig_l3l}
	\end{figure}

\begin{figure}[H]
	\includegraphics[width=0.51\textwidth]{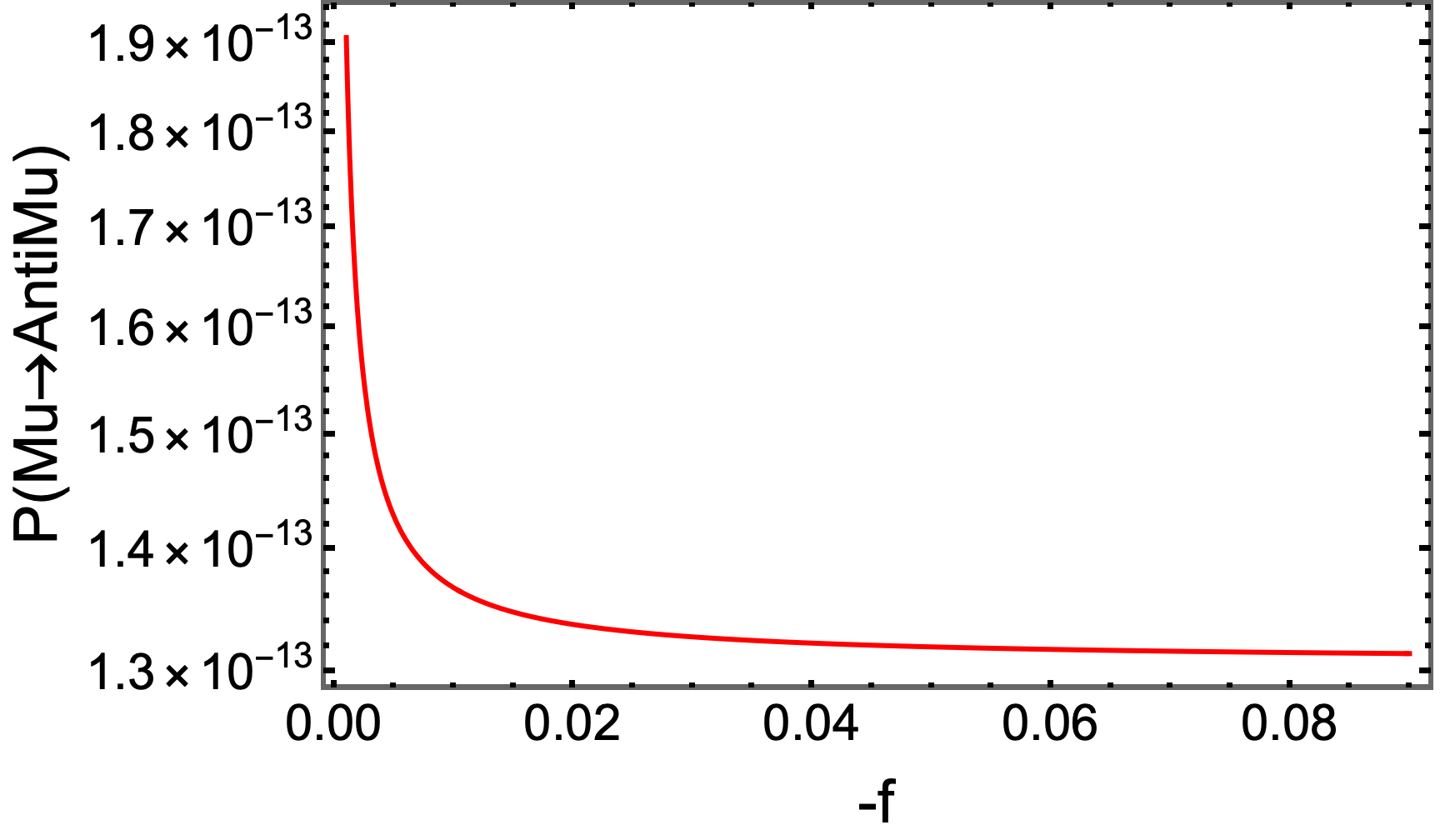}%
	\includegraphics[width=0.51\textwidth]{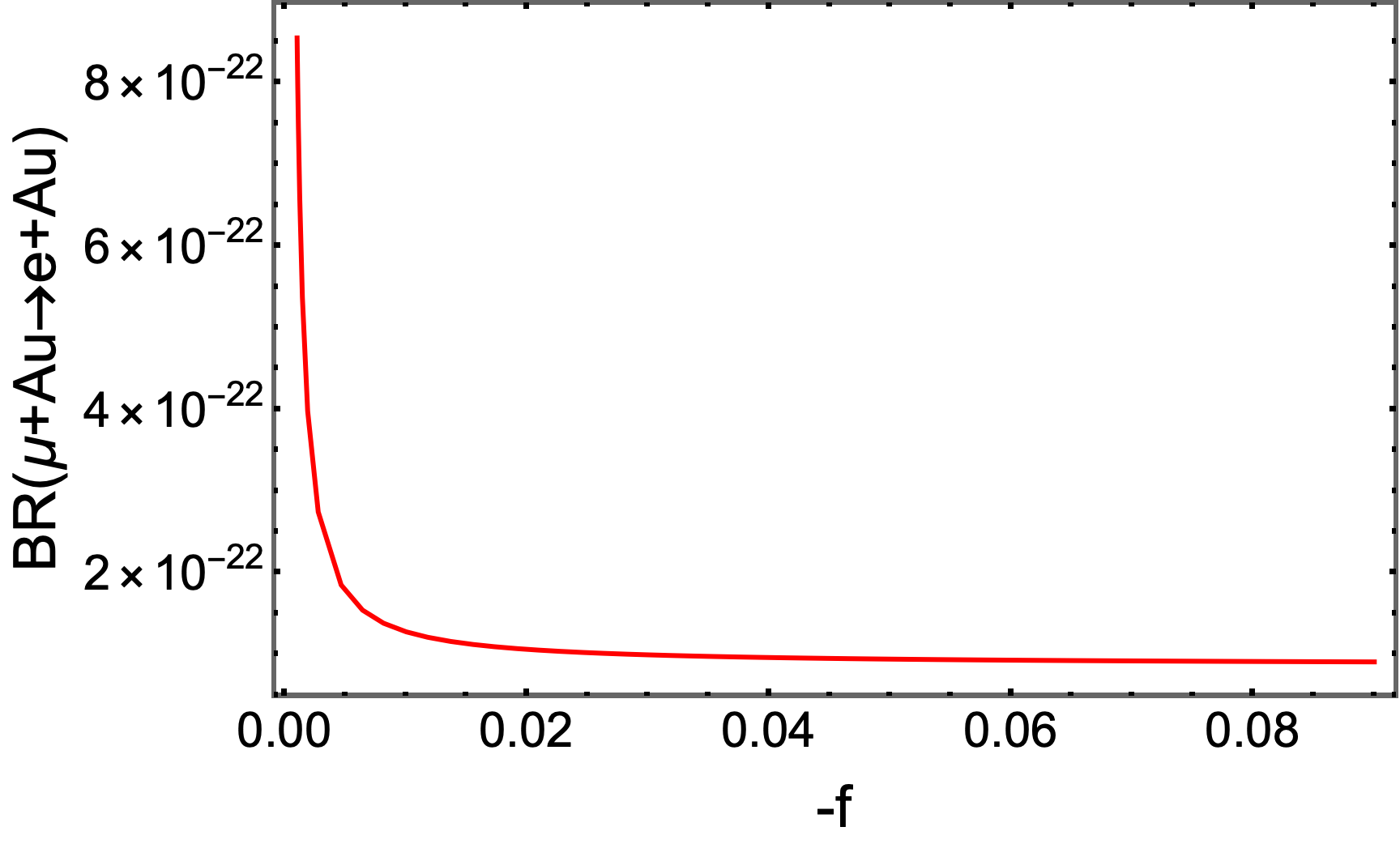}%
	\caption{The left and right panels respectively  show the dependence of Mu-$\overline{\text{Mu}}$ transition in Eq. (\ref{Mu-AntiMu}) and branching ratio BR$(\mu \text{Au}\to e\text{Au})$ in Eq. (\ref{mu-e-N}) as the function of coupling $-f$.} \label{Fig_mutoe_and _mu_antimu}
\end{figure}

	Moreover, we consider the branching ratios of three body leptonic decays BR$(l\to 3l')$, Mu-$\overline{\text{Mu}}$ transition, and coherent conversion $\mu \to e$ in a muonic atom, which only depends on the coupling $f$, as shown in Fig . (\ref{Fig_l3l}) and Fig (\ref{Fig_mutoe_and _mu_antimu}). We observe that the model predicts these observables to be much lower by several orders of magnitude than their corresponding upper experimental bounds, as given in Table (\ref{lepton_exp}). Therefore, the new physics contributions in the considering model for these observables are negligible and can be ignored. 

	\subsection{Quark flavor observables}
\begin{figure}[H]
		\includegraphics[width=0.51\textwidth]{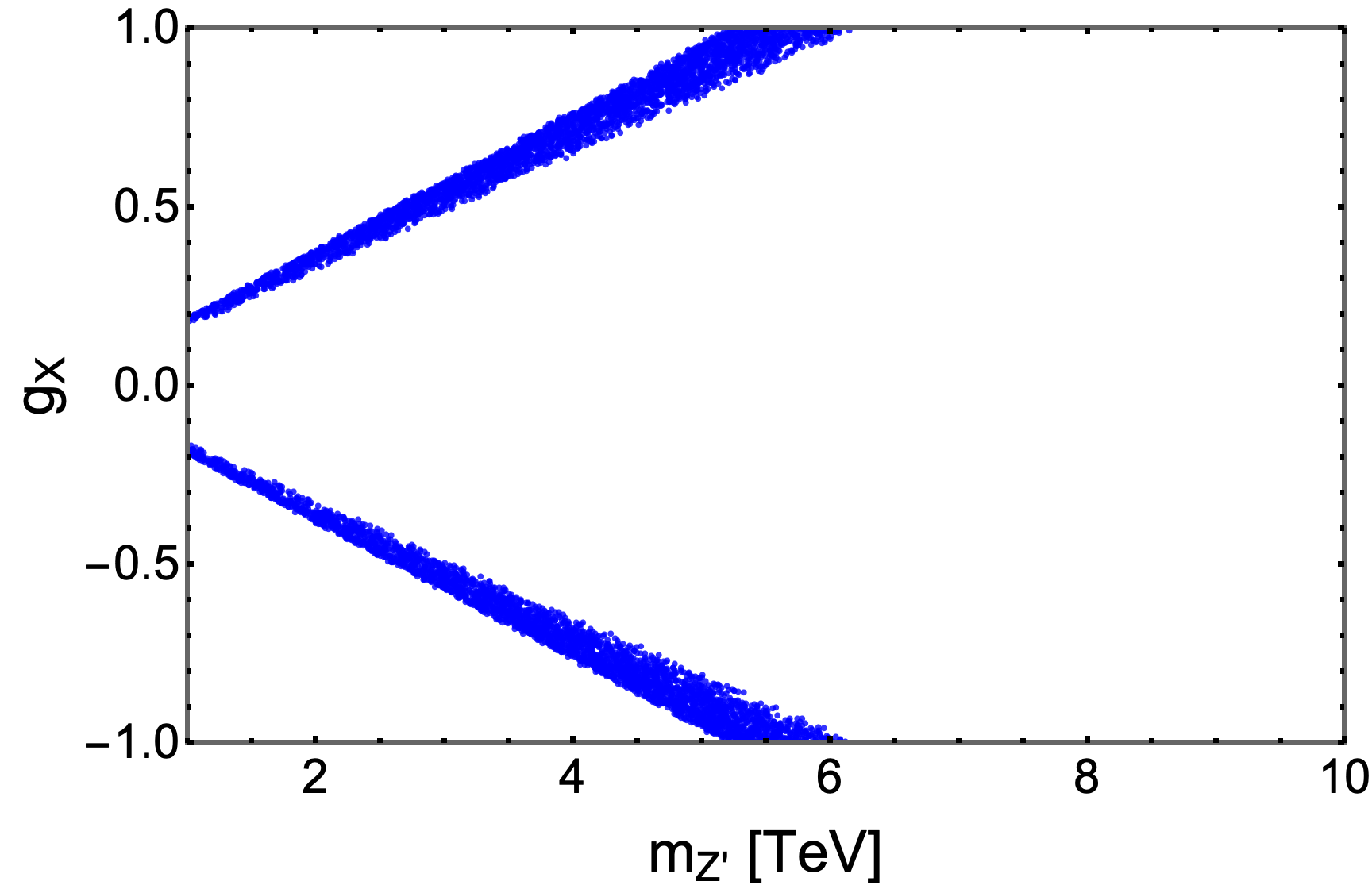}%
		\includegraphics[width=0.51\textwidth]{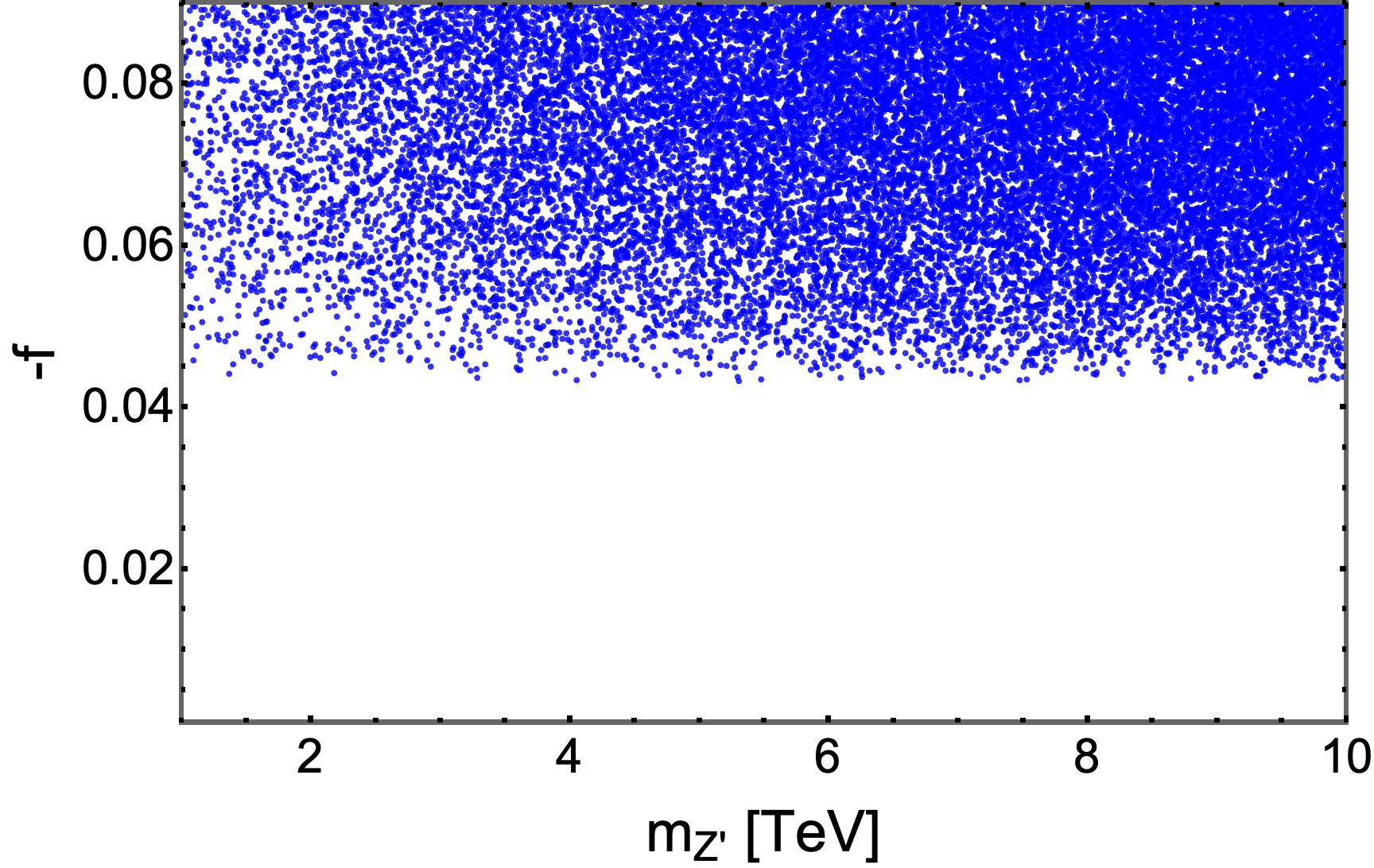}
		\caption{The left and right panels respectively show the correlations of $m_{Z'}-g_X$ and  $m_{Z'}-f$ satisfying the constraints given in Eq. (\ref{Bsmm_constraint}) for BR$(B_s\to \mu^+\mu^-)$ , Eqs. (\ref{DmBq}, \ref{DmK}) for $\Delta m_{K,B_s,B_d}$, and \cite{Workman:2022ynf} for BR$(B^+\to K^+\tau^+\tau^-)$  }\label{Fig_Bsmm-BKtt}
	\end{figure}

	Let's now focus on the phenomenology of flavor quark observables. First, we will consider the FCNC  observables in down type quark transitions $d_a\to d_b$. It is important to note that only meson mass splittings, such as 
	$\Delta m_{K,B_s,B_d}$, BR$(B_s\to \mu^+\mu^-)$ and BR$(B^+\to K^+\tau^+\tau^-)$ do depend on the new neutral gauge boson mass $m_{Z'}$ and the coupling $g_{X}$ of $U(1)_X$ symmetry. Other quark flavor observables, however, are independent of these parameters. Thus, our initial focus is to analyze the constraints on $m_{Z'}$ and $g_{X}$ that satisfy the experimental limits of $\Delta m_{K, B_s, B_d}$, BR$(B_s\to \mu^+\mu^-)$ and BR$(B^+\to K^+\tau^+\tau^-)$ by the Fig. \ref{Fig_Bsmm-BKtt}
	In the left panel, we observe that the range of $m_{Z'}$  varies with the values of $g_X$. For instance, with $|g_X|\simeq 0.2 $, the mass of $Z'$ boson is about $\simeq [1,1.4]$ TeV. However, if $|g_X|> 0.2$, $m_{Z'}$ increases and can reach up to 
	$ 6.5$ TeV for $|g_X|\simeq 1$. However, it should be noted that the LHC searches of the $Z'$ gauge boson {set a} lower mass limit of $m_{Z'}\geq 4$ TeV \cite{Workman:2022ynf}, implying the coupling $g_X$ should satisfy $|g_X|\geq 0.65$. These constraints for $g_X$ and $m_{Z'}$ are stronger than those ones obtained from studying meson oscillations in \cite{Hernandez:2021xet}. 
	
	On the one hand, both observables are also affected by the $f$ coupling and we depict the right panel to demonstrate the correlations between $f$ and $m_{Z'}$. We observe that the correlation here is {weaker} compared to the left panel. This suggests that the (pseudo)scalar WCs $C^{(')}_{S(P)}$, which involve 
	 CP-even or CP-odd Higgs bosons provide very small and subleading contributions to these observables. This can be understood because when we impose the limit, $-f\geq 0.04$ obtained from the analysis of leptonic flavor observables, such WCs become $C^{(')\mu\mu}_{S(P)}\sim \mathcal{O}(10^{-2}-10^{-4})$, which are extremely small compared to WCs $C_{9,10}\sim \mathcal{O}(1)$. It is worth noting that the lower bound $-f\geq 0.045$ obtained in the right panel is relatively higher than that obtained in Fig. (\ref{Fig_lep_1}). 

\begin{figure}[H]
 	\includegraphics[width=0.51\textwidth]{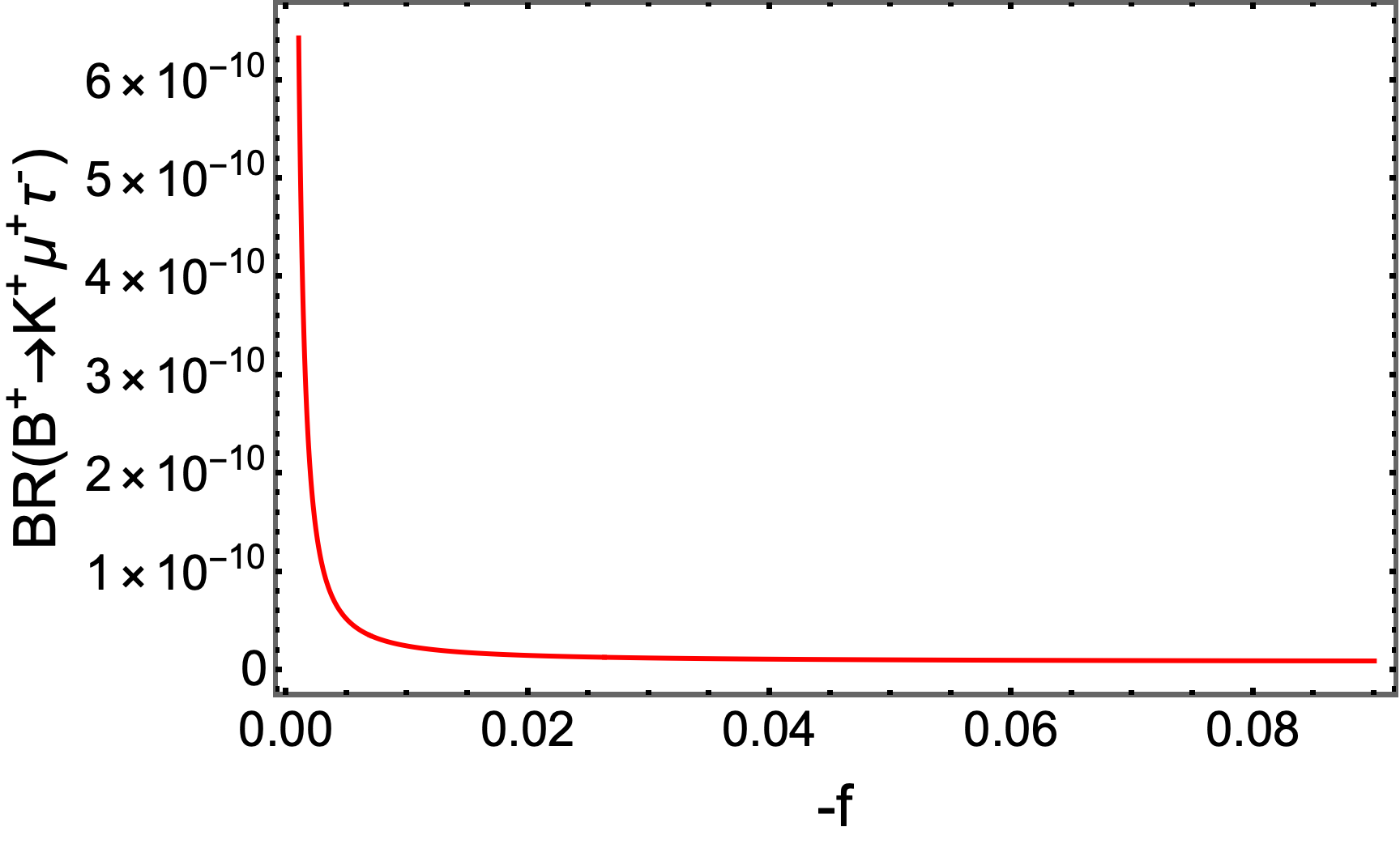}%
 	\includegraphics[width=0.51\textwidth]{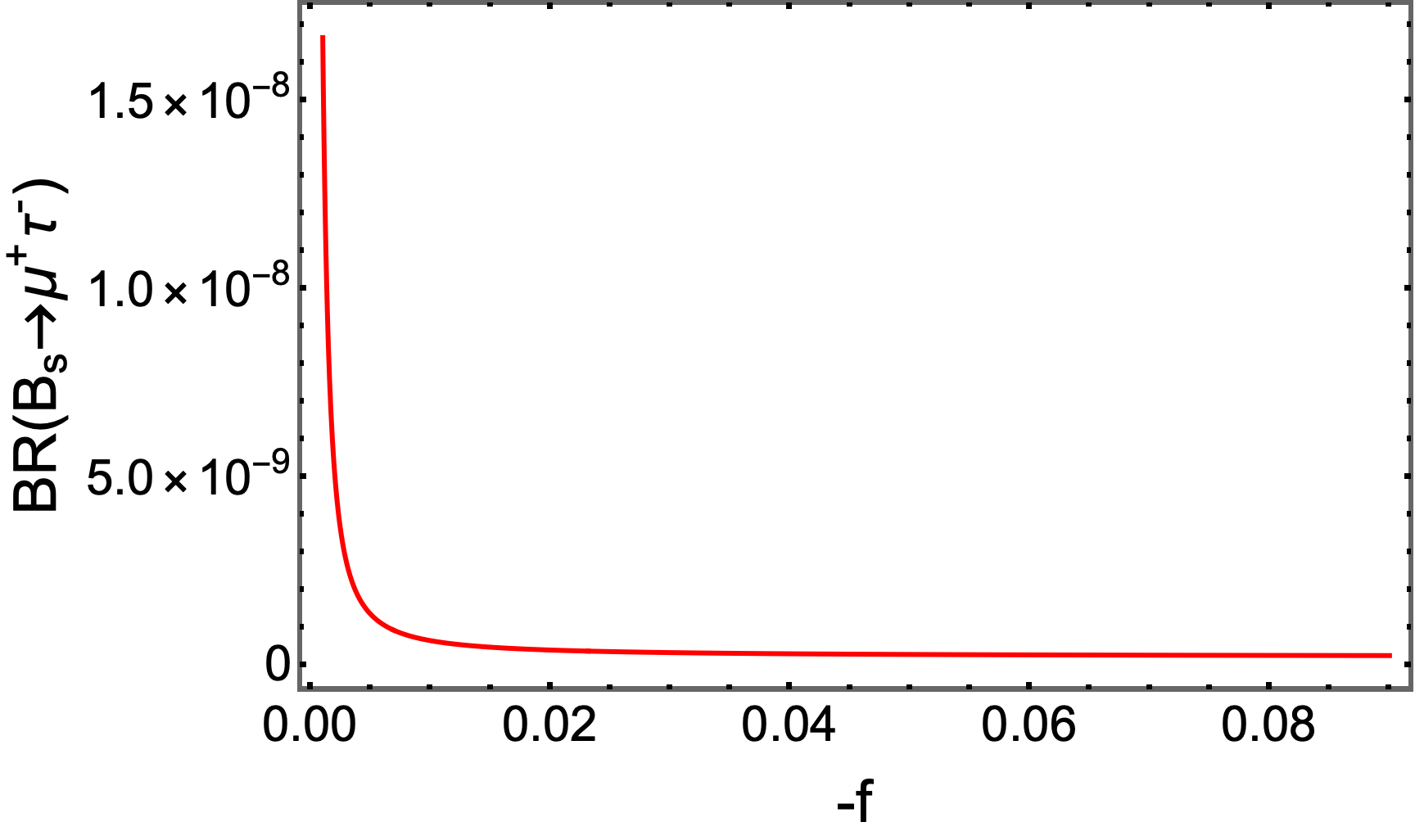}
 	\caption{The left and right panels respectively show the dependence of BR$(B^+\to K^+\tau^+\mu^-)$ and BR$(B_s\to \tau^+\mu^-)$ as the function of coupling $-f$.  }\label{Fig_BctoKcmuutau_Bstomutau}
 \end{figure}

On the other hand, for the branching ratios of (semi)leptonic decays BR$(B^+\to K^+\tau^+\mu^-)$, BR$(B_s\to \tau^+\mu^-)$ influenced by (pseudo)scalar WCs $C_{S(P)}^{(')\tau\mu}$ depending on the parameter $f$, we plot the Fig .\ref{Fig_BctoKcmuutau_Bstomutau} and realize that the model evaluates 
BR$(B^+\to K^+\tau^+\mu^-)_{\text{theory}}\sim 10^{-10}$, BR$(B_s\to \tau^+\mu^-)_{\text{theory}}\sim 10^{-9}-10^{-8}$.
Compared with relative experimental bounds given in Table \ref{SM_exp data}, the predicted values are much lower by several orders of magnitude. 

\begin{figure}[H]
	\includegraphics[width=0.51\textwidth]{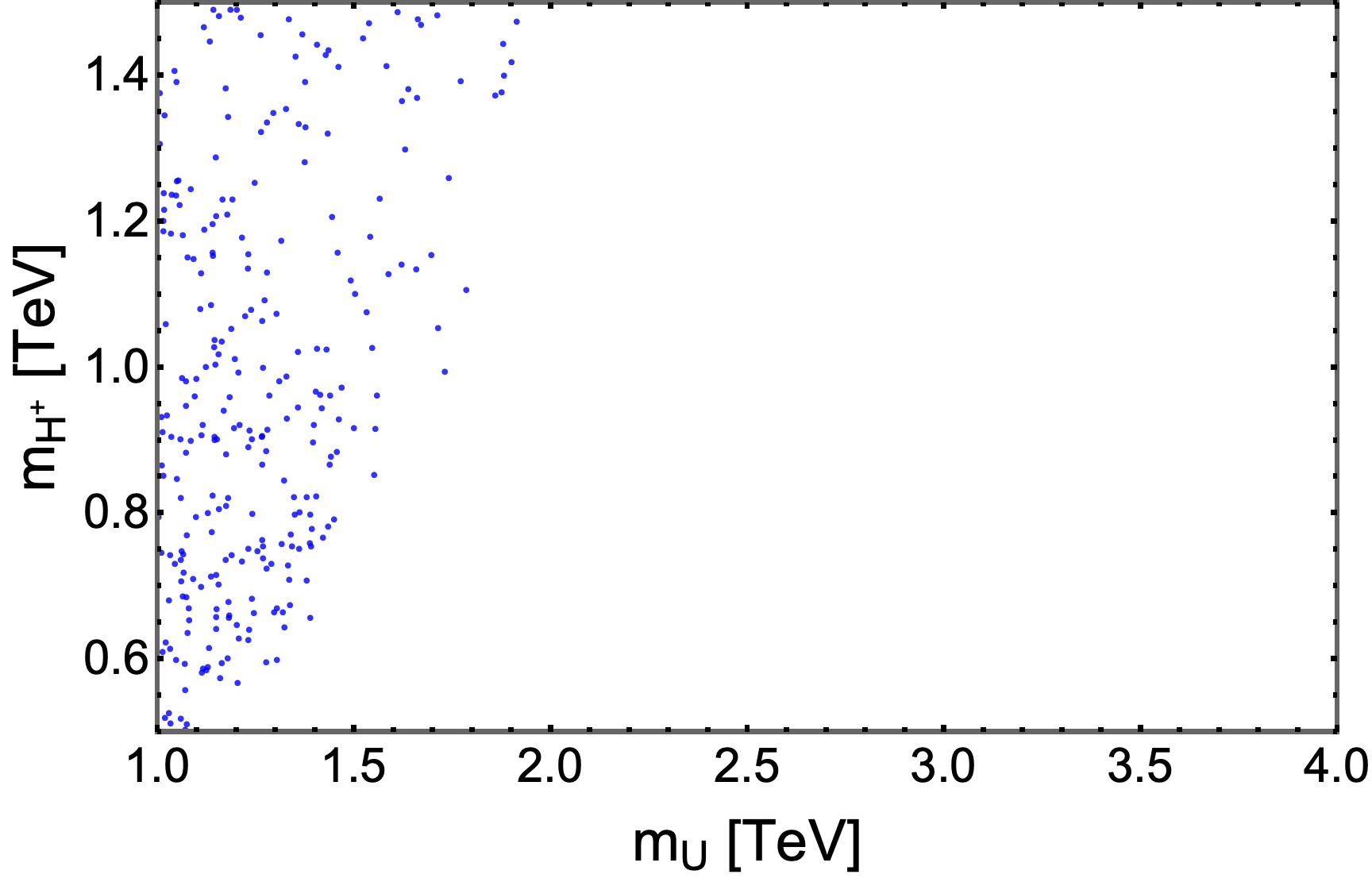}%
	\includegraphics[width=0.51\textwidth]{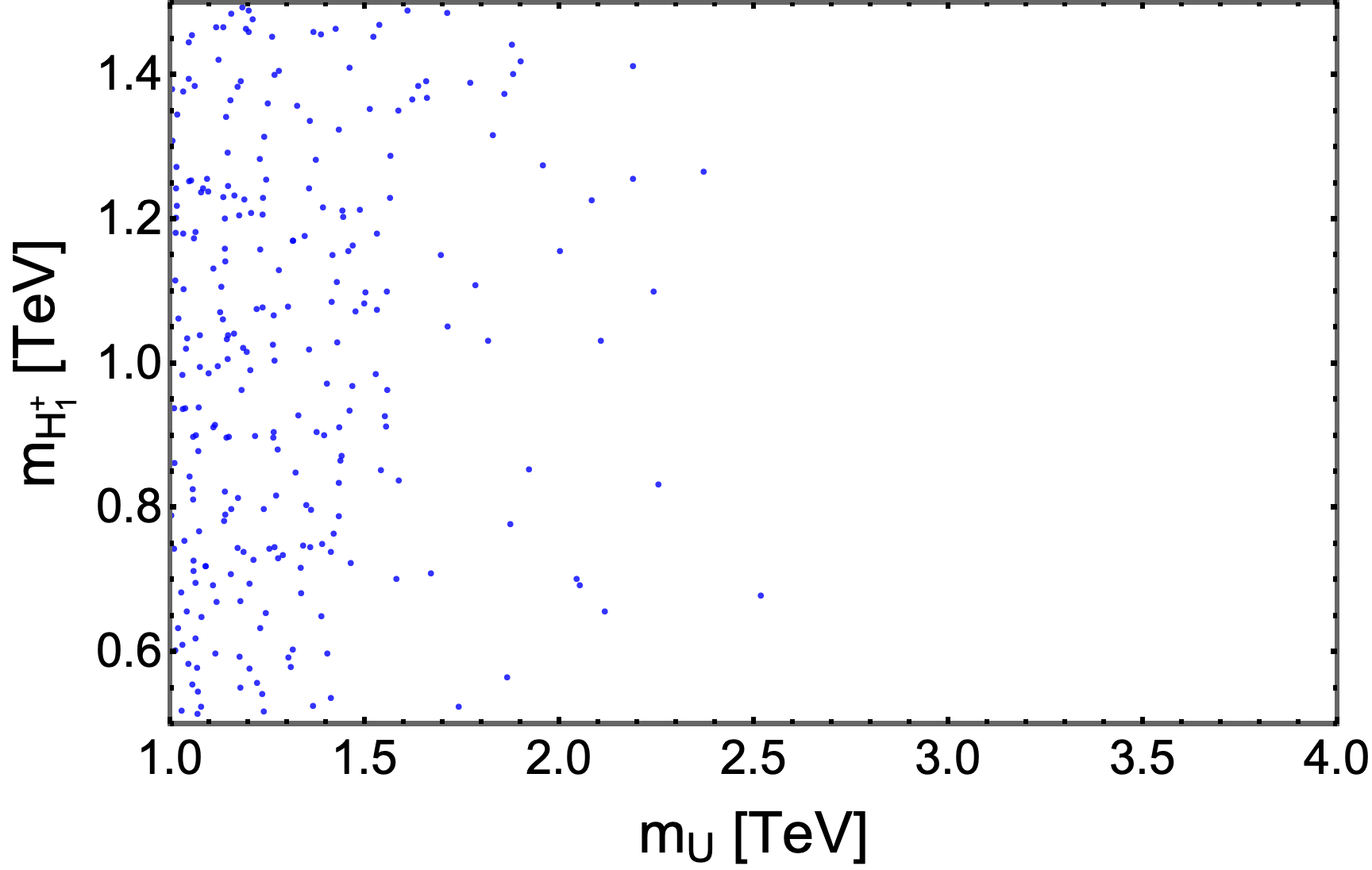}
		 \includegraphics[width=0.51\textwidth]{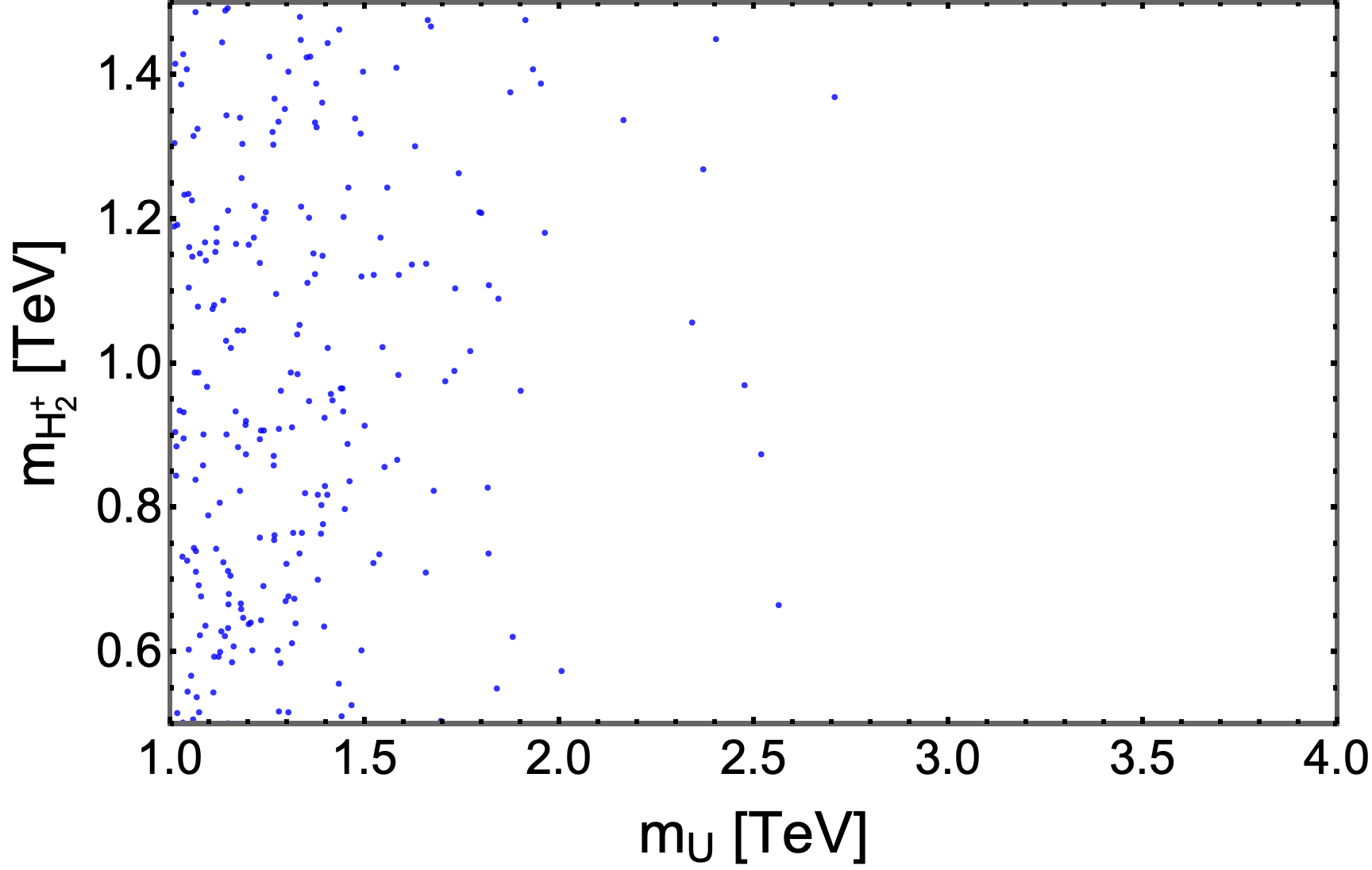}
		 \includegraphics[width=0.51\textwidth]{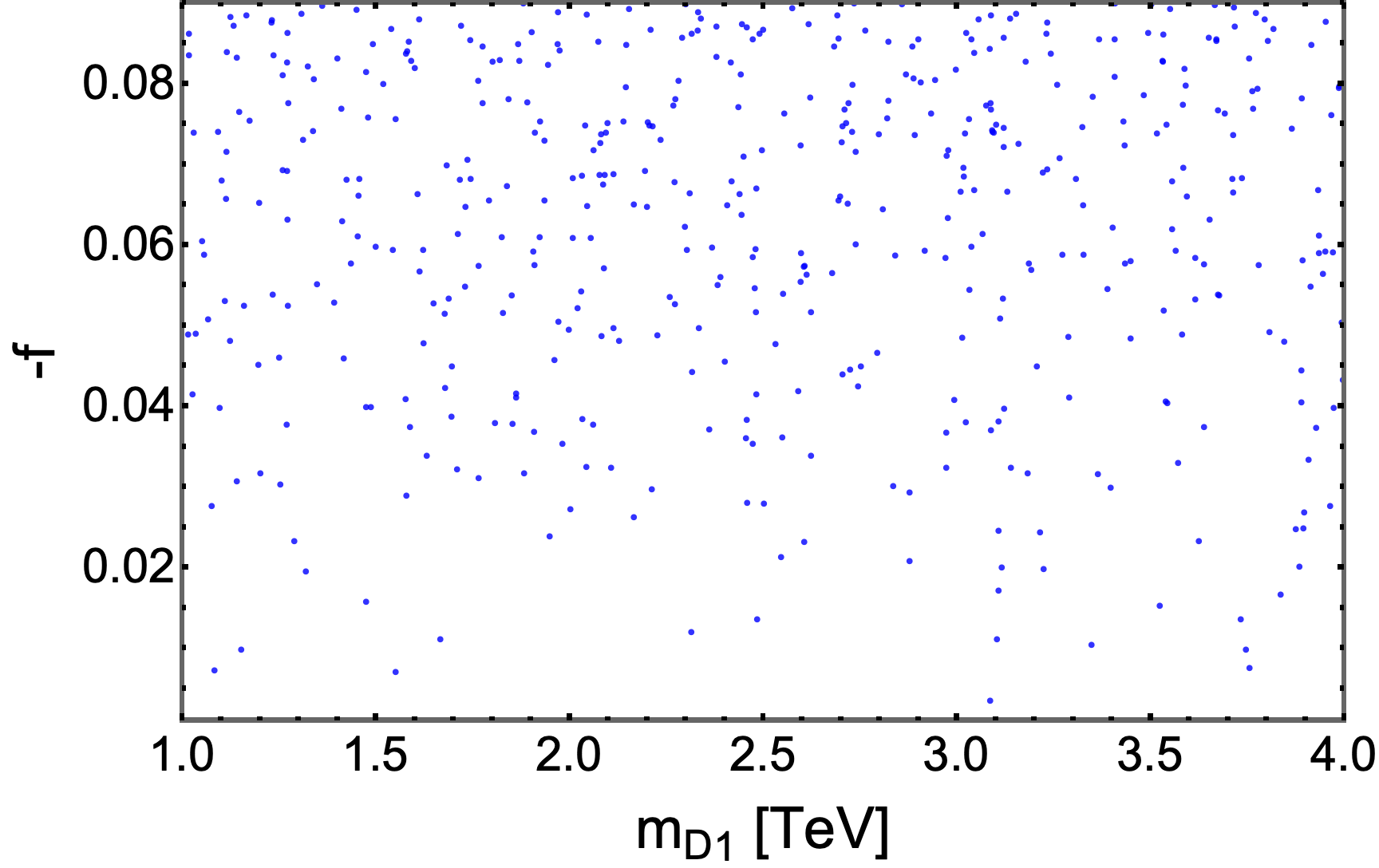}
	\caption{The first three panels (from left to right) and last panel respectively show the correlations between mass of exotic up type quark $m_U$ with masses of charged Higgs bosons $H^{\pm}$, $H_{1,2}^{\pm}$ and coupling $-f$ satisfying the constraints of BR$(\bar{B}\to X_s\ga)$ in Eq. (\ref{bsga_constraint}).}\label{Fig_mU_mHc}
\end{figure}

For remaining observables related to the FCNC $d_a\to d_b$ transitions, namely BR$(\bar{B}\to X_s)\ga$, which are contributed by WCs induced by charged Higgs $H^{\pm}, H_{1,2}^{\pm}$, and neutral Higgs bosons $H(\mathcal{A})$, we plot the Fig. \ref{Fig_mU_mHc} to demonstrate the relationship between parameters fulfilling the constraints given in Eq. (\ref{bsga_constraint}). 
In the first three panels (from left to right) of Fig. \ref{Fig_mU_mHc}, we illustrate the correlation between the mass of new up quark $m_{U}$ and the masses of three charged Higgs bosons $m_{H^{\pm}},m_{H_{1,2}^{\pm}}$ satisfying the constraint of BR$(\bar{B}\to X_s\ga)$ in Eq. (\ref{Bsmm_constraint}). We observe that the entire ranges of $m_{H^{\pm}},m_{H_{1,2}^{\pm}}$ in all panels meet the constraint, whereas the range of $m_U$ is more stringent, ranging up to approximately $\simeq 2.5$ TeV but dominantly distribute with the ranges $m_U\in [1,2]$ TeV. It is worth noting that this observable also depends on $f$ through the WCs induced by the neutral CP even (odd) Higgs bosons such as $C_{7}^{(')H(\mathcal{A})\bar{D}_1d}$. We plot the correlation between $-f$ and mass of down type quark $m_{D_1}$ in the fourth panel of Fig. (\ref{Fig_mU_mHc}). We observe that this correlation is not as strong compared to the others. This can be understood because with $-f\geq 0.04$ and $m_{D_1}\sim \mathcal{O}(1)$ TeV, we numerically estimate the magnitude of WCs as $|C_{7,8}^{H(\mathcal{A})\bar{D}_1d}|\sim \mathcal{O}(10^{-6}-10^{-4})\ll 1$, which is much lower by 3 to 5 orders of magnitude than the corresponding SM predictions $|C_7^{\text{SM}}|\sim \mathcal{O}(10^{-1})$. Consequently, this implies that WCs induced by new quark $U$ and charged Higgs bosons $H^{\pm}, H_{1,2}^{\pm}$ dominantly contribute compared to those containing $D_1$ and $H_p$, and therefore we can ignore these latter contributions in such observables. 

Turning to up quark transition observables $t\to u(c)$, we first study branching ratios of tree-level top quark decays BR$(t\to u(c)h)$, as the couplings of these observables are fixed and do not depend on free parameters. The comparison between the predictions of the model and the upper experimental bounds for such decays is shown in Table \ref{BRtuch}
\begin{table}[H]
	\protect\caption{\label{BRtuch} The comparison between predicted values and experimental limits of BR$(t\to u(c)h)$.}
	\begin{centering}
		\begin{tabular}{|c|c|c|}
			\hline
			Branching ratios & Predicted values & Experimental limits \cite{ATLAS:2023ujo}\tabularnewline
			\hline 
			$t\to u h $ & $ \simeq 3.97\times10^{-10}$ & $<3.8\times 10^{-4}$ \tabularnewline
			$t\to ch $ & $ \simeq 9.71\times 10^{-10}$& $<4.3\times 10^{-4}$  \tabularnewline
			\hline 
		\end{tabular}
		\par
	\end{centering}	
\end{table}
We observe that the obtained 
branching ratios BR$(t\to u(c)h)$ in the model are on the order of  $10^{-9}-10^{-10}$, significantly larger by 
several orders of magnitude than the corresponding SM values. Besides that, these results for the decays satisfy the upper experimental bounds, notably as they are lower than 5 to 6 orders of magnitude than the measurement ones. Thus, the NP contributions of NP in these decays are small and safe under the experimental constraints. 

\begin{figure}[H]
	\includegraphics[width=0.51\textwidth]{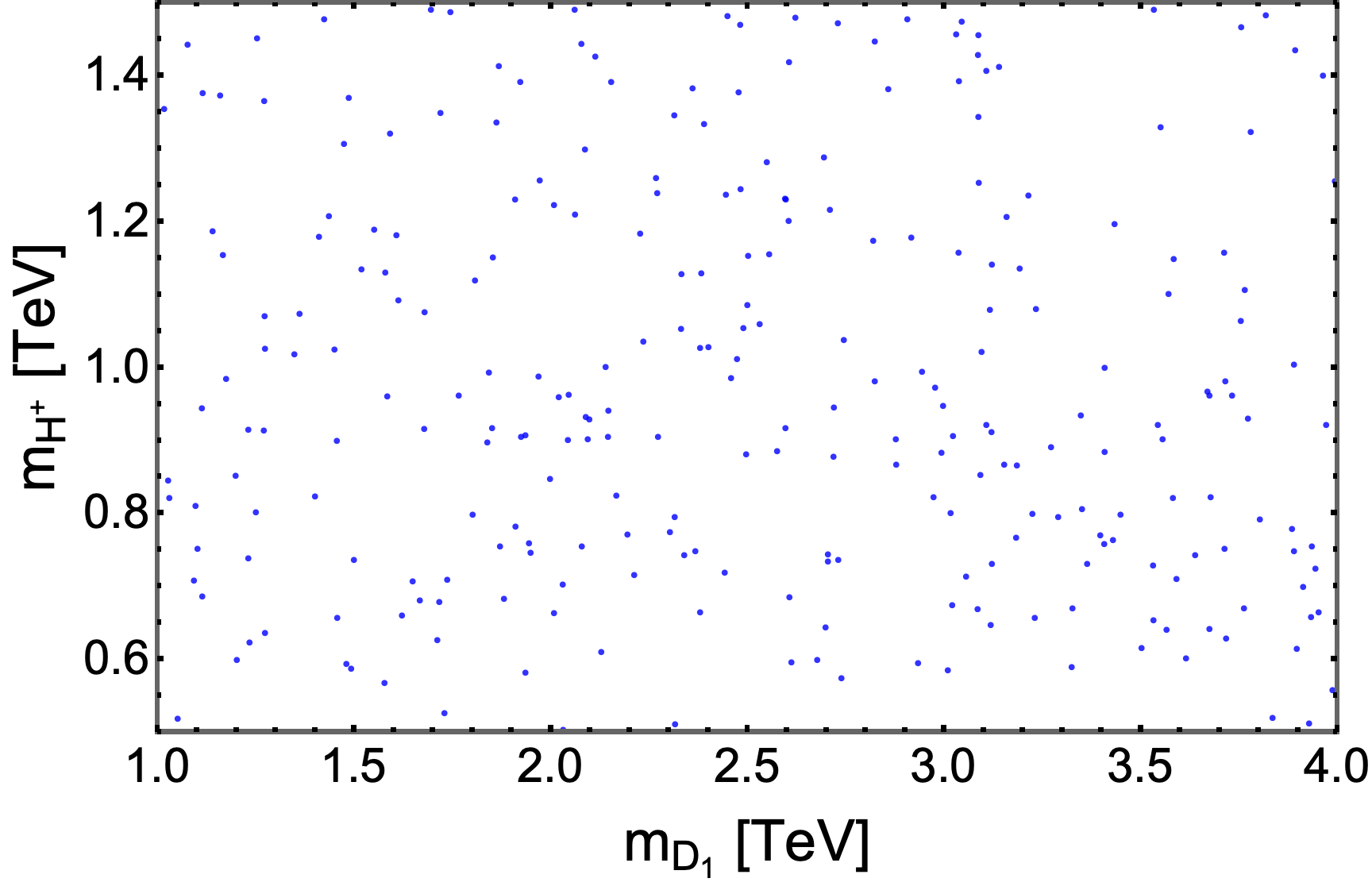}%
	\includegraphics[width=0.51\textwidth]{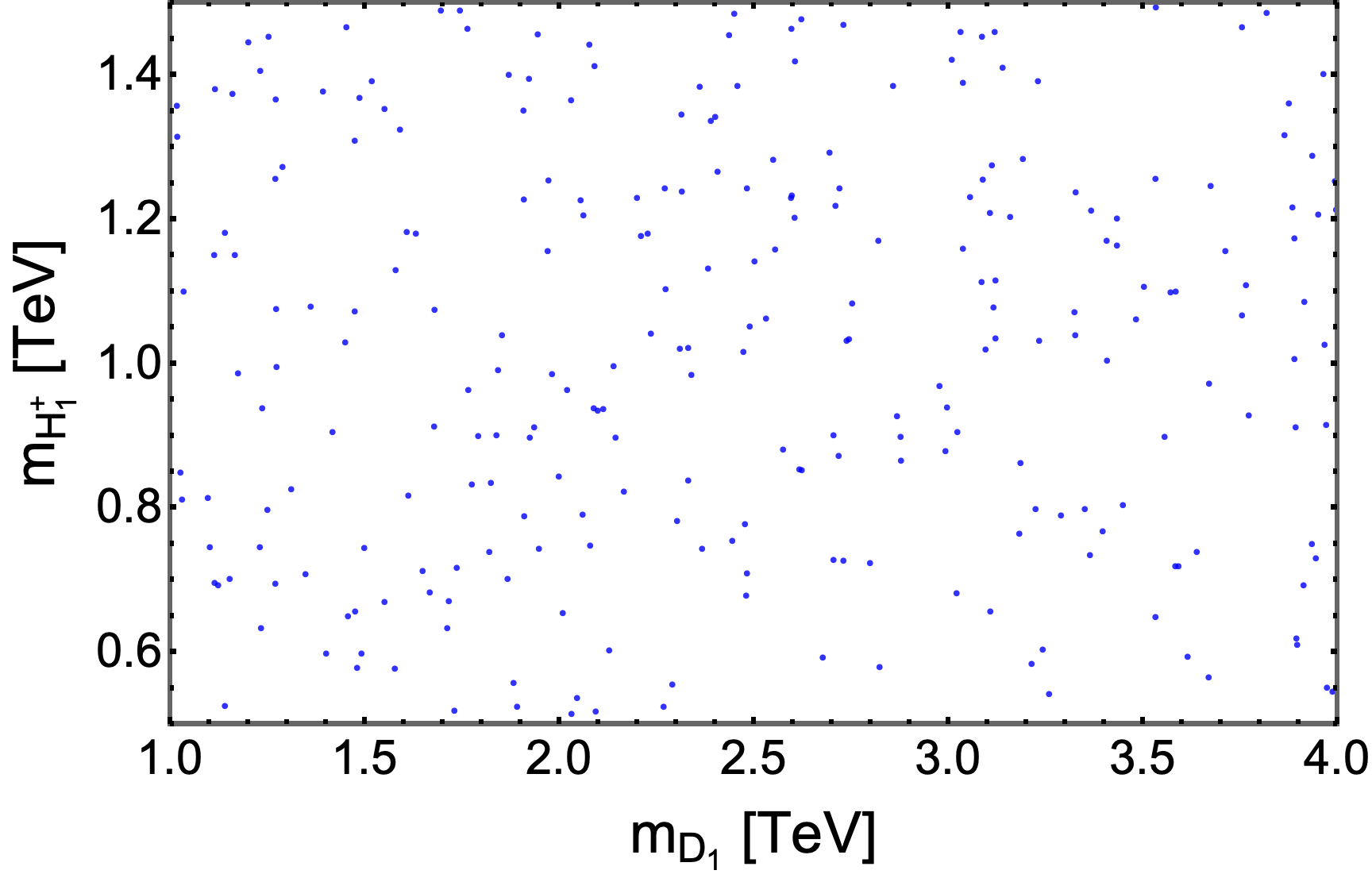}
	\includegraphics[width=0.51\textwidth]{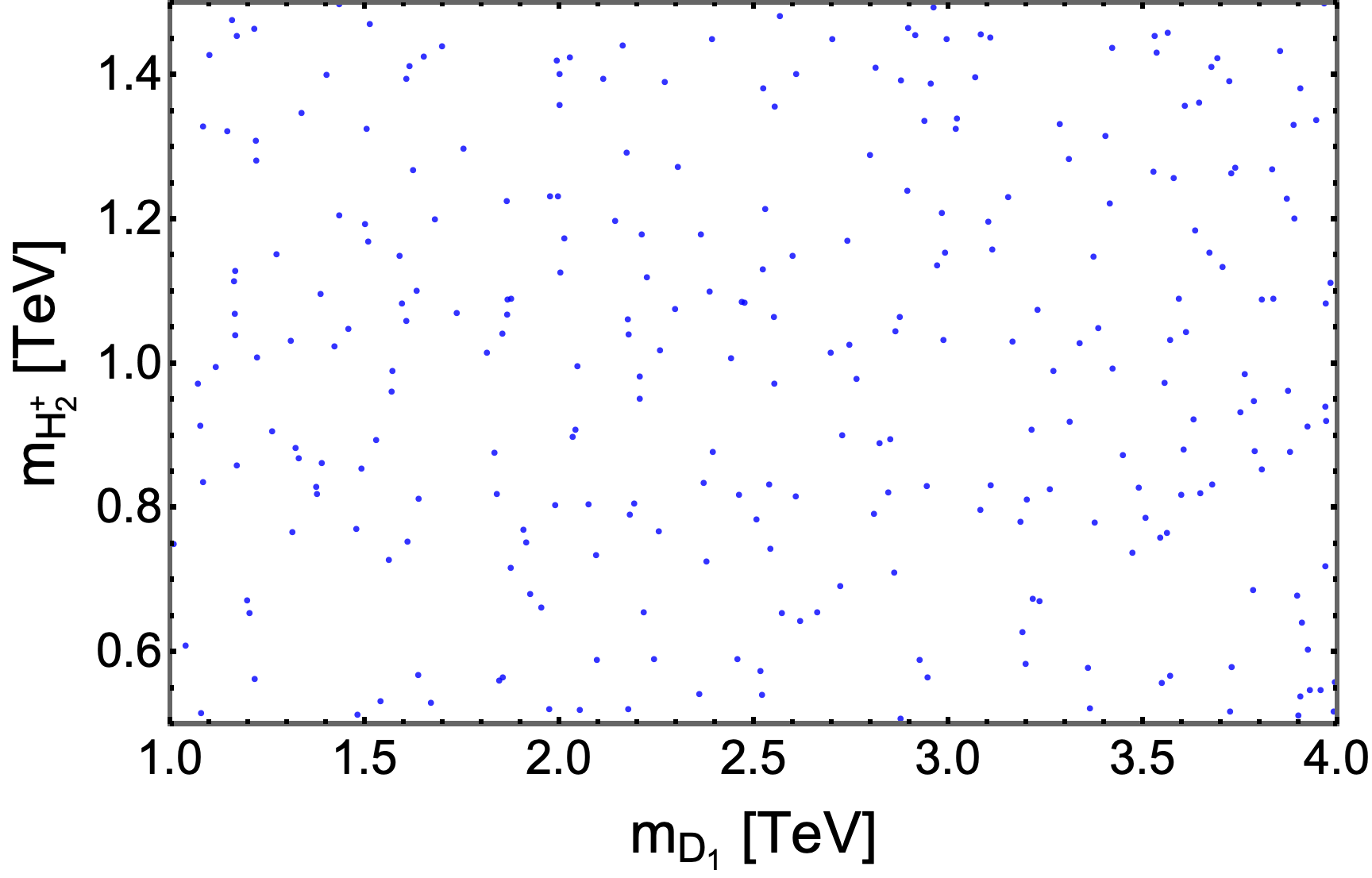}
	\includegraphics[width=0.51\textwidth]{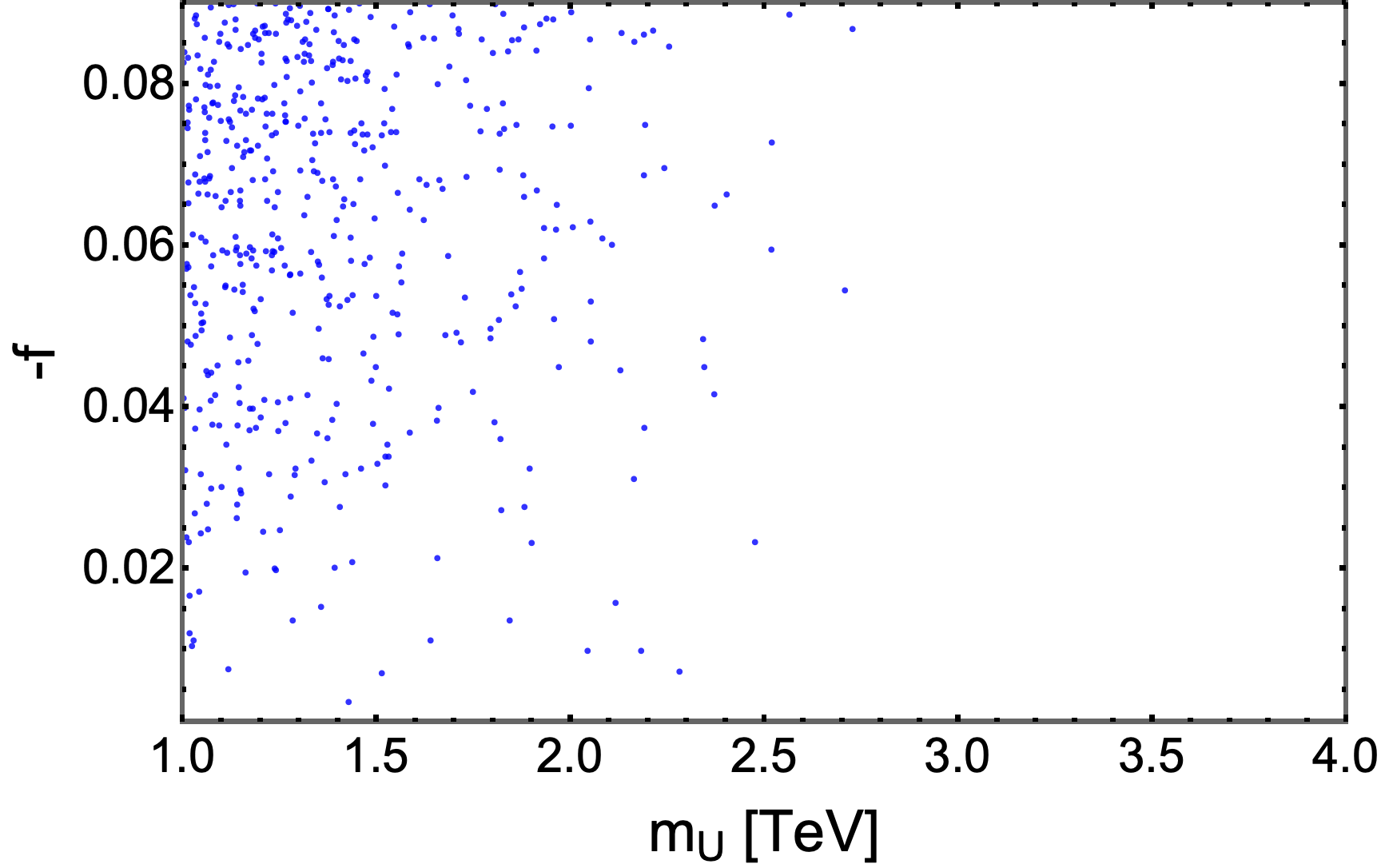}
	\caption{The first three panels (from left to right) and last panel respectively show the correlations between mass of exotic down type quark $m_{D_1}$ with masses of charged Higgs bosons $H^{\pm}$, $H_{1,2}^{\pm}$ and coupling $-f$ satisfying the constraints of BR$(t\to u(c)\ga)$ \cite{ATLAS:2022per}.}\label{Fig_ua_ub_transition}
\end{figure}

On the one hand, with branching ratios of radiative decays BR$(t\to u(c)\gamma)$ {are} induced by loop 
diagrams containing charged Higgs bosons $H^{\pm}, H_{1,2}^{\pm}$ and exotic down-type quark $D_1$, as well as by neutral CP even(odd) Higgs bosons $H(A)$ with up-type quark $U$ or SM quarks $u,c,t$. 
The Figs. (\ref{Fig_ua_ub_transition}) are plotted using parameters obtained from $b\to s$ studies, enhancing the understanding of these transitions. 
 Comparing the two panels, we observe that almost the entire range of the first three panels satisfies the constraint, whereas the last panel demonstrates a tighter correlation. This indicates that, for this kind of observable, the contribution of WCs associated with charged Higgs bosons is not as strong as those associated with neutral CP even (odd) Higgs bosons. This behavior is opposite to the governing
BR$(\bar{B}\to X_s\gamma)$.  
 
\begin{figure}[H]
	\centering \includegraphics[width=0.51\textwidth]{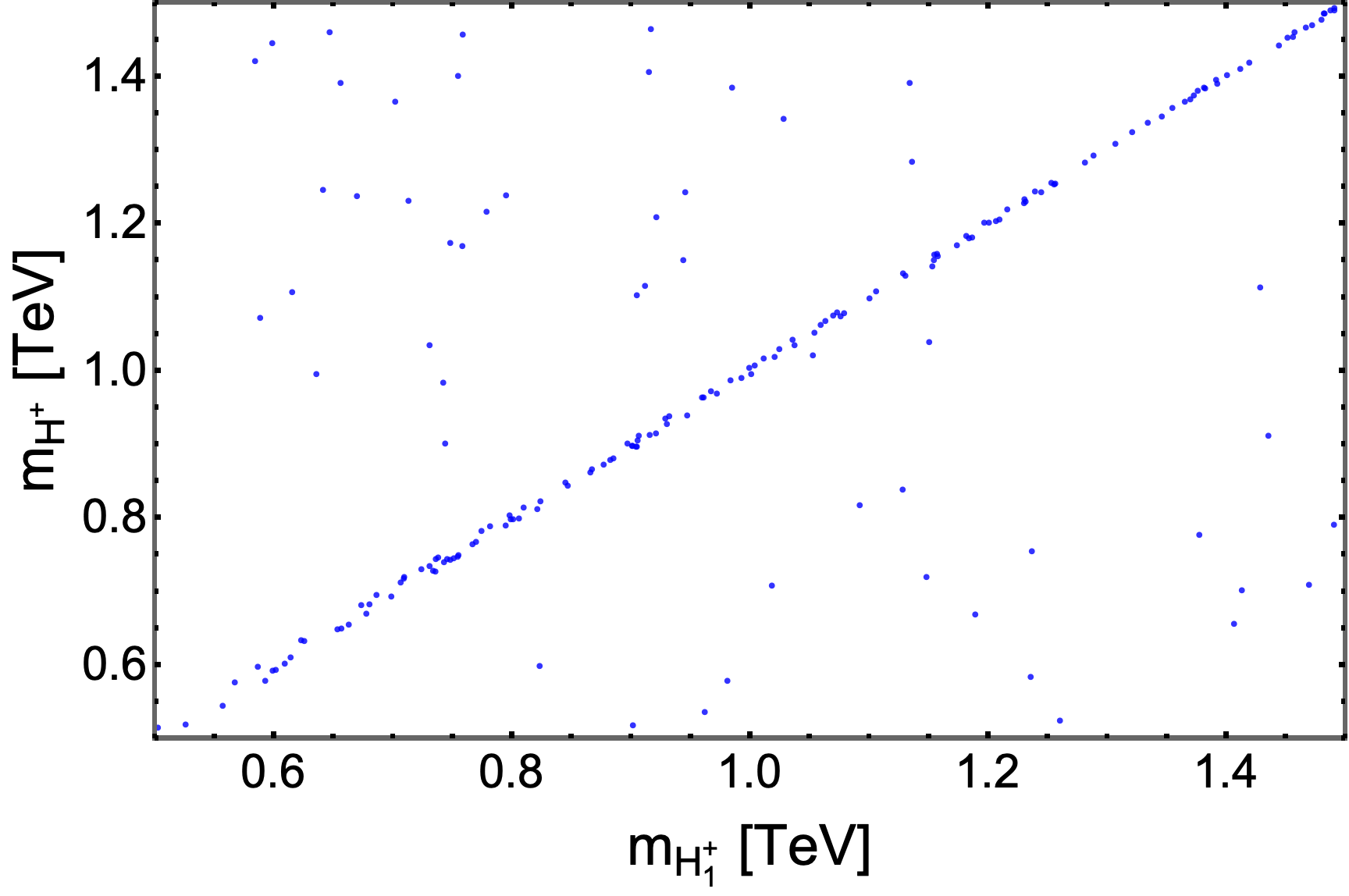}%

	\caption{The figure shows the correlations between mass of charged Higgs bosons $H^{\pm}$,  $H_{1}^{\pm}$ satisfying the constraints of $R_{D^{(*)}}$ in Eq. (\ref{RD_RDs}).}\label{Fig_RD_RDs}
\end{figure}

 Finally, we examine the LFUV ratios $R_{D^{(*)}}$, which are generated at the tree level via the exchange of the SM charged gauge boson $W_{\mu}^{\pm}$ and the charged Higgs bosons  $H^{\pm}, H_{1,2}^{\pm}$. 
 In Fig. (\ref{Fig_RD_RDs}), we plot the correlation between the mass of the charged Higgs
  bosons that satisfy the constraint of of $R_{D^{(*)}}$ in Eq . (\ref{RD_RDs}) with input parameters enhanced by the above studies. The figure shows that satisfied points are dominantly linearly distributed which can be interpreted for the following reasons. Firstly, the ratios are depend on WCs  $\tilde{C}^{(')}_{S}$ induced by charged Higgs $H_{1,2}^{\pm}$ which are proportional $g_{L,R}^{H_{1,2}^+\bar{u}_2d_3}$. In addition, these couplings respectively relate to elements of mixing charged scalar matrix $(U^{\dagger}_c)_{13(14)}$ and  $(U^{\dagger}_c)_{23(24)}\sim \fr{1}{\ep_{1(2)}}\sim \fr{1}{m_{H_{1,2}^{\pm}}^2-m_{H^{\pm}}^2}$ given in Eq. (\ref{Uc_par}). Therefore, the WCs induced by $H_{1,2}^{\pm}$ will be enhanced significantly if there is slight degeneration in charged Higgs masses $m_{H_{1,2}^{\pm}}\sim m_{H^{\pm}}$. This results in almost all points being distributed linearly in Fig . (\ref{Fig_RD_RDs}). For $m_{H_{1,2}^{\pm}}\simeq m_{H^{\pm}}+\mathcal{O}(10^1)$ GeV, and with obtained parameter in above studies, we can estimate the magnitude of these scalar WCs which can attain the maximum value $|\tilde{C}^{(')}_{S}|\sim \mathcal{O}(10^{-4})$, which is much smaller than WC of SM $\tilde{C}^{}_{V}$. This suggests that the charged Higgs boson contributes insignificantly compared to the SM contributions. It is worth mentioning that this scenario of nearly degenerate charged scalar masses is favored by the constraints arising from electroweak precision observables, which is a generic feature of multi-Higgs doublet models \cite{CarcamoHernandez:2015smi}, like the one analyzed in this work.
  
  To close this section, we want to remark that our obtained constraint of charged Higgs masses satisfying $R_{D^{(*)}}$ is different than Refs. \cite{Iguro:2018fni,Iguro:2022uzz,Blanke:2022pjy} for the following reasons. In these works, the charged Higgs interpretation for $R_{D^{(*)}}$ anomalies is indirectly constrained by $\tau \nu$ resonance search at LHC via fast collider simulation $pp\to bc\to \tau \nu$. For instance, the obtained bound of charged Higgs mass is $m_{H^{\pm}}\leq 400$ GeV when the charged resonance $H^+$ decays only to final states $bc/\tau \nu$. However, the considering model provides not only $bc/\tau\nu$ decays by $H^+$ but also more channels due to the existence of couplings of charged Higgs with both SM and exotic fermions, as can be seen in Eqs. (\ref{Lqd},\ref{Lqu}). Particularly, we estimate $\Ga(H^+\to \bar{b}c)\sim  |g_{L}^{H+\bar{u}_2d_3}|^2+|g_{R}^{H+\bar{u}_2d_3}|^2\sim 6.3\times 10^{-11}\ll \Ga(H^+\to \bar{b}t)\sim 3.23\times 10^{-2}$, showing that $bc$ mode is not the dominant one. Moreover, as mentioned above, the ratios $R_{D^{(*)}}$ in our model contain explicitly the dependence on $\tau$ and $\mu$ flavor, as well as there are multi-charged Higgs contributions to $R_{D^{(*)}}$, which are not like \cite{Iguro:2018fni,Iguro:2022uzz,Blanke:2022pjy}. To conclude, the $R_{D^{(*)}}$ in this model is different than previous references, thus the bound for charged Higgs mass in these work is not applied in our case. We also would like to note that since the charged Higgs contributions in our model are shown to be insignificant, the predicted branching ratio BR$(B_c\to \tau\bar{\nu})_{\text{model}}\simeq \text{BR}(B_c\to \tau\bar{\nu})_{\text{SM}}\simeq 2.2 \%$, thus the constraint by BR$(B_c\to \tau\bar{\nu})\leq 63 \%$ \cite{Aebischer:2021ilm} in this model is also relaxed.

	\section{Conclusions \label{Sec6}}
	In this paper, we have investigated in great detail several flavor observables in both the lepton and quark sectors within the context of an extended 2HDM with the inverse seesaw mechanism. We have performed a detailed analysis of the scalar mass spectrum 
	using the perturbation method, based on certain assumptions for the Higgs potential couplings. Additionally, we
	 have obtained the benchmark points fulfilling the SM lepton masses and mixing parameters.Due to the couplings of both Higgs (charged, CP-even and CP-odd) and $Z'$ gauge boson with SM and new fermions at tree-level, the model contains a rich flavor phenomenology in 
lepton and quark sectors. This model has been demonstrated that it may theoretically account for discrepancies between SM and experiment in flavor observables. In particular, we found that the contributions to cLFV decays and anomalous magnetic moments observables, resulting from loop diagrams involving the virtual exchange of CP even (odd) Higgs bosons $H(\mathcal{A})$ and charged exotic leptons,  
	are significantly larger than those arising from the exchange of charged Higgs boson. 
	In addition, the constraints for coupling $f$ and $m_{E_1}$ are derived as $-f\geq 0.04$ and $m_{E_1}\geq 2.2$ TeV. With the obtained bound on $f$, the new physics contributions to the branching ratios of 
	three-body leptonic decays BR$(l\to 3l')$, Mu-$\overline{Mu}$ transition, as well as BR$(\mu \text{Au}\to e \text{Au})$ conversion, are shown to be remarkably smaller than upper experimental limits. On the other hand, the study of quark flavor observables is more complicated. Specifically, for the FCNC $d_a\to d_b$ observables, the WCs induced by charged Higgs bosons $H^{\pm},H_{1,2}^{\pm}$ and exotic quark $U$ exchange contribute the most to BR$(\bar{B_s}\to X_s\ga)$, whereas
	the remaining WCs of $H(\mathcal{A})$ are insignificant. 
	The range of $m_U\geq [1,2]$ TeV is obtained. Furthermore,  
	the meson oscillations are revisited with all contributions and comprehensive quark couplings, combining with studies for BR$(B_s\to \mu^+\mu^-)$ and BR$(B^+\to K^+\tau^+\mu^-)$, we obtain tighter constraints on the $U(1)_X$ coupling $g_X$ 
	and on the $Z'$ gauge boson 
	which read $|g_X|\geq 0.65$ and $m_{Z'}\geq 4$ TeV, respectively. The obtained lower bound of $m_{Z'}$ is consistent with LHC searches of the $Z'$ gauge boson 
\cite{Workman:2022ynf}. 

The numerical values for the up-type quark flavor observables corresponding to 
FCNC decays $t\to u(c)h$ are found to be much lower than their corresponding upper experimental limits \cite{ATLAS:2023ujo}, but several orders of magnitude larger than SM predictions \cite{Eilam:1990zc, Aguilar-Saavedra:2004mfd}.
Concerning radiative decays $t\to u(c)\gamma$,  their obtained branching ratios are primarily influenced by WCs induced by CP even (odd) Higgs bosons interacting with exotic up-type quark $U$ compared to WCs containing charged Higgs bosons. This behavior is opposite to that governing BR$(\bar{B}\to X_s\gamma)$. The FCCC $b\to s$ LFUV ratios $R_{D^{(*)}}$ are calculated to agree with their experimental constraints, with the new physics contributions arising from charged Higgs bosons at the tree-level shown to be negligible.

	The properties of the SM-like Higgs boson $h$ are also discussed. Particularly, the model predicts the deviation factors $a_{h\tau\bar{\tau}}, a_{hb\bar{b}}$ and $a_{ht\bar{t}}$ of the SM Higgs couplings to $\tau\bar{\tau}$, $b\bar{b}$ and $t\bar{t}$ pairs, which are found to agree with their $1\sigma$ 
	experimentally allowed ranges \cite{ATLAS:2022vkf,CMS:2022dwd}. Furthermore, the factor $a_{h\mu\bar{\mu}}$ of the SM Higgs coupling to $\mu\bar{\mu}$ pair is estimated to agree with its corresponding $3\sigma$ bound. 
	This is because the first and second fermion generations acquire  
	masses via the tree and one-loop inverse seesaw mechanisms, respectively. Additionally, the obtained values for the branching ratios for the LFV decays of $h$ such as  $h\to e\mu, h\to e\tau, h\to \mu\tau$, are shown to be consistent 
	with their corresponding experimental upper limits \cite{ATLAS:2023mvd,Workman:2022ynf}.
	
	The model accommodates dark matter due to the presence of the unbroken $Z_2$ symmetry. Assuming that dark matter is represented by a scalar field, which has to be the lightest among the particles having a non trivial charge under the preserved $Z_2$ symmetry, which can be denoted as $\varphi_2$, that scalar dark matter candidate mainly annihilates into $WW$, $ZZ$, $t\overline{t}$, $b\overline{b}$ and $SS$, via a Higgs portal scalar interaction $\lambda_S SS(\varphi_2^{\dagger }\varphi_2)$, where $S$ stands for any of scalars of the model, including the $125$ GeV SM like Higgs boson. Consequently, the constraints arising from the DM relic density and DM direct detection experiments lead to restrictions on the DM mass as well as on its coupling between $\varphi_2$ and the Higgs scalars, including the Standard Model-like Higgs bosons as well as non SM scalars, similarly as, for instance in \cite{Abada:2021yot,Hernandez:2021xet,Bonilla:2023wok}. Thus, the scenario of scalar DM candidate has a good amount of parametric freedom that allows to successfully accommodate the Dark matter relic density while being compatible with the constraints arising from DM direct limits. Based on specific assumptions regarding these scalar coupling bosons, which closely resembles the parameter space used in this study as well as in \cite{Abada:2021yot,Bonilla:2023wok}, we found that aligning with current DM data requires the scalar DM mass to be in the range of a few TeV, as illustrated in Refs. \cite{Abada:2021yot,Hernandez:2021xet,Bonilla:2023wok}. On the other hand, in the scenario of fermionic dark matter (DM) candidate, the DM candidate will corresponds to the lightest among the $Z_2$ odd neutral leptons $\Psi_{1R}$ and $\Psi_{2R}$. Assuming that in that scenario, the DM candidate is $\Psi_{1R}$, it can annihilate into active and heavy neutrinos via the $t$ channel exchange of $Z_2$ odd scalar singlets $\varphi_1$ and $\varphi_2$. Besides that, the fermionic DM candidate can also annihilate into pairs of SM fermions and heavy fermions via the $s$ channel exchange of the heavy $Z'$ gauge boson. Consequently we expect that in the scenario of fermionic DM candidate, the DM constraints will impose limits on the masses of the new gauge boson and the DM itself, similarly as in \cite{Abada:2021yot,Hernandez:2021xet}. A detailed analysis of the Dark matter phenomenology goes beyond the scope of this work.
	
	\section*{\label{acknowledgement}Acknowledgement}
	This research was funded by the Vietnam Academy of Science and Technology, Grant No.CBCLCA.03/25-27. N.T. Duy was funded by the Postdoctoral Scholarship Programme of Vingroup Innovation Foundation (VINIF), code VINIF.2023.STS.65; AECH was funded by Chilean grants ANID-Chile FONDECYT 1210378, FONDECYT 1241855, ANID PIA/APOYO AFB230003, ANID Programa Milenio code ICN2019$\_$044. The authors thank Duong Van Loi for helpful discussions. 
	\
	\appendix
\section{\label{Coefficients_lep_quark} Coefficients in Lagrangian for lepton and quark flavor violating processes}
The coefficients mentioned in Eq. (\ref{lag_lepton}) are given as follows 
\bea 
&& g_{R}^{H^+\bar{\nu}_{aL}l_b}=(U_c^{\dagger})_{22}\sum_{i=1}^3y_i^{(l)}(V^*_{\nu_L})_{ia}(V_{e_R})_{3b}, \hs g_{R}^{H_{1,2}^+\bar{\nu}_{aL}l_b}=(U_c^{\dagger})_{23(24)}\sum_{i=1}^3y_i^{(l)}(V^*_{\nu_L})_{ia}(V_{e_R})_{3b}, \crn  &&
g_R^{H^+\bar{N}_al_b}=(U_c^{\dagger})_{42}\sum_{i=1}^3\sum_{n=1}^2z_{in}^{(l)}(V^*_{N_R})_{ia}(V_{e_R})_{nb},\hs g_R^{H_{1,2}^+\bar{N}_al_b}=(U_c^{\dagger})_{43(44)}\sum_{i=1}^3\sum_{n=1}^2z_{in}^{(l)}(V^*_{N_R})_{ia}(V_{e_R})_{nb},\crn 
&&
g_L^{H^+\bar{\nu}_{aR}l_b}=(U_c^{\dagger})_{22}\sum_{i=1}^3\sum_{j=1}^3 y^{*(\nu)}_{ij}(V^*_{\nu_R})_{ja}(V_{e_L})_{ib}, \hs  g_L^{H_{1,2}^+\bar{\nu}_{aR}l_b}=(U_c^{\dagger})_{23(24)}\sum_{i=1}^3\sum_{j=1}^3 y^{*(\nu)}_{ij}(V^*_{\nu_R})_{ja}(V_{e_L})_{ib},\crn &&
g_{L}^{H_{p}\bar{E}_1l_b}=\fr{1}{\sqrt{2}}\sum_{i=1}^3y_i^{*(E)}(V_{e_L})_{ib}(U_S^{\dagger})_{2p} ,\hs  g_{R}^{H_{p}\bar{E}_1l_b}=\fr{1}{\sqrt{2}}\sum_{n=1}^2x_n^{(E)}(V_{e_R})_{nb}(U_S^{\dagger})_{3p},\crn && g_{L}^{\mathcal{A}_{p}\bar{E}_1l_b}=-\fr{1}{\sqrt{2}}\sum_{i=1}^3y_i^{*(E)}(V_{e_L})_{ib}(U_A^{\dagger})_{2p} ,\hs g_{R}^{\mathcal{A}_{p}\bar{E}_1l_b}=-\fr{1}{\sqrt{2}}\sum_{n=1}^2x_n^{(E)}(V_{e_R})_{nb}(U_A^{\dagger})_{3p} ,\crn &&g_{R}^{H_{p}\bar{l}_al_b}=\fr{1}{\sqrt{2}}\sum_{i=1}^3y_i^{(l)}(V^*_{e_L})_{ia}(V_{e_R})_{3b}(U_S^{\dagger})_{2p}, \hs  g_{R}^{\mathcal{A}_{p}\bar{l}_al_b}=\fr{1}{\sqrt{2}}\sum_{i=1}^3y_i^{(l)}(V^*_{e_L})_{ia}(V_{e_R})_{3b}(U_A^{\dagger})_{2p}.
\eea 
Otherwise, the coefficients used in Eq. (\ref{lag_di_dj}) and Eq. (\ref{lag_ui_uj}) read 
	\bea
&&
g_L^{H^{+}\bar{u}_ad_b}=-(U_c^{\dagger})_{12}\sum_{i=1}^3y_i^{*(u)}(V^*_{u_R})_{ia}(V_{d_L})_{3b}, \hs g_R^{H^{+}\bar{u}_ad_b}=(U_c^{\dagger})_{22}\sum_{i=1}^3y_i^{(d)}(V^*_{u_L})_{3a}(V_{d_R})_{ib}, \crn 
&& g_L^{H_{1(2)}^{+}\bar{u}_ad_b}=-(U_c^{\dagger})_{13(14)}\sum_{i=1}^3y_i^{*(u)}(V^*_{u_R})_{ia}(V_{d_L})_{3b}, \hs g_R^{H_{1(2)}^{+}\bar{u}_ad_b}=(U_c^{\dagger})_{23(24)}\sum_{i=1}^3y_i^{(d)}(V^*_{u_L})_{3a}(V_{d_R})_{ib},\crn
&& g_L^{H^+\bar{U}d_b}=-(U_c^{\dagger})_{22}\sum_{n=1}^2x_n^{*(U)}(V_{d_L})_{nb},\hs g_R^{H^{+}\bar{U}d_b}=(U_c^{\dagger})_{32}\sum_{i=1}^3[w^{(d)}_ki(V_{d_R})_{ib}],\crn 
&&
g_L^{H_{1(2)}^+\bar{U}d_b}=-(U_c^{\dagger})_{23(24)}\sum_{n=1}^2x_n^{*(U)}(V_{d_L})_{nb},\hs g_R^{H_{1(2)}^{+}\bar{U}d_b}=(U_c^{\dagger})_{33(34)}\sum_{i=1}^3[w^{(d)}_ki(V_{d_R})_{ib}],\crn 
&& g_L^{H_p\bar{D}_1d_b}=\fr{1}{\sqrt{2}}\sum_{n=1}^2[x_n^{*(D)}(V_{d_L})_{nb}](U^{\dagger}_S)_{1p},\hs g_R^{H_p\bar{D}_1d_b}=\fr{1}{\sqrt{2}}\sum_{i=1}^3[x_i^{(d)}(V_{d_R})_{ib}](U^{\dagger}_S)_{3p},\crn  
&& g_L^{\mathcal{A}_p\bar{D}_1d_b}=-\fr{1}{\sqrt{2}}\sum_{n=1}^2[x_n^{*(D)}(V_{d_L})_{nb}](U^{\dagger}_A)_{1p},\hs g_R^{\mathcal{A}_p\bar{D}_1d_b}=\fr{1}{\sqrt{2}}\sum_{i=1}^3[x_i^{(d)}(V_{d_R})_{ib}](U^{\dagger}_A)_{3p},\crn 
&& 
g_R^{H_p\bar{d}_ad_b}=\fr{1}{\sqrt{2}}\sum_{i=1}^3[y_i^{(d)}(V^*_{d_L})_{3a}(V_{d_R})_{ib}](U_S^{\dagger})_{2p}, \hs g_R^{\mathcal{A}_p\bar{d}_ad_b}=\fr{1}{\sqrt{2}}\sum_{i=1}^3[y_i^{(d)}(V^*_{d_L})_{3a}(V_{d_R})_{ib}](U_A^{\dagger})_{2p}, \label{gLRd}
\eea 
\bea
&& 
g_L^{H^-\bar{D}_1u_b}=(U_c^{\dagger})_{12}\sum_{n=1}^2x_n^{*(D)}(V_{u_L})_{nb},\hs g_R^{H^{-}\bar{D}_1u_b}=(U_c^{\dagger})_{32}\sum_{i=1}^3[w^{(U)}_i(V_{u_R})_{ib}],\crn 
&&
g_L^{H_{1(2)}^-\bar{D}_1u_b}=(U_c^{\dagger})_{13(14)}\sum_{n=1}^2x_n^{*(D)}(V_{u_L})_{nb},\hs g_R^{H_{1(2)}^{-}\bar{D}_1u_b}=(U_c^{\dagger})_{33(34)}\sum_{i=1}^3[w^{(U)}_i(V_{u_R})_{ib}],\crn  
&& g_L^{H_p\bar{U}u_b}=\fr{1}{\sqrt{2}}\sum_{n=1}^2[x_n^{*(U)}(V_{u_L})_{nb}](U^{\dagger}_S)_{1p},\hs g_R^{H_p\bar{U}u_b}=\fr{1}{\sqrt{2}}\sum_{i=1}^3[x_i^{*(u)}(V_{u_R})_{ib}](U^{\dagger}_S)_{3p},\crn  
&& g_L^{\mathcal{A}_p\bar{U}u_b}=-\fr{1}{\sqrt{2}}\sum_{n=1}^2[x_n^{*(U)}(V_{u_L})_{nb}](U^{\dagger}_A)_{1p},\hs g_R^{\mathcal{A}_p\bar{U}u_b}=-\fr{1}{\sqrt{2}}\sum_{i=1}^3[x_i^{*(u)}(V_{u_R})_{ib}](U^{\dagger}_A)_{3p},\crn 
&& 
g_R^{H_p\bar{u}_au_b}=\fr{1}{\sqrt{2}}\sum_{i=1}^3[y_i^{(u)}(V^*_{u_L})_{3a}(V_{u_R})_{ib}](U_S^{\dagger})_{1p}, \hs g_R^{\mathcal{A}_p\bar{u}_au_b}=\fr{1}{\sqrt{2}}\sum_{i=1}^3[y_i^{(u)}(V^*_{u_L})_{3a}(V_{u_R})_{ib}](U_A^{\dagger})_{1p} \label{gLRu}
\eea  
	\section{\label{Loopfuncs}Loop functions }
	The loop functions used in Eq .(\ref{WCs_quark1}) are given as follow 
	\bea
	&& 
	 f_{\ga}^{W}(x)=\fr{-4x^3+45x^2-33x+10}{12(x-1)^3}-\fr{3x^3}{2(x-1)^4}\ln{x}  \crn	
	&& f_{\gamma}(x)=\fr{(-1+5x+2x^2)}{12(x-1)^3}-\fr{x^2}{2(x-1)^4} \ln{x}, \hs 
	f'_{\gamma}(x)=\fr{-2-5x+x^2}{12(x-1)^3}+\fr{x}{2(x-1)^4} \ln{x}, \crn 
	&& f''_{\gamma}(x)=\fr{-7+5x+8x^2}{36(x-1)^3}+\fr{(2-3x)x}{6(x-1)^4} \ln{x}, \hs 
	f'''_{\gamma}(x)=\fr{5-10x-7x^2}{36(x-1)^3}+\fr{x(3x-1)}{6(x-1)^4} \ln{x}, \crn 
	&&
	h'_{\ga}(x)=\fr{x-3}{2(x-1)^2}+\fr{1}{(x-1)^3}\ln{x} ,\hs h''_{\ga}(x)=\fr{-3+5x}{6(x-1)^2}+\fr{(2-3x)}{3(x-1)^3}\ln{x},\hs h'''_{\ga}(x)=-\fr{2x}{3(x-1)^2}+\fr{3x-1}{3(x-1)^3}\ln{x},\crn 
	&&  f'_g(x)=f''_{g}(x)=\fr{-2-5x+x^2}{12(x-1)^3}+\fr{x}{2(x-1)^4} \ln{x}, \hs 
	h'_g(x)=h''_{g}(x)=\fr{x-3}{2(x-1)^2}+\fr{1}{(x-1)^3}.
	\eea 
	\section{\label{Anomalycheck} Anomaly checking}
In this appendix, we show explicitly how the anomaly cancellation is satisfied with fermion content given in Tables \ref{quarks} and \ref{leptons}. 
\bea 
[SU(3)_C]^2U(1)_X&=&\sum_{\text{quarks}}X_{q_L}-X_{q_R}\crn 
&=&2\times2X_{q_{nL}}+2X_{q_{3L}}+X_{U_L}+X_{D_{1L}}+X_{D_{2L}}-3X_{u_{iR}}-3X_{d_{iR}}-X_{U_R}-X_{D_{1R}}-X_{D_{2R}}\crn 
&=& 2\times 0+2\times 1/3+1/3+0+0-3\times 2/3-3\times (-1/3)-2/3-(-1/3)-(-1/3)=0, \eea 
\bea 
[SU(3)_C]^2U(1)_Y&=&\sum_{\text{quarks}}Y_{q_L}-Y_{q_R}\crn
&=&2\times2Y_{q_{nL}}+2Y_{q_{3L}}+Y_{U_L}+Y_{D_{1L}}+Y_{D_{2L}}-3Y_{u_{iR}}-3Y_{d_{iR}}-Y_{T_R}-Y_{D_{1R}}-Y_{D_{2R}}\crn 
&=& 2\times 2(1/6)+2(1/6)+ 2/3 -1/3-1/3-3(2/3)-3 (-1/3)-2/3-(-1/3)-(-1/3)=0, \eea 
\bea 
[SU(2)_L]^2U(1)_X&=&\sum_{\text{doublets}}X_{f_L} =2\times 3X_{q_{nL}}+3X_{q_{3L}}+3\times X_{l_{iL}}\crn 
&=& 2\times 0+3\times 1/3+3\times -1/3=0, \eea 
\bea 
[SU(2)_L]^2U(1)_Y&=&\sum_{\text{doublets}}Y_{f_L}-X_{q_R}=2\times 3Y_{q_{nL}}+3Y_{q_{3L}}+3Y_{l_{iL}}\crn 
&=& 2\times 3 \times 1/6+3\times 1/6+1/3+3\times -1/2 =0, \eea
\bea 
[\text{Gravity}]^2U(1)_X&\sim&\sum_{\text{fermions}}X_{f_L}-X_{f_R}\crn 
&=& 2\times 3\times 2X_{q_{nL}}+ 3\times 2X_{q_{3L}}+3\times 2X_{l_{iL}}+3X_{U_L}+3X_{D_{1L}}+3X_{D_{2L}}+X_{E_{1L}}-3\times 3X_{u_{iR}}-3\times 3X_{d_{iR}}\crn &&-3X_{U_R}-3X_{D_{1R}}-3X_{D_{2R}}-2X_{l_{nR}}-X_{l_{3R}}-X_{E_{1R}}-3X_{\nu_{iR}}-3X_{N_{iR}}-2X_{\psi_{nR}}-2X_{\Omega_{nR}}\crn 
&=&2\times 3\times 0+3\times2(1/3)+3\times 2(-1/3)+3(1/3)+3(0)+3(0)+1(-1)-3\times 3(2/3)-3\times 3(-1/3)\crn 
&&-2/3-(-1/3)-(-1/3)-2(-1)-(-1)-(-1)-3(1/3)-3(0)-2(1)-2(-1)=0
\eea 
\bea 
[\text{Gravity}]^2U(1)_Y&\sim&\sum_{\text{fermions}}Y_{f_L}-Y_{f_R}\crn 
&=& 2\times 3\times 2Y_{q_{nL}}+ 3\times 2Y_{q_{3L}}+3\times 2Y_{l_{iL}}+3Y_{U_L}+3Y_{D_{1L}}+3Y_{D_{2L}}+Y_{E_{1L}}-3\times 3Y_{u_{iR}}-3\times 3Y_{d_{iR}}\crn &&-3Y_{U_R}-3Y_{D_{1R}}-3Y_{D_{2R}}-2Y_{l_{nR}}-Y_{l_{3R}}-Y_{E_{1R}}-3Y_{\nu_{iR}}-3Y_{N_{iR}}-2Y_{\psi_{nR}}-2Y_{\Omega_{nR}}\crn
&=&2\times 3\times 2(1/6)+3\times 2(1/6)+3\times 2(-1/2)+3(2/3)+3(-1/3)+3(-1/3)+(-1)-3\times 3(2/3)\crn 
&&- 3\times 3(-1/3)-3(2/3)-3(-1/3)-3(-1/3)-2(-1)-(-1)-(-1)\crn &&-3(0)-3(0)-2(0)-2(0)=0
\eea
\bea 
[U(1)_Y]^2U(1)_X&=&\sum_{\text{fermions}}(Y_{f_L}^2X_{f_L}-Y_{f_R}^2X_{f_R})\crn
&=& 3\times 2\times 2Y_{q_{nL}}^2X_{q_{nL}}+3\times 2Y_{q_{3L}}^2X_{q_{3L}}+3Y_{U_{L}}^2X_{U_{L}}+3Y_{D_{1L}}^2X_{D_{1L}}+3Y_{D_{2L}}^2X_{D_{2L}}+3\times 2 Y_{l_{iL}}^2X_{l_{iL}}\crn 
&&+Y_{E_{1L}}^2X_{E_{1L}}-3\times 3Y_{u_{iR}}^2X_{u_{iR}}-3\times 3Y_{d_{iR}}^2X_{d_{iR}}-3Y_{U_{R}}^2X_{U_{R}}-3Y_{D_{1R}}^2X_{D_{1R}}-3Y_{D_{2R}}^2X_{D_{2R}}-2Y_{l_{nR}}^2X_{l_{nR}}\crn &&-Y_{l_{3R}}^2X_{l_{3R}}-Y_{E_{1R}}^2X_{E_{1R}}-3Y_{\nu_{iR}}^2X_{\nu_{iR}}-3Y_{N_{iR}}^2X_{N_{iR}}-2Y_{\psi_{nR}}^2X_{\psi_{nR}}-2Y_{\Omega_{nR}}^2X_{\Omega_{nR}}\crn
&=&3\times 2\times 2(1/6)^2(0)+3\times 2(1/6)^2(1/3)+3(2/3)^2(1/3)+3(-1/3)^2(0)+3(-1/3)^2(0)\crn
&&+3\times 2(-1/2)^2(-1/3)+(-1)^2(-1)-3\times 3(2/3)^2(2/3)-3\times 3(-1/3)^2(-1/3)-3(2/3)^2(2/3)\crn
&&-3(-1/3)^2(-1/3)-3(-1/3)^2(-1/3)-2(-1)^2(-1)-(-1)^2(-1)-(-1)^2(-1)-3(0)^2(1/3)\crn
&&-3(0)^2(0)-2(0)^2(1)-2(0)^2(-1)=0
\eea 
\bea 
[U(1)_X]^2U(1)_Y&=&\sum_{\text{fermions}}(X_{f_L}^2Y_{f_L}-X_{f_R}^2Y_{f_R})\crn
&=& 3\times 2\times 2X_{q_{nL}}^2Y_{q_{nL}}+3\times 2X_{q_{3L}}^2Y_{q_{3L}}+3X_{U_{L}}^2Y_{U_{L}}+3X_{D_{1L}}^2Y_{D_{1L}}+3X_{D_{2L}}^2Y_{D_{2L}}+3\times 2 X_{l_{iL}}^2Y_{l_{iL}}\crn 
&&+X_{E_{1L}}^2Y_{E_{1L}}-3\times 3X_{u_{iR}}^2Y_{u_{iR}}-3\times 3X_{d_{iR}}^2Y_{d_{iR}}-3X_{U_{R}}^2Y_{U_{R}}-3X{D_{1R}}^2Y_{D_{1R}}-3X_{D_{2R}}^2Y_{D_{2R}}-2X_{l_{nR}}^2Y_{l_{nR}}\crn &&-X_{l_{3R}}^2Y_{l_{3R}}-X_{E_{1R}}^2Y_{E_{1R}}-3X_{\nu_{iR}}^2Y_{\nu_{iR}}-3X_{N_{iR}}^2Y_{N_{iR}}-2X_{\psi_{nR}}^2Y_{\psi_{nR}}-2X_{\Omega_{nR}}^2Y_{\Omega_{nR}}\crn
&=&3\times 2\times 2(0)^2(1/6)+3\times 2(1/3)^2(1/6)+3(1/3)^2(2/3)+3(0)^2(-1/3)+3(0)^2(-1/3)\crn
&&+3\times 2(-1/3)^2(-1/2)+(-1)^2(-1)-3\times 3(2/3)^2(2/3)-3\times 3(-1/3)^2(-1/3)-3(2/3)^2(2/3)\crn
&&-3(-1/3)^2(-1/3)-3(-1/3)^2(-1/3)-2(-1)^2(-1)-(-1)^2(-1)-(-1)^2(-1)-3(-1/3)^2(0)\crn
&&-3(0)^2(0)-2(1)^2(0)-2(-1)^2(0)=0
\eea 
\bea 
[U(1)_Y]^3&=&\sum_{\text{fermions}}(Y_{f_L}^3-Y_{f_R}^3)\crn 
&=&3\times 2Y_{q_{3L}}^3+3\times 2\times 2Y_{q_{nL}}^3+3Y_{U_L}^3+3Y_{D_{1L}}^3+3Y_{D_{2L}}^3+3\times 2Y_{l_{iL}}^3+Y_{E_{1L}}^3-3\times 3Y_{u_{iR}}^3-3\times 3Y_{d_{iR}}^3\crn 
&&-3Y_{T_R}^3-3Y_{D_{1R}}^3-3Y_{D_{2R}}^3-2Y_{l_{nR}}^3-Y_{l_{3R}}^3-Y_{E_{1R}}^3-3Y_{\nu_{iR}}^3-3Y_{N_{iR}}^3-2Y_{\psi_{nR}}^3-2Y_{\Omega_{nR}}^3\crn 
&&=3\times 2(1/6)^3+3\times 2\times 2(1/6)^3+3(2/3)^3+3(-1/3)^3+3(-1/3)^3+3\times 2(-1/2)^3+(-1)^3\crn 
&&-3\times 3(2/3)^3-3\times(-1/3)^3-3(2/3)^2-3(-1/3)^3-3(-1/3)^3-2(-1)^3-(-1)^3-(-1)^3\crn 
&&-3(0)^3-3(0)^3-2(0)^2-2(0)^2=0
\eea 
\bea 
[U(1)_X]^3&=&\sum_{\text{fermions}}(X_{f_L}^3-X_{f_R}^3)\crn 
&=&3\times 2X_{q_{3L}}^3+3\times 2\times 2X_{q_{nL}}^3+3X_{U_L}^3+3X_{D_{1L}}^3+3X_{D_{2L}}^3+3\times 2X_{l_{iL}}^3+X_{E_{1L}}^3-3\times 3X_{u_{iR}}^3-3\times 3X_{d_{iR}}^3\crn 
&&-3X_{U_R}^3-3X_{D_{1R}}^3-3X_{D_{2R}}^3-2X_{l_{nR}}^3-X_{l_{3R}}^3-X_{E_{1R}}^3-3X_{\nu_{iR}}^3-3X_{N_{iR}}^3-2X_{\psi_{nR}}^3-2X_{\Omega_{nR}}^3\crn 
&=&3\times 2(0)^3+3\times 2\times 2(1/3)^3+3(1/3)^3+3(0)^3+3(0)^3+3\times 2(-1/3)^3+(-1)^3\crn
&&-3\times 3(2/3)^3-3\times (-1/3)^3-3(2/3)^3-3(-1/3)^3-3(-1/3)^3-2(-1)^3-(-1)^3-(-1)^3\crn
&&-3(1/3)^3-3(0)^3-2(1)^3-2(-1)^3=0
\eea 
Here we ignore the contribution from vector-like fermions $T_{L,R},B_{L,R}$ and $E_{2L,R}$ since they have identical quantum numbers for both chiral parts.
\section{{\label{Z couplings}}The couplings of neutral bosons, $Z_\mu, Z_\mu^\prime$, to SM fermions $\bar{f}_a-f_b$}
Interactions of neutral currents with fermions are written in the mass eigenstates have the following form
\bea
\mathcal{L}^{\text{NC}}= \bar{f}_a \ga^\mu \left\{ g_{V}^{Z,(Z^\prime)}\left( f_af_b\right) -g_{A}^{Z,(Z^\prime)}\left( f_af_b\right) \ga_5 \right\}f_b Z_\mu (Z^{\prime}_\mu),
\eea
where the expressions of $g_{V}^{Z}\left( f_af_b\right)$; $g_{A}^{Z}\left( f_af_b\right) $ are given in Table \ref{coupling}.\\
\begin{table}[h]
	\centering
	\begin{tabular}{|c|c|c|}
		\hline
	$\bar{f}_af_b$ &$12 \frac{g_{V}^Z\left( f_af_b\right)}{g}$& $ -12 \frac{g_{A}^Z\left( f_a f_b\right)}{g}$\\
		\hline
	$\bar{\nu}_{a}\nu_{b}$ & $ \left(3 \frac{c_{\theta_{ZZ^\prime}}}{c_W}+2t_X  s_{\theta_{ZZ^\prime}}\right) \delta_{ab}$ & $ \left(3 \frac{c_{\theta_{ZZ^\prime}}}{c_W}+2t_X  s_{\theta_{ZZ^\prime}}\right) \delta_{ab}$  \\
		\hline
	$\bar{e}_ae_b$& $\left( -3c_{\theta_{ZZ^\prime}}c_W+9c_{\theta_{ZZ^\prime}}s_Wt_W+8t_X s_{\theta_{ZZ^\prime}}\right)\delta_{ab}$ & $\left(3\frac{c_{\theta_{ZZ^\prime}}}{ c_W}+4t_X s_{\theta_{ZZ^\prime}}\right)\delta_{ab}$ \\
		\hline  $\bar{u}_au_{b}$& $\delta_{ab}\left(-1+4 c_{2 \theta_W} \right) \frac{c_{\theta_{ZZ^\prime}}}{c_W} -4\delta_{ab}t_X s_{\theta_{ZZ^\prime}}-2t_X s_{\theta_{ZZ^\prime}}\left( V_{u_L}^*\right)_{3a}\left( V_{u_L}\right)_{3b}$ & $  \left(3\frac{c_{\theta_{ZZ^\prime}}}{c_W} +4t_X s_{\theta_{ZZ^\prime}} \right) \delta_{ab}+2t_X s_{\theta_{ZZ^\prime}}\left( V_{u_L}^*\right)_{3a}\left( V_{u_L}\right)_{3b} $\\ \hline
		$\bar{d}_ad_b$ &  $-\delta_{ab}(1 + 2 c_{2 \theta_W})  \frac{c_{\theta_{ZZ^\prime}}}{c_W} - 2 \delta_{ab}t_X s_{\theta_{ZZ^\prime}}-2t_X  s_{\theta_{ZZ^\prime}}\left( V_{d_L}^*\right)_{3a}\left( V_{d_L}\right)_{3b} $ & $  \left(3  \frac{c_{\theta_{ZZ^\prime}}}{c_W} + 2 t _X s_{\theta_{ZZ^\prime}} \right)\delta_{ab}+2t_X  s_{\theta_{ZZ^\prime}}\left( V_{d_L}^*\right)_{3a}\left( V_{d_L}\right)_{3b} $  \\ \hline
	\end{tabular}
	\caption{ The couplings of $Z_\mu-$ bosons to SM fermions}
	\label{coupling}\end{table}
Here, we define the following notations: $c_{W} = \cos \theta_W$, $s_{W} = \sin \theta_W$, $t_W = \frac{g'}{g}$, $t_X = \frac{g_X}{g}$, $c_{\theta_{ZZ'}} = \cos \theta_{ZZ'}$ and $s_{\theta_{ZZ'}} = \sin \theta_{ZZ'}$. The mixing angle $\theta_{ZZ'}$ is determined by the following equation:
\bea 
\tan \theta_{ZZ'} = \frac{12 \sqrt{1 + t_X^2} t_X (v_1^2 + 2 v_2^2)}{(9 + 9 t_X^2 - 4 t_X^2) v_1^2 + (9 + 9 t_X^2 - 16 t_X^2) v_2^2 - 4 t_X^2 \Lambda_{\text{new}}^2},
\eea 
where $\Lambda_{\text{new}}^2 = v_\sigma^2 + 4 v_\chi^2 + 9 v_\eta^2 + 36 v_\rho^2 + 36 v_\kappa^2$.
The vector coupling of new gauge boson $Z'$  $g_{V}^{Z^\prime}\left( f_af_b\right)$ is related to $g_{V}^Z\left( f_af_b\right)$ by a rotation in the $Z$-$Z^\prime$ basis, specifically through the transformation $s_{\theta_{ZZ^\prime}}  \to - c_{\theta_{ZZ^\prime}} $ and $c_{\theta_{ZZ^\prime}}  \to s_{\theta_{ZZ^\prime}} $. The axial-vector coupling $g_{A}^{Z^\prime}\left( f_if_j\right)$ undergoes a similar rotation, with the substitutions $s_{\theta_{ZZ^\prime}}  \to c_{\theta_{ZZ^\prime}} $ and $c_{\theta_{ZZ^\prime}} \to -s_{\theta_{ZZ^\prime}}$.
\newpage 
	\centerline{\bf{REFERENCES}}\vspace{-0.4cm} 
	\bibliographystyle{utphys}
	\bibliography{combine}
	
\end{document}